\documentclass[10pt,journal,twocolumns]{IEEEtran}
\IEEEoverridecommandlockouts
\usepackage{cite}
\usepackage[cmex10]{amsmath}
\usepackage{amsthm}
\usepackage{amssymb}
\usepackage{winsnotation}
\usepackage[bookmarks,colorlinks]{hyperref} 
\hypersetup{colorlinks,citecolor= red,filecolor= blue,linkcolor= blue,urlcolor=blue}
\usepackage{acronym}  
\usepackage{psfrag}
\usepackage{color}  
\usepackage[usenames,dvipsnames]{xcolor}
\usepackage{epstopdf}
\usepackage{algorithm}
\usepackage{algpseudocode}
\usepackage[utf8]{inputenc}
\usepackage[english]{babel}
\usepackage{physics}
\usepackage[percent]{overpic}
\usepackage{tikz}
\usepackage{pgfplots}
\usepackage{pstricks}
\usepackage[free-standing-units]{siunitx}
\usepackage{circuitikz}
\usepackage{float}
\usepackage{schemabloc}
\usepackage{mathtools}
\usepackage{subfig}
\usepackage{multirow}
\usepackage{accents}
\usepackage[font=small]{caption}
\usepackage{dblfloatfix}    
\usepackage{xcolor}
\usepackage{pict2e}
\usepackage{listings}
\usetikzlibrary{matrix}
\usetikzlibrary{circuits}
\usetikzlibrary{arrows,positioning}
\usepackage[font={small}]{caption}
\usepackage{amsmath}
\usepackage{wgroup_switch,graphicx,amsmath,amssymb}

\definecolor{0}{HTML}{54FF00}
\definecolor{1}{HTML}{FFFFFF}
\definecolor{2}{HTML}{FF0000}
\definecolor{3}{HTML}{0048FF}
\definecolor{4}{HTML}{EA00FF}
\definecolor{5}{HTML}{DC00F0}
\definecolor{6}{HTML}{FFFF00}
\definecolor{7}{HTML}{1E90FF}
\definecolor{8}{HTML}{FF1493}
\definecolor{9}{HTML}{00FFFF}
\definecolor{10}{HTML}{C0C0C0}

\usepackage[font=small,labelfont=bf]{caption}
\def\BibTeX{{\rm B\kern-.05em{\sc i\kern-.025em b}\kern-.08em
    T\kern-.1667em\lower.7ex\hbox{E}\kern-.125emX}}

\usetikzlibrary{shapes,arrows}
\usetikzlibrary{shapes.misc}
\tikzstyle{line} = [draw, -latex']

\tikzstyle{block} = [draw, fill=white, rectangle,
    minimum height=4.5em, minimum width=6em]
\tikzstyle{sum} = [draw, fill=white, circle, node distance=1cm]
\tikzstyle{input} = [coordinate]
\tikzstyle{output} = [coordinate]
\tikzstyle{pinstyle} = [pin edge={to-,thin,black}]

\tikzset{%
  wireless/.pic={
      \draw [-] (0,0) -| (.5,#1);
      \foreach \r in {}
      \draw (.6,#1) ++ (60:\r) arc (60:-60:\r);
  },
  vdots/.pic={
    \foreach \i in {-.1,0,.1}
      \fill (.25,\i) circle [radius=.75pt];
  },
  block/.style={
    shape=rectangle,
    minimum width=6em,
    minimum height=4em,
    draw
  },
  RF chain/.style 2 args={
    block,
    node contents=RF chain,
    append after command={
      \pgfextra{\pgfnodealias{@}{\tikzlastnode}}
      (@.north #1) [yshift=-.36cm] pic [#2] {wireless=.83}
    }
  },
  MIMO RF chain east/.style={RF chain={east}{xscale=1}},
  MIMO RF chain west/.style={RF chain={west}{xscale=-1}},
}

\newcommand{\bd}{\begin{description}}
\newcommand{\ed}{\end{description}}
\newcommand{\be}{\begin{enumerate}}
\newcommand{\ee}{\end{enumerate}}
\newcommand{\bi}{\begin{itemize}}
\newcommand{\ei}{\end{itemize}}
\newcommand{\bl}{\begin{list}}
\newcommand{\el}{\end{list}}
\newcommand{\bt}{\begin{tabbing}}
\newcommand{\et}{\end{tabbing}}

\definecolor{BLUE}{rgb}{0,0,1}

\usepackage{acronym}  
\acrodef{pdf}[PDF]{probability density function}
\acrodef{amp}[AMP]{approximate message passing}
\acrodef{csi}[CSI]{channel state information}
\acrodef{ga}[GA]{genetic algorithm}
\acrodef{bs}[BS]{base station}
\acrodef{iot}[IoT]{Internet-of-Things}
\acrodef{cdma}[CDMA]{code division multiple access}
\acrodef{bs}[BS]{base station}
\acrodef{bpsk}[BPSK]{binary phase-shift-keying}
\acrodef{dsss}[DS-SS]{direct-sequence spread spectrum}
\acrodef{sssr}[SSSR]{simultaneous sparse signal reconstruction}
\acrodef{tdd}[TDD]{time-delay diversity}
\acrodef{tddssa}[TDD-SSA]{time-delay diversity sparse signal approximation}
\acrodef{kkt}[KKT]{Karush-Kuhn-Tucker}
\acrodef{map}[MAP]{maximum a posteriori probability}
\acrodef{somp}[S-OMP]{Simultaneous Orthogonal Matching Pursuits}
\acrodef{ls}[LS]{least square}
\acrodef{mud}[MUD]{multiuser detection}
\acrodef{mse}[MSE]{mean squared error}
\acrodef{ber}[BER]{bit error rate}
\acrodef{wcss}[WCSS]{within-cluster sum of squares}
\acrodef{sb}[SB]{Symbol-Based}
\acrodef{rd}[RD]{ridge regression}
\acrodef{ssr}[SSR]{sparse signal reconstruction}
\acrodef{iid}[i.i.d.]{independent and identically distributed}
\acrodef{ls}[LS]{least-squares}
\acrodef{mse}[MSE]{mean-squared error}
\acrodef{2mc}[2-MC]{2-mean clustering}
\acrodef{sae}[SAe]{sparsity-aware}
\acrodef{pb}[PB]{Packet-Based}
\acrodef{pmf}[PMF]{probability mass function}
\acrodef{cv}[CV]{cross-validation}
\acrodef{mmse}[MMSE]{minimum mean squared error}
\acrodef{snr}[SNR]{signal-to-noise ratio}
\acrodef{cvpl}[CVPL]{cross-validated partial likelihood}
\acrodef{sdl}[SDL]{sparse dictionary learning}
\acrodef{mod}[MOD]{method of optimal directions}
\acrodef{stls}[S-TLS]{sparse-total least square}
\acrodef{cv}[CV]{cross validation}
\acrodef{mlr}[MLR]{maximum likelihood ratio}
\acrodef{gcv}[GCV]{generalized cross validation}
\acrodef{cdf}[CDF]{cumulative distribution function}
\acrodef{pls}[P-LS]{penalized-LS}
\acrodef{ldpc}[LDPC]{low-density parity-check}
\acrodef{dc}[DC]{Decision Combining}
\acrodef{ec}[EC]{Estimate Combining}
\acrodef{roc}[ROC]{receiver operating characteristic}
\acrodef{mtc}[MTC]{machine-type communications}
\acrodef{ma}[MA]{multiple access}
\acrodef{mac}[MAC]{media access control}
\acrodef{phy}[PHY]{physical}
\acrodef{ra}[RA]{random access}
\acrodef{fcc}[CSI]{channel state information}
\acrodef{cfo}[CFO]{carrier frequency offset}
\acrodef{ble}[BLE]{Bluetooth low-energy}
\acrodef{rfid}[RFID]{radio frequency identification}
\acrodef{csma}[CSMA]{carrier sensing multiple access}
\acrodef{ca}[CA]{collision avoidance}
\acrodef{lte}[LTE]{Long-Term Evolution}
\acrodef{rfid}[RFID]{radio frequency identification}
\acrodef{fsa}[FSA]{frame slotted ALOHA }
\acrodef{lpwa}[LPWA]{low-power wide area}
\acrodef{rftdm}[R-FTDM]{random frequency-time division multiplexing}
\acrodef{rpma}[RPMA]{random phase multiple access}
\acrodef{cdma}[CDMA]{code division multiple access}
\acrodef{lbt}[LBT]{listen-before-talk}
\acrodef{ss}[SS]{spread spectrum}
\acrodef{bch}[BCH]{Bose–Chaudhuri–Hocquenghem}
\acrodef{ap}[AP]{access point}
\acrodef{3gpp}[3GPP]{3rd Generation Partnership Project}
\acrodef{nb}[NB]{narrowband}
\acrodef{fdma}[FDMA]{frequency division multiple access}
\acrodef{rach}[RACH]{Random Access Channel}
\acrodef{fa}[FA]{fixed assignment}
\acrodef{tdma}[TDMA]{time-division multiple access}
\acrodef{ds}[DS]{direct sequence}
\acrodef{dsa}[DSA]{device sparsity-aware}
\acrodef{pdsa}[PDSA]{packet-device sparsity-aware}
\acrodef{pts}[PTS]{packet transmission state}
\acrodef{ml}[ML]{maximum likelihood}
\acrodef{cb}[CB]{contention-based}
\acrodef{css}[CSS]{chirp spread spectrum}
\acrodef{mf}[MF]{matched filter}
\acrodef{ltem}[LTE-M]{Long Term Evolution for Machines}
\acrodef{cp}[CP]{carrier phase}
\acrodef{per}[PER]{packet error rate}
\acrodef{dcd}[DCD]{differentially coherent decorrelation}
\acrodef{ca}[CA]{code-aided}
\acrodef{bic}[BIC]{Bayesian information criterion}
\acrodef{sic}[SIC]{successive interference cancellation}
\acrodef{pic}[PIC]{parallel interference cancellation}
\acrodef{mmtc}[mMTC]{massive machine-type communications}
\acrodef{gf}[GF]{grant-free}
\acrodef{rg}[RG]{request-grant}
\acrodef{htc}[HTC]{human-type communication}
\acrodef{mimo}[MIMO]{multiple-input and multiple-output}

\makeatletter
\def\BState{\State\hskip-\ALG@thistlm}
\makeatother

\DeclareMathAlphabet{\pazocal}{OMS}{zplm}{m}{n}

\newtheorem{innercustomgeneric}{\customgenericname}
\providecommand{\customgenericname}{}
\newcommand{\newcustomtheorem}[2]{%
  \newenvironment{#1}[1]
  {%
   \renewcommand\customgenericname{#2}%
   \renewcommand\theinnercustomgeneric{##1}%
   \innercustomgeneric
  }
  {\endinnercustomgeneric}
}

\newcustomtheorem{customthm}{Theorem}
\newcustomtheorem{customlemma}{Lemma}

\usepackage[yyyymmdd,hhmmss]{datetime} 
\newdateformat{monthyeardate}{\monthname[\THEMONTH] \THEDAY, \THEYEAR} 

\begin{document}

\renewcommand{\figurename}{Fig.}
{
	\twocolumn
	
	

	
\title{\vspace{-0.2em} Massive Uncoordinated Multiple Access
\\ for Beyond 5G}


\author{
\vspace{-0.5em}
    Mostafa~Mohammadkarimi,~\IEEEmembership{Member,~IEEE},
    Octavia~A.~Dobre,~\IEEEmembership{Fellow,~IEEE}, \\ and
    Moe~Z.~Win,~\IEEEmembership{Fellow,~IEEE}
    \vspace{-2em}
	\thanks{This research was
supported by the Natural Sciences and Engineering Research Council
of Canada (NSERC) through its Discovery program. }
    \thanks{
        M.~Mohammadkarimi is with the Faculty of Electrical Engineering, Mathematics and Computer Science, Delft University of Technology, Delft, Netherlands
        (e-mail: \texttt{m.mohammadkarimi@tudelft.nl}).

        O.~A. Dobre is with the
        Faculty of Engineering and Applied Science, Memorial University, St. John's, NL, Canada,
        (e-mail: \texttt{odobre@mun.ca}).

         M.~Z.~Win is with the Laboratory for Information and Decision
        Systems (LIDS), Massachusetts Institute of Technology, Boston, MA,
        USA (e-mail: \texttt{moewin@mit.edu}).
	}
\thanks{This paper published in IEEE Transactions on Wireless Communications (DOI: 10.1109/TWC.2021.3117256).}

%
}
\maketitle
	
	\setcounter{page}{1}
\begin{abstract}
Existing wireless communication systems have been mainly designed to provide substantial gain in terms of data rates. However, 5G and Beyond will depart from this
scheme, with the objective not only to provide services with higher data rates.
One of the main goals is to support \ac{mmtc} in the \ac{iot} applications.
Supporting
massive uplink communications for devices with sporadic traffic pattern and short-packet size, as it is in many \ac{mmtc} use cases, is a challenging task, particularly when the control signaling is not negligible in size compared to the payload.
In addition, channel estimation becomes challenging for sporadic and short-packet transmission due to the limited number of employed pilots.
In this paper, a new uplink \ac{ma} scheme is proposed for \ac{mmtc}, which can support a large number of uncoordinated \ac{iot} devices with short-packet and sporadic traffic.
The proposed uplink \ac{ma} scheme removes the overheads associated with the device identifier as well as pilots and preambles related to channel
estimation. An alternative mechanism for device identification (DI) is employed, where
a unique spreading code is dedicated to each \ac{iot} device as identifier. This unique code is simultaneously used for the spreading purpose and DI.
Two \ac{iot} DI algorithms which employ sparse signal reconstruction methods are proposed to determine the active \ac{iot} devices prior to data detection.
Specifically, the Bayesian information criterion model order selection method  is employed to develop an \ac{iot} DI algorithm for unknown and time-varying activity rate.
Our proposed \ac{ma} scheme benefits from
a new non-coherent nonlinear multiuser detection algorithm designed on the basis of unsupervised machine learning techniques to enable data detection without {\it {a priori}} knowledge on channel state information.
For performance improvement, an extension to multiple receive antennas through hard decision combining is proposed.
The effectiveness of the proposed \ac{ma} scheme for known and unknown activity rate and high overloading factor is supported by simulation results.
\end{abstract}

\vspace{-0.2em}	

\acresetall		
\vspace{-1em}	
\section{Introduction}\label{sec:intro}
\IEEEPARstart{M} {Assive uplink} connectivity is the key factor in the realization of the \ac{iot}, as part of 5G and Beyond wireless communication systems \cite{bockelmann2016massive}.
In many \ac{iot} applications, \ac{mmtc} services are required, where  a large number of devices transmit very short packets sporadically. Typically, the number of \ac{iot} devices assigned to each \ac{bs} in \ac{mmtc} is in orders
of magnitude above what current communication networks
are capable to support. Moreover, \ac{iot} devices do not transmit continuously, rather updates are infrequently transmitted to the \ac{bs}, whenever a measured value changes. Hence, small packets are expected to carry critical payload in \ac{mmtc} \cite{zhong20195g}.

The design of the current wireless communication systems relies on the assumption that the
control signaling related to \ac{phy} and \ac{mac} layers is of negligible size compared
to the payload.
Thus, heuristic design of control signaling is acceptable and does not affect the overall system
performance. However, in \ac{mmtc} with short-packet transmission, the control signaling can be similar in size with the payload; thus, inefficient design of control signaling
leads to highly suboptimal transmission schemes.
Excessive control signaling, e.g., the overheads, preambles, and pilots associated with device identifier, exploited for channel estimation, and used for
random access procedure,
hinders massive connectivity
\cite{durisi2016toward}.
Thus, efficient \ac{ma} schemes with highly limited (or non-existent) control signaling are required. 

Moreover, channel estimation is another challenge for sporadic and short-packet transmission, especially for a massive number of non-orthogonal transmissions. Existing channel estimation approaches are often based on the assumption that devices are active over long periods so that channel estimation through pilots and preambles is feasible. However, if an \ac{iot} device only transmits occasionally, such an assumption cannot longer be valid. Instead, channel estimation has to rely on a single transmission that may be very short, which constrains the number of orthogonal pilots \cite{zanella2013m2m}.
 Channel estimation becomes more challenging in the grant-free uplink \ac{ma} scheme, where resources are randomly selected by devices \cite{mohammadkarimi2018signature}.

Motivated by these facts, a new uplink \ac{ma} scheme for short-packet and sporadic traffic in \ac{mmtc} is proposed in this paper.
The main idea behind the proposed \ac{ma} scheme is to reduce the control signaling while simultaneously supporting a massive number of uncoordinated \ac{iot} devices with a single \ac{bs}.
The proposed \ac{ma} scheme is designed based on asynchronous \ac{dsss} with non-orthogonal spreading codebook, and is
capable of supporting undetermined \ac{dsss} systems
in static networks, where the \ac{bs} and \ac{iot} devices are immobile.

To remove the control signaling associated with the \ac{iot} device identifier,
a unique spreading code is dedicated to each \ac{iot} device which is simultaneously used for the spreading purpose and device identification (DI).
In a nutshell, instead of allocating a fragment of the \ac{iot} packet to the signaling associated with the \ac{mac} address (device identifier), the unique spreading code is used as \ac{iot} device identifier.
Our \ac{ma} scheme also relies on an unsupervised machine learning technique to enable non-coherent data detection, thus removing the need of preambles and pilots used for channel estimation. The lack of preambles and pilots further reduces the control signaling.

Our proposed approach for removing the device identifier relies on sparsity-aware \ac{iot} DI at the \ac{bs} to determine the active \ac{iot} devices before data detection. Based on the sporadic traffic pattern of the \ac{iot} devices as well as lack of knowledge about the \ac{csi} of the \ac{iot} devices, the squared $\ell_2$-norm \ac{ssr} and \ac{bic} $\ell_1-\ell_2$  mixed-norm \ac{sssr} \ac{iot} DI algorithms are developed. In the former algorithm, the \ac{iot} identification problem is formulated as an \ac{ssr} using the generalized cross-validation (GCV) approach followed by parallel hypothesis testing. The latter algorithm formulates the \ac{iot} DI  problem as a \ac{bic} model order selection \ac{sssr} problem.

The proposed uplink \ac{ma} scheme is also equipped with a new non-coherent nonlinear \ac{mud} algorithm to detect data of the
active \ac{iot} devices, applied after the \ac{iot} DI algorithms.
We propose the non-coherent \ac{2mc}-\ac{mud} algorithm based on \ac{2mc} unsupervised machine learning and differential coding to detect data without channel estimation at the \ac{bs}.

\vspace{-0.8em}
\subsection{Related Works}
\ac{iot} device activity and data detection techniques can be categorized into three groups: 1) regularized, 2) greedy, and 3) iterative-thresholding based methods \cite{di2020detection}.
The regularized methods apply a regularization parameter into the cost
function which balances both approximation error and sparsity level of the solution.
Sparse maximum a posteriori probability
(S-MAP) and its relaxed versions were the first algorithms for activity and data detection that took
 into account regularization parameter \cite{zhu2011exploiting}.
To achieve an acceptable performance
with lower computational complexity compared to the optimal S-MAP detector,
several detectors were then proposed, such as sparsity-aware successive interference cancellation (SA-SIC)
\cite{knoop2013sparsity}, SA-SIC with sorted QR decomposition  \cite{ahn2018sparsity}, activity-aware multiple feedback SIC \cite{di2019activity},
activity-aware recursive least squares with decision feedback \cite{di2019adaptive},
and direction method of multipliers
(ADMM)\cite{cirik2017multi}.

In the greedy techniques, the goal is to solve the sparse representation with the $\ell_0$-norm minimization. Because of the fact that this problem is NP-hard, the greedy technique provides an approximate solution to alleviate this difficulty.  The  greedy  strategy  searches  for the best local optimal solution  in  each  iteration with  the  goal of achieving  the  optimal  holistic solution.
The pioneering work in this category  applied
the  orthogonal least squares and orthogonal  matching  pursuit (OMP) algorithms to perform  joint  detection  of  data and  device  activity
for \ac{mmtc} \cite{schepker2011sparse}. Other detection algorithms
in this category are:
group OMP (GOMP) \cite{schepker2012compressive},
compressive
sample matching pursuit, detection-based OMP \cite{xiong2014sparse},
weighted GOMP
\cite{schepker2013improving}, detecting-based GOMP,
block-correlation SIC  \cite{schepker2015efficient}, simultaneous OMP
with extrinsic information transfer, and the threshold aided block
sparsity adaptive subspace pursuit \cite{du2018block}.

Iterative-thresholding based methods are alternatives to convex optimization for large-scale problems.
These methods are inspired by belief propagation (BP) in graphical models, where
their
foundation is Gaussian loopy BP with
simplified message passing that assumes high dimensional
signal vectors in order to factorize  a multivariate distribution. There are several approaches in this category, such as }
\ac{amp} with non-separable denoiser \cite{tang2020device}, vector \ac{amp} with MMSE denoiser \cite{liu2018massive}, bilinear generalized AMP
 \cite{ding2019sparsity},
joint expected maximization and \ac{amp} \cite{wei2016approximate}, and mixture of compressive sensing and message passing \cite{wang2015compressive}.
A blind BP detection algorithm for non-coherent non-orthogonal MA with massive receive antennas was also proposed in \cite{wang2019non}.

Most of the mentioned approaches either assume perfect \ac{csi} or its estimate through pilot at the \ac{bs}.
While using pilot results in lower spectral efficiency and higher latency in these approaches compared to our non-coherent method, some of these approaches,  such as \cite{liu2018massive}, can offer lower \ac{per} when a sufficient
number of pilots is used for joint activity detection and channel estimation. On the other hand, the single phase joint activity detection, channel estimation, and data detection
approaches without requiring  pilot, such as
the structured  sparsity learning MUD algorithm in \cite{ding2019sparsity}, exhibit higher \ac{per} compared to our non-coherent method due to lack of spreading.
Moreover, unlike most of the above-mentioned approaches, which considered coordinated uplink MA with perfectly synchronized transmissions, our proposed MA scheme supports uncoordinated \ac{iot} devices. It is worth mentioning that the algorithm in \cite{ding2019sparsity} is very promising for symbol-level synchronized transmission, and \cite{liu2018massive} can analytically
characterize channel estimation error.
Besides, existing regularized algorithms do not propose any solution to set the tuning parameter in the optimization problem when the activity rate is unknown and time-varying. The degree of sparsity, and thus, the false alarm and correct identification rates depend on the value of the tuning parameter.

\subsection{Contributions}
The main contributions of this work are as follows:
\begin{itemize}
\item {A new uplink \ac{ma} scheme is proposed for \ac{mmtc}.
 The proposed \ac{ma} scheme exhibits the following advantages:}
\begin{itemize}
It is capable to support thousands of uncoordinated IoT devices;
\item It supports sporadic traffic pattern and short-packet;
\item {It significantly reduces packet time on-air since it is designed for underdetermined \ac{dsss} (number of devices is larger than the spreading factor);}
\item It removes the control signaling associated with the device identifier
as
well as pilots and preambles employed for channel estimation to reduce uplink overhead;

\item{It exhibits high scalability in terms of adding new \ac{iot} devices (high overloading factor) without
 negatively affecting the system performance.
}
\end{itemize}
\item {A new mechanism for the \ac{iot} DI at the \ac{bs} is developed instead of using device identifier. Since the active \ac{iot} devices in the network do not use a device identifier in order to identify themselves to the \ac{bs}, the squared $\ell_2$-norm \ac{ssr} and the \ac{bic} $\ell_1-\ell_2$  mixed-norm \ac{sssr} \ac{iot} identification algorithms are proposed to detect active \ac{iot} devices. The proposed algorithms exhibit the following advantages:}
    \begin{itemize}
\item{They can detect active \ac{iot} devices without knowledge of the \ac{csi};}
\item {They remove the need for matched-filter (MF) implementation for all spreading codes; thus reducing the complexity of the receiver;}
\item {{The \ac{bic} $\ell_1-\ell_2$ mixed-norm \ac{sssr} algorithm can identify active \ac{iot} devices when the activity rate is unknown and time-varying;}}
\item { {They take the advantage of optimal tuning parameter;}}
\item {{The \ac{bic} $\ell_1-\ell_2$ mixed-norm \ac{sssr} algorithm can identify active \ac{iot} devices for  non-identical activity rate due to the \ac{bic} model selection;}}
    \item {{There is control over the false alarm and correct identification rates of the individual \ac{iot} devices in the $\ell_2$-norm \ac{ssr} \ac{iot} DI algorithm.}}
\end{itemize}
\item{{The statistical performance analysis of the squared $\ell_2$-norm \ac{ssr} \ac{iot} DI algorithm is
       presented, and theoretical expressions for the
       correct identification and false alarm rates are derived.}}
\item {A new non-coherent nonlinear \ac{mud} algorithm, i.e., \ac{2mc}-\ac{mud} in combination with differential coding is designed for short packet transmission. The proposed \ac{2mc}-\ac{mud} algorithm exhibits the following advantages: }
\begin{itemize}
\item {It supports both coordinated and uncoordinated \ac{dsss} transmission irrespective of the traffic pattern;}
\item{It does not require knowledge of the CSI at the BS.}
\end{itemize}
\item
An extension to multiple receive antennas through hard decision combining is proposed. This combination offers the following advantages:
\begin{itemize}
\item
The performance of the proposed uplink \ac{ma} boosts because of spatial diversity;
\item
Higher overloading factor can be supported.
\end{itemize}
\end{itemize}

\subsection{Notations}

The identity matrix and zero vector are shown by $\bf{I}$ and ${\bf{0}}$, and the indicator function is defined as $\mathbb{I}\big{\{}x\big{\}}=1$ $\text{if}\ x \ \text{is true}$; otherwise, $\mathbb{I}\big{\{}x\big{\}}=0$.
The cardinality of a set, which measures the number of elements of the set, is denoted by $\mathbf{card}(\cdot)$.
The $\ell_0$ quasi-norm of vector ${\bf{a}}_j= [a_{0,j} \ a_{1,j}  \ \dots \ a_{m-1,j}]^\dag $ and the $\ell_0-\ell_0$ quasi-norm of matrix ${\bf{A}}\triangleq [{\bf{a}}_0 \ {\bf{a}}_1 \ \dots \ {\bf{a}}_{n-1}]$ are respectively defined as
${\|}{\bf{a}}_j{\|}_0\triangleq \mathbf{card} \big{(}\big{\{}i \in \mathcal{I} \big{|}{a}_{i,j}\neq 0\big{\}} \big{)}$,
and ${\|}{\bf{A}} {\|}_0\triangleq \mathbf{card} \big{(}\big{\{}i \in \mathcal{I}
\big{|} \exists j,  j = 0,1, \dots ,n-1,  {a}_{i,j}\neq 0 \big{\}} \big{)}$,
where $\mathcal{I}\triangleq  \big{\{}0,1, \dots,m-1\big{\}}$.
We use ${\text{tr}}({\bf{B}})$, ${\bf{B}}^{-1}$ and $\det({\bf{B}})$ to show the trace, inverse, and determinant of an square matrix $\bf{B}$.
We also employ $\text{diag}({\bf{B}})$ to represent the diagonal elements of $\bf{B}$ in vector form.
Throughout the paper, $(\cdot)^*$, $(\cdot)^{\dag}$, and $(\cdot)^{\rm{H}}$ show the complex conjugate, transpose, and Hermitian transpose, respectively. Also,
 $| \cdot |$, $\left\lfloor \cdot \right\rfloor$,
 and $\otimes$  represent the absolute value operator, floor function, and Kronecker product, respectively.
$\mathbb{E}\{\cdot\}$ is the statistical expectation, $\hat{{x}}$ is an estimate of $x$.
The complex Gaussian distribution with mean vector $\bm{\mu}$ and covariance matrix $\bf{\Sigma}$ is denoted by ${\cal{C}\cal{N}}\big{(}{\bm{\mu}},{\bf{\Sigma}}\big{)}$.

\begin{figure}
\hspace{1cm}
  \includegraphics[width=8cm]{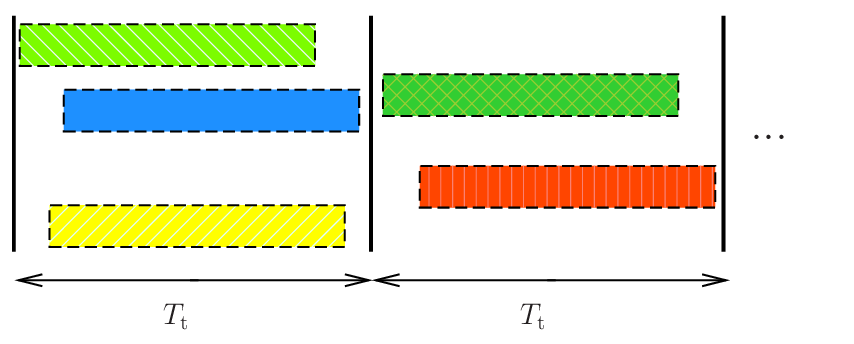}
   \vspace{-1.5em}
   \caption{Received packets at the \ac{bs}.}\label{fig:Fnk}
   \vspace{-1em}
\end{figure}

\begin{figure*}[!t]\label{fig:TX}
    \centering
\begin{tikzpicture}
[cross/.style={path picture={
  \draw[black]
(path picture bounding box.south east) -- (path picture bounding box.north west) (path picture bounding box.south west) -- (path picture bounding box.north east);
}}]
[auto, node distance=1cm,>=latex']
    \node [input, name=T1] {};
    \node [input, name=C1] (C1) at (13.1,-1.5){${\mathbf{c}}_k$};
    \node [block, right of=T1,node distance=2.5cm,rounded corners=0.15cm] (T2) {Channel coding};
    \node [block, right of=T2,node distance=4cm,rounded corners=0.15cm] (T3) {Differential coding};
    \node [block, right of=T3,node distance=4.2cm,rounded corners=0.15cm] (T4) {BPSK modulation};
    \node [draw,circle,cross,minimum width=0.6 cm,rounded corners=0.15cm](T5) at (13.1,0){};
    \node (T6) at (15,0) [MIMO RF chain east,rounded corners=0.15cm];
    \node [block, below of=T5,node distance=2.2cm,text width=2.4cm,rounded corners=0.15cm] (T7) {DS-SS sequence};
    \node [input, name=C2,below of=T6,node distance=2cm] (C2) {${{s}}_k$};

     \draw [->] (T1) -- node[above]{${\mathbf{d}}_k$}(T2) ;
     \draw [->] (T2) -- node[above]{${\mathbf{b}}_k^{\rm{c}}$}(T3);
     \draw [->] (T3) -- node[above]{${\mathbf{b}}_k^{\rm{d}}$}(T4);
     \draw [->] (T4) -- node[above]{${\mathbf{b}}_k$}(T5);
     \draw [->] (T5) -- node[above]{}(T6);
     \draw [->] (C1) -- node[below right]{\vspace{0.3em} ${{s}}_k(t)$}(T5);
     \draw [->] (C2) -- node[below=0.6cm]{${\omega_{\rm{c}}}$}(T6);
    \end{tikzpicture}
\caption{Block diagram that illustrates packet transmission at \ac{iot} devices.}
\label{fig:iot_tx}
\vspace{0.15em}
\hrule
\vspace{-0.4em}
 \end{figure*}

The remaining of the paper is organized as follows. Section
\ref{sec:sm} introduces the system model. Section \ref{sec:actyb12} describes the
proposed \ac{iot} DI algorithms and presents their analytical performance evaluation. In Section \ref{sec:sm00000m}, the data detection problem is discussed, and the nonlinear \ac{2mc}-\ac{mud} algorithm is proposed.
An extension to multiple receive antennas is discussed in \ref{mimo_dsi}.
Simulation results
are provided in Section \ref{nn:iop}, and conclusions are drawn in
Section \ref{sec:conclude}.
\vspace{-0.25em}
\section{System Model}\label{sec:sm}
Consider $K_{\rm{u}}$ \ac{iot} devices communicating with a single \ac{iot} \ac{bs} in a single-hop communication.
It is considered that the \ac{iot} devices transmit data in short packets over independent doubly block fading channel, where the fading channel is block fading in time and in frequency.
The activity rate for each \ac{iot} device is assumed to be $P_{\rm{a}}$.\footnote{Both known and unknown activity rate are studied in this paper.}
The \ac{iot} devices transmit their packet after receiving a beacon signal transmitted by the \ac{iot} \ac{bs}.
This signal is periodically transmitted with period
 $T_{\rm{t}}= N_{\rm{s}}T_{\rm{s}}+\tau_{\rm{max}}$, where $N_{\rm{s}}$ is the number of symbols per \ac{iot} packet, $T_{\rm{s}}$ is the symbol duration,
and $\tau_{\rm{max}}$ is the known maximum delay of the single-hop \ac{iot} network.
It is assumed that $T_{\rm{t}}$ equals the coherence time of the fading channel.

We denote $\mathcal{X}_{\rm{u}} \triangleq \{0,1,\dots,K_{\rm{u}}-1\}$ and $\mathcal{X}_{\rm{a}}$ the total and active \ac{iot} devices in the network, respectively.
The round-trip delay of the $k$th \ac{iot} device is shown by ${\tau}_{k} \triangleq 2d_k/c$, ${\tau}_k \in [0,\tau_{{\rm{max}}}]$, where $c$ is the speed of light, and $d_k$ is the distance between the $k$-th \ac{iot} device and the \ac{bs}.
We consider that $\tau_k$, $k, \in \mathcal{X}_{\rm{u}}$, is known at the receiver.
Fig. \ref{fig:Fnk} illustrates the received \ac{iot} packets at the \ac{bs}.

As illustrated in Fig. \ref{fig:iot_tx}, for each \ac{iot} device, the payload bits ${\bf{d}}_{k}$, $k \in \mathcal{X}_{\rm{a}}$, are encoded by the channel encoder to increase the reliability of packet transmission. Then, the encoded data is passed through the differential encoding block.
Differential encoding is employed to remove the need of channel estimation in the \ac{mud} at the \ac{bs} to enable non-coherent detection.
After differential encoding, the data is \ac{bpsk} modulated.
Finally, the modulated signal is multiplied by a unique spreading waveform and then transmitted.
It is considered that the spreading waveforms of the \ac{iot} devices do not change over time.

The impulse response of the doubly block fading channel for the $k$th \ac{iot} device is given as $g_k(t) \approx {\breve{{g}}}_k  \delta(t-\tau_k)$,
where ${\breve{{g}}}_k$ is the fading coefficient of the $k$th \ac{iot} device, which
is constant during a packet but changes to an independent value for the next packet.
Doubly block fading channel is a suitable model for sporadic traffic and short packet transmission.
The received
baseband signal over doubly block fading
channel in each transmission period with respect to the timing reference of the \ac{bs} is modeled as
\vspace{-0.2em}
\begin{align}\label{E4} \nonumber
r(t) \hspace{-0.1em} &= \hspace{-0.2em} \sum_{k=0}^{K_{\rm{u}}-1}\sum_{n=0}^{N_{\rm{s}}-1}{\breve{{g}}_k}\sqrt{\eta_k p_k} e^{j{\phi}_k} {b}_{k,n}
s_k(t-nT_{\rm{s}}-{\tau}_k)+{w}(t) \\
&=\sum_{k=0}^{K_{\rm{u}}-1}\sum_{n=0}^{N_{\rm{s}}-1}{{{g}}_k}{b}_{k,n}
s_k(t-nT_{\rm{s}}-{\tau}_k)+{w}(t),
\end{align}
where $t \in [0,T_{\rm{t}}]$, $g_k \triangleq  \breve{g}_k
\sqrt{\eta_k p_k} e^{j{\phi}_k}$, and ${\breve{g}}_k$, ${\phi}_k$, and $\big{\{}{b}_{k,n}, $ $
n=0,1,\dots,N_{\rm{s}}-1\big{\}}$
respectively denote the fading channel coefficient,
carrier phase (CP),
and symbol stream of the $k$th \ac{iot} device, which are unknown at the \ac{bs}. Also, $\eta_k=\big{(}\frac{\lambda_{\rm c}}{4\pi d_k}\big{)}^2$ and  $p_k$  denote the pathloss and  transmit power of the $k$th \ac{iot} device, respectively, where $\lambda_{\rm c}$ is the wavelength of the carrier signal.
It is considered that $\breve{g}_k \sim {\cal{C}\cal{N}}\big{(}\mu_k,\sigma_k^2\big{)}$, and
the envelope of the CSI, i.e., $|\breve{g}_k|$ has a Rician distribution with $K$-factor $|\mu_k|^2/(\sigma_k^2)$.
The symbol stream for the inactive \ac{iot} devices is modeled as transmitting zeros
during the packet, i.e., ${b}_{k,n}=0$, $n=0,1,\dots,N_{\rm{s}}-1$,
while active IoT devices employ \ac{bpsk} modulation  with $\mathbb{E}\{|{b}_{k,n}|^2\}$ $=1$. The \ac{dsss} signaling waveform of the $k$th IoT device,
$s_k\left(t\right)$, is given by
\begin{equation}\label{E2}
        s_k(t)=\sum_{m=0}^{N_{\rm{c}}-1}c_k^{(m)}\psi(t-mT_{\rm{c}}), \,\,\,\,\,\,\,\,\,\,\,\,\,\,\,\,\,\,t\in \left[0,T_{\rm{s}}\right],
        \vspace{-0.6em}
\end{equation}
where $T_{\rm{c}}$ is the chip duration, ${\bf c}_k= \big{[}c_k^{(0)} \  c_k^{(1)} \ \hdots \  c_k^{({N_{\rm{c}}-1})}\big{]}^\dag$ is
 the spreading sequence of $\left\{+1,-1\right\}$ assigned to the $k$th \ac{iot} device, $N_{\rm{c}}$ is the spreading factor, and $\psi(t)$ is the chip waveform with unit power. It is assumed that $\psi(t)$ is
 a rectangular pulse confined within
$\left[0,T_{\rm{c}}\right]$.
To support massive connectivity, $K_{\rm{u}}> N_{\rm{c}}$, which leads to non-orthogonal transmission.
 The baseband additive complex Gaussian noise at the output of the receive filter with bandwidth $1/T_{\rm{c}}$ is denoted by $w(t)$ in \eqref{E4}.

\begin{figure*}[!t]\label{fig:RX}
{
\begin{tikzpicture}
     \node (R1) at (0,0) [MIMO RF chain west,rounded corners=0.2cm];
     \node [block, right of=R1,node distance=3.5cm,text width=2cm,rounded corners=0.2cm] (R2) {\ \ \ Chip MF};
     \node [block, right of=R2,node distance=5.3cm,text width=2.5cm,rounded corners=0.2cm] (R3) {\,\,\,\,\,\,\,\ \ac{iot} device \textcolor{white}{--} \textcolor{white}{-i} identification};
     \node [block, right of=R3,node distance=3.9cm,text width=2.2cm,rounded corners=0.2cm] (R4) {\ \ \ac{2mc}-\ac{mud}};
     \node [block, below of=R4,node distance=3cm,text width=3.2cm,rounded corners=0.2cm] (R6) {\hspace{0.02em} Differential decoding};
     \node [block, left of=R6,node distance=6.5cm,text width=2.6cm,rounded corners=0.2cm] (R7) {Channel  decoding};%

       \draw [->] (R1) -- node[above]{${r}(t)$}(R2) ;
       \draw [-] (4.62,0) -- node[name=u]{}(5.7,0) ;
       \draw (5.7,0)to[short,o-](5.7,0);
       \draw[ ->, ](5.7,-0.01)-- +(35:0.72);
       \draw (6.5,0)to[short,o-](6.5,0);
       \draw (6.54,0)to(R3);
       \draw (6,0.4) node[above] {$T_{\rm{c}}$};
       \draw [->] (R3) -- node[above]{$\hat{\mathcal{X}}_{\rm{a}}$}(R4);
       \draw  (2.1,0) to (2.1,1.8);
       \path [line] (2.1,1.8) -| (R4);
       \draw (R4) to (15.8,0);
       \path [line] (15.8,0)  node[above] {\hspace{-3em} $\big{\{}\hat{\mathbf{b}}_{k_0}^{\rm{m}}, \dots,
       \hat{\mathbf{b}}_{k_{\hat{K}_{\rm{a}}-1}}^{\rm{m}} \big{\}}$} |- (R6);
       \path [line] (R6) -- node[above] {$\big{\{}\hat{\mathbf{b}}_{k_0}^{\rm{c}}, \dots,
       \hat{\mathbf{b}}_{k_{\hat{K}_{\rm{a}}-1}}^{\rm{c}} \big{\}}$} (R7);
       \path [line] (4.8,-2.5)-- node[above] {${\hat{\mathbf{d}}}_{k_0}$}(3.5,-2.5);
       \path [line] (4.8,-2.5)-- node[below] {\vdots}(3.5,-2.5);
       \path [line] (4.8,-3.5)--node[below] {$\hat{\mathbf{d}}_{k_{\hat{K}_{\rm{a}}-1}}$}(3.5,-3.5);
       \draw  (7,0) to (7,-0.001);
\end{tikzpicture}
}
\vspace{-1em}
\caption{Block diagram of the proposed receiver at the \ac{bs}.}\label{fig:iot_RX}
\vspace{0.2em}
\hrule
\vspace{-0.7em}
    \end{figure*}

Fig. \ref{fig:iot_RX} shows the block diagram of the proposed receiver at the \ac{iot} \ac{bs}. As seen, the received baseband signal
is passed through the chip MF and sampled at the chip rate. The output of the sampled chip MF for the $i$th chip at the $j$th observation symbol is obtained as
\begin{align}\label{E5}
{r}_j^{(i)}&  \triangleq  \int_{jT_{\rm{s}}+iT_{\rm{c}}}^{jT_{\rm{s}}+(i+1)T_{\rm{c}}}
{r}(t)\psi\left(t-jT_{\rm{s}}-iT_{\rm{c}}\right)dt  \\ \nonumber
&=\sum_{k=0}^{K_{\rm{u}}-1}g_k{u}_{k,j}^{(i)}+w_j^{(i)}\,\,\,\,\,\,\,\,\,\ i=0,1,\dots,N_{\rm{c}}-1,
\end{align}
where
\begin{align}\label{E6}
w_j^{(i)} \triangleq   \int_{jT_{\rm{s}}+iT_{\rm{c}}}^{jT_{\rm{s}}+(i+1)T_{\rm{c}}}
w(t)\psi\left(t-jT_{\rm{s}}-iT_{\rm{c}}\right)dt,
\end{align}
and
\begin{align}\label{E7}
u_{k,j}^{(i)} \triangleq \int_{jT_{\rm{s}}+iT_{\rm{c}}}^{jT_{\rm{s}}+(i+1)T_{\rm{c}}}
\sum_{n=0}^{N_{\rm{s}}-1}&b_{k,n}
s_k(t-nT_{\rm{s}}-{\tau}_k)\\ \nonumber
& \ \times \psi(t-jT_{\rm{s}}-iT_{\rm{c}})dt.
\end{align}
By employing \eqref{E6}, one can show that the joint \ac{pdf} of the corresponding noise vector associated with the $j$th observation vector, i.e., ${\bf{w}}_j \triangleq \big{[}{w}_j^{(0)} \ {w}_j^{(1)} \ \hdots \ $ $ {w}_j^{(N_{\rm{c}}-1)}\big{]}^\dag$ is
characterized by ${\bf{w}}_j  \sim {\cal{C}\cal{N}}\left({\bf{0}}_{N_{\rm{c}}},\sigma_{\rm{w}}^2{\bf{I}} \right)$ with $\sigma_{\rm{w}}^2\triangleq{N_0}/{T_{\rm{c}}}$, where $N_0/2$ is the power spectral density of the white noise.
The integral in \eqref{E7} represents the area under
 the received signal waveform of the $k$th \ac{iot} device during the $i$th chip-matched filtering duration at the $j$th observation symbol.

Let us write the delay of the $k$th \ac{iot} as
\begin{align}\label{E9}
{\tau}_k \triangleq {\alpha}_k{T_{\rm{s}}}+{\beta}_k{T_{\rm{c}}} + {\xi}_k,
\end{align}
with ${\alpha}_k\triangleq\left\lfloor{\tau_k}/{T_{\rm{s}}}\right\rfloor$, ${\beta}_k\triangleq\left\lfloor{\tau_k}/{T_{\rm{c}}}\right\rfloor-{\alpha}_kN_{\rm{c}}$, and ${\xi}_k \in [0,{T_{\rm{c}}})$.

Based on the values of ${\alpha}_k$, ${\beta}_k$, and ${\xi}_k$, ${u}_{k,j}^{(i)}$ in \eqref{E7} is expressed as a function of ${b}_{k,j-{\alpha}_k}$ and ${b}_{k,j-{\alpha}_k-1}$ as \cite{pham2007nonlinear}
\begin{align}\label{Eyy}
{u}_{k,j}^{(i)} & \triangleq \sum_{n=0}^{N_{\rm{s}}-1}\sum_{m=0}^{N_{\rm{c}}-1}c_{k}^{(m)}{b}_{k,n}   \\ \nonumber
&\times \int_{jT_{\rm{s}}+iT_{\rm{c}}}^{jT_{\rm{s}}+(i+1)T_{\rm{c}}}
\hspace{-1.2em}\psi(t-nT_{\rm{s}}-mT_{\rm{c}}-{\tau}_k) \psi(t-jT_{\rm{s}}-iT_{\rm{c}})dt \\ \nonumber
& =  b_{k,j-{\alpha}_k-1}{x}_k^{(i)}{(1-{{\xi}_{k}})}+ b_{k,j-{\alpha}_k}{x}_k^{(i)}{({{\xi}_{k}})},
\end{align}
where
\begin{align}
\hspace{-1em}{x}_k^{(i)}{(\nu)} \hspace{-0.1em} \triangleq \hspace{-0.1em}
\sum_{m=0}^{N_{\rm{c}}-1}c_k^{(m)}\int_{iT_{\rm{c}}}^{(i+1)T_{\rm{c}}}
\hspace{-1.5em}\psi(t-mT_{\rm{c}}-{\nu}T_{\rm{c}})\psi(t-iT_{\rm{c}})dt,
\end{align}
with $\nu \in [0,1)$. We can write
 \eqref{Eyy} in vector form as follows
\begin{align}\label{E10}
{\bf{u}}_{k,j}={b}_{k,j-{\alpha}_k-1}{\bf{x}}_{k,0}+b_{k,j-{\alpha}_k}{\bf{x}}_{k,1}
\end{align}
where ${b}_{k,j}=0$ when $j \notin [0,N_{\rm{s}}-1]$, and
\begin{subequations}\label{E18}
\begin{align}
{\bf{u}}_{k,j}& \triangleq
\Big{[}
{u}_{k,j}^{(0)} \  {u}_{k,j}^{(1)} \  \hdots \
{u}_{k,j}^{(N_{\rm{c}}-1)}
\Big{]}^\dag, \\ \label{kkk3}
{\bf{x}}_{k,1}&\triangleq
\Big{[}
{x}_k^{(0)}({\xi}_k) \  {x}_k^{(1)}({\xi}_k) \  \hdots \   {x}_k^{(N_{\rm{c}}-1)}({\xi}_k)
\Big{]}^\dag, \\ \nonumber
{\bf{x}}_{k,0}&\triangleq
\Big{[}
{x}_k^{(0)}(1-{\xi}_k)  \  {x}_k^{(1)}(1-{\xi}_k) \  \hdots \ {x}_k^{(N_{\rm{c}}-1)}(1-{\xi}_k)
\Big{]}^\dag.
\end{align}
\end{subequations}

For the rectangular chip waveform $\psi(t)$, we can obtain
\begin{equation}\label{E11}
\begin{bmatrix}
\vspace{0.75em}
    \mathbf{x}_{k,1}\\
    \mathbf{x}_{k,0}
    \end{bmatrix}
=
(1-{\xi}_k)\begin{bmatrix}
    {\bf{0}}_{{\beta_k}}\\
    \mathbf{c}_k \\
    {\bf{0}}_{N_{\rm{c}}-{\beta_k}}
    \end{bmatrix}
 +{\xi}_k
    \begin{bmatrix}
    {\bf{0}}_{{\beta_k}+1}\\
    \mathbf{c}_k \\
    {\bf{0}}_{N_{\rm{c}}-{\beta_k}-1}
    \end{bmatrix}.
\end{equation}
Let us define $\mathbf{X}_k \triangleq \big{[} \mathbf{x}_{k,0} \  \mathbf{x}_{k,1}\big{]}$.  By employing \eqref{E5} and \eqref{E10}, the $j$th observation vector, i.e., $\mathbf{r}_{j}\triangleq\big{[}\mathbf{r}_j^{(0)} \ \mathbf{r}_j^{(1)} \ \hdots \ \mathbf{r}_{j}^{(N_{\rm{c}}-1)}\big{]}^\dag$, is written as follows
\begin{equation}\label{E13}
    \mathbf{r}_j=
    \mathbf{X}\mathbf{G}\mathbf{b}_j+\mathbf{w}_j=\mathbf{X}\mathbf{h}_j+\mathbf{w}_j,
\end{equation}
where
\begin{equation}\label{E14}
    \mathbf{X} \triangleq
    \begin{bmatrix}
    \mathbf{X}_0 & \mathbf{X}_1 & \hdots & \mathbf{X}_{K_{\rm{u}}-1}
    \end{bmatrix},
\end{equation}
\begin{equation}\label{E15}
\mathbf{G}\triangleq
   \begin{bmatrix}
   {{g}}_0 &  & \multicolumn{2}{c}{\multirow{2}{*}{\scalebox{1.2}{$0$}}} \\
    & {{g}}_1 & &  \\
    \multicolumn{2}{c}{\multirow{2}{*}{\scalebox{1.2}{$0$}}} & \ddots &  \\
    & &  &  {{g}}_{K_{\rm{u}}-1}
  \end{bmatrix}\otimes \mathbf{I}_2,
 \end{equation}
 \begin{align}\label{E16}
  \mathbf{b}_{j} & \triangleq \Big{[} {b}_{0,j-{\alpha}_0-1} \,\ {b}_{0,j-{\alpha}_0} \,\ {b}_{1,j-{\alpha}_1-1} \,\ {b}_{1,j-{\alpha}_1} \,\  \hdots \ \\ \nonumber
& \quad \quad \quad \quad \quad \quad {b}_{K_{\rm{u}}-1,j-{\alpha}_{(K_{\rm{u}}-1)}-1} \,\ {b}_{K_{\rm{u}}-1,j-{\alpha}_{(K_{\rm{u}}-1)}} \Big{]}^{\dag},
\end{align}
and
\begin{align}\label{eq16x}
 \hspace{-0.4em} \mathbf{h}_{j} \hspace{-0.2em}  \triangleq \hspace{-0.2em}  \Big{[} {h}_{0,j,0} \ {h}_{0,j,1} \ {h}_{1,j,0} \ {h}_{1,j,1} \  \hspace{-0.21em} \hdots  \hspace{-0.2em} \
   {h}_{K_{\rm{u}}-1,j,0} \ {h}_{K_{\rm{u}}-1,j,1}\Big{]}^\dag
\end{align}
with
\begin{align}\label{eq:opip}
{h}_{k,j,f}\triangleq {g}_k{b}_{k,j-{\alpha}_k-1+f}, \,\,\,\  f \in \{{0,1}\}.
\end{align}
Finally, by staking the $N_{\rm{t}}$ observation vectors, the observation matrix is written as follows
\begin{equation}\label{E17}
{\mathbf{R}}_{\rm{T}} =\mathbf{X}{\mathbf{G}}{\mathbf{B}}_{\rm{T}}+{\mathbf{W}}_{\rm{T}} = \mathbf{X}{\mathbf{H}}_{\rm{T}}+{\mathbf{W}}_{\rm{T}} ,
\end{equation}
where ${\mathbf{R}}_{\rm{T}} \triangleq [\mathbf{r}_0  \ \mathbf{r}_1 \ \hdots \ \mathbf{r}_{N_{\rm{t}}-1}]$, ${\mathbf{B}}_{\rm{T}}  \triangleq [\mathbf{b}_0  \ \mathbf{b}_1 \ \hdots \ \mathbf{b}_{N_{\rm{t}}-1}]$, ${\mathbf{W}}_{\rm{T}}  \triangleq [\mathbf{w}_0  \ \mathbf{w}_1 \ \hdots \ \mathbf{w}_{N_{\rm{t}}-1}]$, and
${\mathbf{H}}_{\rm{T}} \triangleq [\mathbf{h}_0  \ \mathbf{h}_1 \ \hdots \ \mathbf{h}_{N_{\rm{t}}-1}]$.
In \eqref{E17}, $\mathbf{X}$ is referred to as dictionary.

As seen in Fig. \ref{fig:iot_RX}, after chip-matched filtering and sampling, the \ac{iot} DI algorithm is applied to the measurement matrix ${\mathbf{R}}_{\rm T}$ to detect the active \ac{iot} devices. The outcome of the \ac{iot} DI algorithm is a set of \ac{iot} devices $\hat{\mathcal{X}}_{\rm{a}}$. Then, the \ac{mud} algorithm is applied to detect data of the \ac{iot} devices in $\hat{\mathcal{X}}_{\rm{a}}$.
After \ac{mud}, the bit streams related to the active \ac{iot} devices pass through differential and channel decoders, respectively.
In the remaining of the paper, we propose different algorithms to realize the system in Fig.~\ref{fig:iot_RX}.
\vspace{-0.4em}
\section{\ac{iot} DI}\label{sec:actyb12}
DI is the first step in uplink \ac{ma} schemes
that devices do not use control signaling in order to identify themselves to the \ac{bs}. In this case, the \ac{bs} needs to determine the active devices before
data detection. In this section, different \ac{iot} DI algorithms are developed.
\vspace{-0.5em}
\subsection{\ac{iot} DI: Problem Formulation}
For the sake of decreasing the complexity, a portion of the observation window can be employed for \ac{iot} DI.
Let us consider a truncated observation window of length $L$ as
\begin{equation}\label{u8i02xngo}
{\mathbf{R}}=\mathbf{X}{\mathbf{G}}{\mathbf{B}}+{\mathbf{W}}= \mathbf{X}{\mathbf{H}}+{\mathbf{W}},
\end{equation}
where ${\mathbf{R}} \triangleq [\mathbf{r}_{\bar{\alpha}}  \ \mathbf{r}_{\bar{\alpha}+1} \ \hdots \ \mathbf{r}_{{\bar{\alpha}}+L-1}]$,
${\mathbf{B}} \triangleq [\mathbf{b}_{\bar{\alpha}}  \ \mathbf{b}_{\bar{\alpha}+1} \ \hdots \ $ $\mathbf{b}_{{\bar{\alpha}}+L-1}]$,
${\mathbf{W}} \triangleq [\mathbf{w}_{\bar{\alpha}}  \ \mathbf{w}_{\bar{\alpha}+1} \ \hdots \ \mathbf{w}_{{\bar{\alpha}}+L-1}]$, and
${\mathbf{H}} \triangleq [\mathbf{h}_{\bar{\alpha}}  \ \mathbf{h}_{\bar{\alpha}+1} \ $ $ \hdots \ \mathbf{h}_{{\bar{\alpha}}+L-1}]$
with $1 \leq L\leq N_{\rm{s}}+\alpha_{\rm{min}}-\bar{\alpha}$, where $\bar{\alpha}$ is an arbitrary positive integer,
 $\bar{\alpha} > \alpha_{\rm{max}}\triangleq {\text{max}}\{\alpha_0,\alpha_1,\dots, \alpha_{K_{\rm{u}}-1}\}$, and $\alpha_{\rm{min}}\triangleq{\text{min}}\{\alpha_0,\alpha_1,\dots,\alpha_{K_{\rm{u}}-1}\}$.\footnote{The number of non-zero elements of the observation window for the active \ac{iot} devices is considered the same to facilitate theoretical analysis.}
Fig.~\ref{fig:under} shows the underdetermined system of
linear equations in \eqref{E17}, and Fig. \ref{fig:undervr} illustrates truncated observation windows for \ac{iot} DI in \eqref{u8i02xngo}.

\begin{figure*}
\vspace{-1em}
\centering
\tikzset{table/.style={
matrix of nodes,
row sep=-\pgflinewidth,
column sep=-\pgflinewidth,
nodes={rectangle,draw=black,fill=0,minimum size=0.35cm,align=center},
nodes in empty cells
}
}
\begin{tikzpicture}[every node/.style={scale=0.6}]
\matrix[anchor=north] (mat1) [table]
{
|[fill=2]| & |[fill=9]| & |[fill=7]| & |[fill=6]| & |[fill=3]| & |[fill=6]| & |[fill=3]| & |[fill=10 ]| \\
|[fill=3]| & & & |[fill=5]| & |[fill=6]| & |[fill=2]| & |[fill=8]| & |[fill=9]| \\
|[fill=5]| & |[fill=10]| & |[fill=3]| & |[fill=7]| & |[fill=3]| & |[fill=9]| & |[fill=2]| & \\
|[fill=3]| & |[fill=4]| & & |[fill=1]| & |[fill=6]| & |[fill=2]| & |[fill=6]| & |[fill=9]| \\
|[fill=6]| & |[fill=9]| & |[fill=7]| & |[fill=6]| & |[fill=3]| & |[fill=6]| & |[fill=3]| & |[fill=10 ]| \\
};
\node [right of=mat1,node distance=1.9cm] (eq) {\huge{${=}$}};
\matrix [right of=mat1,node distance=5cm] (mat2) [table]
{|[fill=2]| & |[fill=5]| & |[fill=4]|& |[fill=2]| & |[fill=6]| & & |[fill=2]| & |[fill=9]|& & |[fill=3]| & &
|[fill=2]| & &  \\
& |[fill=3]| & & |[fill=6]| & |[fill=4]| & |[fill=7]| & |[fill=3]| & |[fill=8]| & |[fill=2]| & |[fill=9]| & &
|[fill=5]| & |[fill=10]| & |[fill=5]|  \\
& |[fill=3]| & |[fill=4]| & & & |[fill=5]| & |[fill=5]| & |[fill=7]| & |[fill=8]| & |[fill=5]| & |[fill=6]| &
|[fill=2]| & |[fill=5]| & |[fill=3]|   \\
|[fill=3]| & |[fill=3]| & |[fill=4]| & |[fill=3]| & |[fill=5]| & |[fill=5]| & |[fill=7]| & |[fill=8]| & |[fill=5]| &
|[fill=6]| & |[fill=2]| & |[fill=5]| & &  \\
& |[fill=3]| & |[fill=2]| & |[fill=7]| & |[fill=4]| & |[fill=6]| & |[fill=3]| & |[fill=2]| & |[fill=5]| &
|[fill=3]| & |[fill=10]| & |[fill=2]| & & |[fill=6]|  \\
};
\matrix[right= 3.5cm of mat2.north west, anchor=north west ] (mat3) [table]
{|[fill=1]| & |[fill=9]| & |[fill=9]| & |[fill=9]| & |[fill=9]| & |[fill=9]| & |[fill=9]| & |[fill=1]| \\
|[fill=9]|& |[fill=9]| & |[fill=9]|& |[fill=9]| & |[fill=9]| & |[fill=9]| & |[fill=1]| & |[fill=1]| \\
|[fill=1]| & |[fill=1]| & |[fill=1]| & |[fill=1]| & |[fill=1]| & |[fill=1]| & |[fill=1]| & |[fill=1]|\\
|[fill=1]| & |[fill=1]| & |[fill=1]| & |[fill=1]| & |[fill=1]| & |[fill=1]| & |[fill=1]| & |[fill=1]|\\
|[fill=1]| & |[fill=1]| & |[fill=1]| & |[fill=1]| & |[fill=1]| & |[fill=1]| & |[fill=1]| & |[fill=1]|\\
|[fill=1]| & |[fill=1]| & |[fill=1]| & |[fill=1]| & |[fill=1]| & |[fill=1]| & |[fill=1]| & |[fill=1]|\\
|[fill=1]| & |[fill=1]| & |[fill=1]| & |[fill=1]| & |[fill=1]| & |[fill=1]| & |[fill=1]| & |[fill=1]|\\
|[fill=1]| & |[fill=1]| & |[fill=1]| & |[fill=1]| & |[fill=1]| & |[fill=1]| & |[fill=1]| & |[fill=1]|\\
|[fill=1]| & |[fill=1]| & |[fill=7]| & |[fill=7]| & |[fill=7]| & |[fill=7]| & |[fill=7]| & |[fill=7]|\\
|[fill=1]| & |[fill=7]| & |[fill=7]| & |[fill=7]| & |[fill=7]| & |[fill=7]| & |[fill=7]| & |[fill=1]|\\
|[fill=1]| & |[fill=1]| & |[fill=1]| & |[fill=1]| & |[fill=1]| & |[fill=1]| & |[fill=1]| & |[fill=1]|\\
|[fill=1]| & |[fill=1]| & |[fill=1]| & |[fill=1]| & |[fill=1]| & |[fill=1]| & |[fill=1]| & |[fill=1]|\\
|[fill=1]| & |[fill=1]| & |[fill=1]| & |[fill=1]| & |[fill=1]| & |[fill=1]| & |[fill=1]| & |[fill=1]|\\
|[fill=1]| & |[fill=1]| & |[fill=1]| & |[fill=1]| & |[fill=1]| & |[fill=1]| & |[fill=1]| & |[fill=1]|\\
};

\node [right of=eq,node distance=9.7cm] (R1) {\huge{${=}$}};
\matrix[right= 2.3cm of mat3.north west, anchor=north west ] (mat4) [table]
{
|[fill=2]| & |[fill=5]| & |[fill=4]|& |[fill=2]| & |[fill=6]| & & |[fill=2]| & |[fill=9]|& & |[fill=3]| & &
|[fill=2]| & &  \\
& |[fill=3]| & & |[fill=6]| & |[fill=4]| & |[fill=7]| & |[fill=3]| & |[fill=8]| & |[fill=2]| & |[fill=9]| & &
|[fill=5]| & |[fill=10]| & |[fill=5]|  \\
& |[fill=3]| & |[fill=4]| & & & |[fill=5]| & |[fill=5]| & |[fill=7]| & |[fill=8]| & |[fill=5]| & |[fill=6]| &
|[fill=2]| & |[fill=5]| & |[fill=3]| \\
|[fill=3]| & |[fill=3]| & |[fill=4]| & |[fill=3]| & |[fill=5]| & |[fill=5]| & |[fill=7]| & |[fill=8]| & |[fill=5]| &
|[fill=6]| & |[fill=2]| & |[fill=5]| & & \\
& |[fill=3]| & |[fill=2]| & |[fill=7]| & |[fill=4]| & |[fill=6]| & |[fill=3]| & |[fill=2]| & |[fill=5]| &
|[fill=3]| & |[fill=10]| & |[fill=2]| & & |[fill=6]| \\
};

\matrix [right= 3.5cm of mat4.north west, anchor=north west ]  (mat5) [table]
{
|[fill=2]| & |[fill=1]| & |[fill=1]| & |[fill=1]| & |[fill=1]| & |[fill=1]| & |[fill=1]| & |[fill=1]| &
|[fill=1]| & |[fill=1]| & |[fill=1]| & |[fill=1]| & |[fill=1]| & |[fill=1]| \\
|[fill=1]| & |[fill=2]| & |[fill=1]| & |[fill=1]| & |[fill=1]| & |[fill=1]| & |[fill=1]| & |[fill=1]| &
|[fill=1]| & |[fill=1]| & |[fill=1]| & |[fill=1]| & |[fill=1]| & |[fill=1]| \\
|[fill=1]| & |[fill=1]| & |[fill=3]| & |[fill=1]| & |[fill=1]| & |[fill=1]| & |[fill=1]| & |[fill=1]| &
|[fill=1]| & |[fill=1]| & |[fill=1]| & |[fill=1]| & |[fill=1]| & |[fill=1]| \\
|[fill=1]| & |[fill=1]| & |[fill=1]| & |[fill=3]| & |[fill=1]| & |[fill=1]| & |[fill=1]| & |[fill=1]| &
|[fill=1]| & |[fill=1]| & |[fill=1]| & |[fill=1]| & |[fill=1]| & |[fill=1]| \\
|[fill=1]| & |[fill=1]| & |[fill=1]| & |[fill=1]| & |[fill=4]| & |[fill=1]| & |[fill=1]| & |[fill=1]| &
|[fill=1]| & |[fill=1]| & |[fill=1]| & |[fill=1]| & |[fill=1]| & |[fill=1]| \\
|[fill=1]| & |[fill=1]| & |[fill=1]| & |[fill=1]| & |[fill=1]| & |[fill=4]| & |[fill=1]| & |[fill=1]| &
|[fill=1]| & |[fill=1]| & |[fill=1]| & |[fill=1]| & |[fill=1]| & |[fill=1]| \\
|[fill=1]| & |[fill=1]| & |[fill=1]| & |[fill=1]| & |[fill=1]| & |[fill=1]| & |[fill=6]| & |[fill=1]| &
|[fill=1]| & |[fill=1]| & |[fill=1]| & |[fill=1]| & |[fill=1]| & |[fill=1]| \\
|[fill=1]| & |[fill=1]| & |[fill=1]| & |[fill=1]| & |[fill=1]| & |[fill=1]| & |[fill=1]| & |[fill=6]| &
|[fill=1]| & |[fill=1]| & |[fill=1]| & |[fill=1]| & |[fill=1]| & |[fill=1]| \\
|[fill=1]| & |[fill=1]| & |[fill=1]| & |[fill=1]| & |[fill=1]| & |[fill=1]| & |[fill=1]| & |[fill=1]| &
|[fill=8]| & |[fill=1]| & |[fill=1]| & |[fill=1]| & |[fill=1]| & |[fill=1]| \\
|[fill=1]| & |[fill=1]| & |[fill=1]| & |[fill=1]| & |[fill=1]| & |[fill=1]| & |[fill=1]| & |[fill=1]| &
|[fill=1]| & |[fill=8]| & |[fill=1]| & |[fill=1]| & |[fill=1]| & |[fill=1]| \\
|[fill=1]| & |[fill=1]| & |[fill=1]| & |[fill=1]| & |[fill=1]| & |[fill=1]| & |[fill=1]| & |[fill=1]| &
|[fill=1]| & |[fill=1]| & |[fill=9]| & |[fill=1]| & |[fill=1]| & |[fill=1]| \\
|[fill=1]| & |[fill=1]| & |[fill=1]| & |[fill=1]| & |[fill=1]| & |[fill=1]| & |[fill=1]| & |[fill=1]| &
|[fill=1]| & |[fill=1]| & |[fill=1]| & |[fill=9]| & |[fill=1]| & |[fill=1]| \\
|[fill=1]| & |[fill=1]| & |[fill=1]| & |[fill=1]| & |[fill=1]| & |[fill=1]| & |[fill=1]| & |[fill=1]| &
|[fill=1]| & |[fill=1]| & |[fill=1]| & |[fill=1]| & |[fill=10]| & |[fill=1]| \\
|[fill=1]| & |[fill=1]| & |[fill=1]| & |[fill=1]| & |[fill=1]| & |[fill=1]| & |[fill=1]| & |[fill=1]| &
|[fill=1]| & |[fill=1]| & |[fill=1]| & |[fill=1]| & |[fill=1]| & |[fill=10]| \\
};

\matrix [right= 3.3cm of mat5.north west, anchor=north west ](mat6) [table]
{|[fill=1]| & & & & & & & |[fill=1]| \\
& & & & & & |[fill=1]| & |[fill=1]| \\
|[fill=1]| & |[fill=1]| & |[fill=1]| & |[fill=1]| & |[fill=1]| & |[fill=1]| & |[fill=1]| & |[fill=1]|\\
|[fill=1]| & |[fill=1]| & |[fill=1]| & |[fill=1]| & |[fill=1]| & |[fill=1]| & |[fill=1]| & |[fill=1]|\\
|[fill=1]| & |[fill=1]| & |[fill=1]| & |[fill=1]| & |[fill=1]| & |[fill=1]| & |[fill=1]| & |[fill=1]|\\
|[fill=1]| & |[fill=1]| & |[fill=1]| & |[fill=1]| & |[fill=1]| & |[fill=1]| & |[fill=1]| & |[fill=1]|\\
|[fill=1]| & |[fill=1]| & |[fill=1]| & |[fill=1]| & |[fill=1]| & |[fill=1]| & |[fill=1]| & |[fill=1]|\\
|[fill=1]| & |[fill=1]| & |[fill=1]| & |[fill=1]| & |[fill=1]| & |[fill=1]| & |[fill=1]| & |[fill=1]|\\
|[fill=1]| & |[fill=1]| & |[fill=3]| & |[fill=3]| & |[fill=3]| & |[fill=3]| & |[fill=3]| & |[fill=3]|\\
|[fill=1]| & |[fill=3]| & |[fill=3]| & |[fill=3]| & |[fill=3]| & |[fill=3]| & |[fill=3]| & |[fill=1]|\\
|[fill=1]| & |[fill=1]| & |[fill=1]| & |[fill=1]| & |[fill=1]| & |[fill=1]| & |[fill=1]| & |[fill=1]|\\
|[fill=1]| & |[fill=1]| & |[fill=1]| & |[fill=1]| & |[fill=1]| & |[fill=1]| & |[fill=1]| & |[fill=1]|\\
|[fill=1]| & |[fill=1]| & |[fill=1]| & |[fill=1]| & |[fill=1]| & |[fill=1]| & |[fill=1]| & |[fill=1]|\\
|[fill=1]| & |[fill=1]| & |[fill=1]| & |[fill=1]| & |[fill=1]| & |[fill=1]| & |[fill=1]| & |[fill=1]|\\
};
\node[text width=2cm] at (3.5,-1.5) {\large $\mathbf{X}$};
\node[text width=2cm] at (0.5,-1.5) {\large ${\mathbf{R}}_{\rm T}$};
\node[text width=2cm] at (6.27,-3.4) {\large ${\mathbf{H}}_{\rm T}$};
\node[text width=2cm] at (9.4,-1.5) {\large $\mathbf{X}$};
\node[text width=2cm] at (12.7,-3.4) {\large $\mathbf{G}$};
\node[text width=2cm] at (15.41,-3.4) {\large ${\mathbf{B}}_{\rm T}$};
\end{tikzpicture}

\caption{Underdetermined system of
linear equations for $K_{\rm{u}}=7$, $K_{\text{\rm{a}}}=2$, $N_{\text{\rm{t}}}=8$, $N_{\text{\rm{c}}}=5$, $\alpha_{\rm{max}}=1$, and
$N_{\text{\rm{s}}}=6$. Due to the asynchronicity among the \ac{iot} devices, the matrix of the transmitted symbols ${\mathbf{B}}_{\rm T}$ includes two rows for each \ac{iot} device.}\label{fig:under}
\vspace{0.2em}
\hrule
\vspace{-0.6em}
\end{figure*}

The activity of an IoT device is defined for an entire packet, i.e, the rows of $\mathbf{H}$ corresponding to the active and inactive IoT devices
are non-zero and zero, respectively. Thus, the problem of \ac{iot} DI for the $k$th \ac{iot} device, $k \in \mathcal{X}_{\rm{u}}$, can be expressed as the following binary hypothesis testing:
\begin{align}\label{eq:testcv}
&H_{1k}:\,\,\,\,\ {\mathbf{h}}_{k}^{\{\bar{\alpha},L\}} \neq\mathbf{0}  \\ \nonumber
&H_{0k}:\,\,\,\,\ {\mathbf{h}}_{k}^{\{\bar{\alpha},L\}}  = \mathbf{0},
\end{align}
\vspace{-0.3em}
where
\begin{subequations}\label{q:sarabvv0}
\begin{align}
{\mathbf{h}}_{k}^{\{\bar{\alpha},L\}} &\triangleq  \Big{[}\mathbf{h}_{k,\bar{\alpha}}^\dag
\mathbf{h}_{k,\bar{\alpha}+1}^\dag \ \dots \ \mathbf{h}_{k,\bar{\alpha}+L-1}^\dag\Big{]}^\dag, \\ \label{eq:mop097}
\mathbf{h}_{k,j} &\triangleq \Big{[}{h}_{k,j,0} \ {h}_{k,j,1} \Big{]}^\dag,
\end{align}
\end{subequations}
and $H_{0k}$ and $H_{1k}$  are the null and alternative hypotheses denoting that the $k$th \ac{iot} device is active and inactive, respectively.
As seen in \eqref{eq:testcv}, the \ac{iot} DI problem is formulated as $K_{\rm{u}}$ parallel binary hypothesis testing problems.
The first step in \ac{iot} DI is to reconstruct ${\mathbf{h}}_{k}^{\{\bar{\alpha},L\}}$, $k \in \mathcal{X}_{\rm{u}}$, from the truncated observation matrix in \eqref{u8i02xngo}.
.

Let us denote the number of active \ac{iot} devices by the random variable ${k}_{\rm{a}}=\mathbf{card}(\mathbf{X}_{\rm{a}})$.
For $P_{\mathrm{a}}\ll 1$,  $\mathbb{P}\{{k}_{\rm{a}} \ll K_{\rm{u}}\}=1$, and thus,
$\mathbf{B}$ and $\mathbf{H}$ in \eqref{u8i02xngo} are sparse matrices.
Moreover, the columns of $\mathbf{H}$$(\mathbf{B})$ share the same sparsity profiles. This sparse structure is referred to as block-sparse. The block-sparse structure of $\mathbf{H}$ can be observed in Fig. \ref{fig:under}.

The sparse structure of $\mathbf{H}$ can be employed to reconstruct the columns of $\mathbf{H}$ from the underdetermined linear observation model in \eqref{u8i02xngo}.
When each column of $\mathbf{H}$ is individually reconstructed from its corresponding column in $\mathbf{R}$, it is referred to as \ac{ssr}.
The \ac{ssr} for the columns of $\mathbf{H}$, i.e., $\mathbf{h}_j$, $\bar{\alpha}\leq j \leq \bar{\alpha}+L-1$, is formulated as follows
\vspace{-0.2em}
\begin{align}\label{EEqz1}
& \hat{\mathbf{h}}_j=\text{arg}\underset{\hspace{0.5em} {\mathbf{h}}_j} {\ \text{min}}
&  \hspace{-2em}\frac{1}{2}\big{\|}\mathbf{r}_j-\mathbf{X}{\mathbf{h}}_j\big{\|}_{\rm{2}}^2+\lambda_{\ell_0} \big{\|}{\mathbf{h}}_j\big{\|}_0,
\end{align}
where $\lambda_{\ell_0}$ is the tuning parameter which balances both
approximation error and sparsity level of the solution.

The $\ell_0$-minimization in \eqref{EEqz1} is both numerically unstable and NP-hard since the
$\ell_0$ quasi-norm is a discrete-value function.
One approach to the \ac{ssr} is to replace the $\ell_0$ quasi-norm by
a convex function with common sparsity profile that leads to a solution very close to the one of the original problem.
Different convex functions can be employed
to relax $\|\mathbf{h}_j\|_0$ in \eqref{EEqz1}. A common family of
convex functions is the $\ell_q$ norm, given as
\vspace{-0.4em}
\begin{equation}\label{q}
\big{\|}{\mathbf{h}}_j\big{\|}_q  = \left(\sum_{k=0}^{K_{\rm{u}}-1}\sum_{f=0}^{1}\Big{|}{h}_{k,j,f}\Big{|}^q\right)^{\frac{1}{q}}.
\vspace{-0.4em}
\end{equation}
The recovered vectors by the $\ell_q$ norm minimization can be employed to infer the
 active \ac{iot} set $\mathcal{X}_{\rm{a}}$.

On the other hand, the block-sparse structure of $\mathbf{H}$ can be employed to improve the reconstruction of $\mathbf{H}$ in \eqref{EEqz1}. This method of signal reconstruction is referred to as \ac{sssr}.
Opposite to \ac{ssr}, the \ac{sssr} simultaneously exploits the column sparsity along with the block-sparse structure
in the optimization problem in order to reconstruct the matrix $\mathbf{H}$.
The
\ac{sssr} of $\mathbf{H}$,
given $\mathbf{R}$ and the dictionary $\mathbf{X}$,
is expressed  as
\begin{align}\label{EE}
  \hat{\mathbf{H}}=\text{arg}\underset{\ \mathbf{H}} {\ \text{min}}
\ \ \frac{1}{2}\big{\|}\mathbf{R}-\mathbf{X}\mathbf{H}\big{\|}_{\rm{F}}^2+\lambda_{\ell_0}^{\ell_0} \big{\|}\mathbf{H}\big{\|}_0,
\end{align}
where $\lambda_{\ell_0}^{\ell_0}$ is the tuning parameter which balances both
approximation error and sparsity level of the solution.
Similar to the $\ell_0$-minimization in \eqref{EEqz1}, the $\ell_0-\ell_0$-minimization in \eqref{EE} is unstable and NP-hard. Therefore, the quasi-norm $\big{\|}{\mathbf{H}}\big{\|}_0$ is replaced with the $\ell_p-\ell_q$ $(p,q\geq1)$ mixed-norm as
\vspace{-0.4em}
\begin{equation}\label{pq}
J_{p,q}(\mathbf{H})=
\sum_{k=0}^{K_{\rm{u}}-1}\big{\|}{\mathbf{h}}_{k}^{\{\bar{\alpha},L\}}\big{\|}_q^p
\end{equation}
to convert the combinatorial
problem in \eqref{EE} into a convex optimization problem. The vector ${\mathbf{h}}_{k}^{\{\bar{\alpha},L\}}$ in \eqref{pq} is defined in \eqref{q:sarabvv0}.
The recovered matrix by the relaxed \ac{sssr} can also be employed to infer the active \ac{iot} set.

\vspace{-0.5em}
\subsection{\ac{iot} DI for Known Activity Rate}
Here, we
propose an algorithm for \ac{iot} DI
when the probability of activity $P_{\rm{a}}$ is known at the \ac{bs}.
Convex relation through squared $\ell_2$-norm followed by a threshold setting mechanism is employed for \ac{iot} DI.
\begin{figure}[!t]
\vspace{-1.2em}
\centering
\tikzset{table/.style={
matrix of nodes,
row sep=-\pgflinewidth,
column sep=-\pgflinewidth,
nodes={rectangle,draw=black,fill=0,minimum size=0.35cm,align=center},
nodes in empty cells
}
}
\begin{tikzpicture}[every node/.style={scale=0.6}]
\vspace{-0.5em}
\matrix[anchor=north] (mat1) [table]
{|[fill=1]| & |[fill=4]| & |[fill=4]| & |[fill=4]| & |[fill=4]| & |[fill=4]| & |[fill=4]| & |[fill=4]| & |[fill=1]| & |[fill=1]| \\
|[fill=4]| & |[fill=4]| & |[fill=4]|& |[fill=4]| & |[fill=4]| & |[fill=4]| & |[fill=4]| & |[fill=1]| & |[fill=1]| & |[fill=1]| \\
|[fill=1]| & |[fill=1]| & |[fill=1]| & |[fill=4]| & |[fill=4]| & |[fill=4]| & |[fill=4]| & |[fill=4]| & |[fill=4]| & |[fill=4]|  \\
|[fill=1]| & |[fill=1]| & |[fill=4]| & |[fill=4]| & |[fill=4]| & |[fill=4]| & |[fill=4]|  & |[fill=4]| & |[fill=4]| & |[fill=1]|\\
|[fill=1]| & |[fill=1]| & |[fill=4]| & |[fill=4]| & |[fill=4]| & |[fill=4]| & |[fill=4]| & |[fill=4]| & |[fill=4]| & |[fill=1]|  \\
|[fill=1]| & |[fill=4]| & |[fill=4]| & |[fill=4]| & |[fill=4]| & |[fill=4]| & |[fill=4]|  & |[fill=4]| & |[fill=1]| & |[fill=1]|\\
|[fill=1]| & |[fill=4]| & |[fill=4]| & |[fill=4]| & |[fill=4]| & |[fill=4]| & |[fill=4]| & |[fill=4]| & |[fill=1]| & |[fill=1]|  \\
|[fill=4]| & |[fill=4]| & |[fill=4]| & |[fill=4]| & |[fill=4]| & |[fill=4]| & |[fill=4]|  & |[fill=1]| & |[fill=1]| & |[fill=1]|\\
|[fill=1]| & |[fill=1]| & |[fill=4]| & |[fill=4]| & |[fill=4]| & |[fill=4]| & |[fill=4]| & |[fill=4]| & |[fill=4]| & |[fill=1]|\\
|[fill=1]| & |[fill=4]| & |[fill=4]| & |[fill=4]| & |[fill=4]| & |[fill=4]| & |[fill=4]| & |[fill=4]| & |[fill=1]| & |[fill=1]|\\
|[fill=1]| & |[fill=4]| & |[fill=4]| & |[fill=4]| & |[fill=4]| & |[fill=4]| & |[fill=4]| & |[fill=4]| & |[fill=1]| & |[fill=1]|  \\
|[fill=4]| & |[fill=4]| & |[fill=4]| & |[fill=4]| & |[fill=4]| & |[fill=4]| & |[fill=4]|  & |[fill=1]| & |[fill=1]| & |[fill=1]|\\
|[fill=1]| & |[fill=1]| & |[fill=1]| & |[fill=4]| & |[fill=4]| & |[fill=4]| & |[fill=4]| & |[fill=4]| & |[fill=4]| & |[fill=4]|  \\
|[fill=1]| & |[fill=1]| & |[fill=4]| & |[fill=4]| & |[fill=4]| & |[fill=4]| & |[fill=4]|  & |[fill=4]| & |[fill=4]| & |[fill=1]|\\
};
\node[text width=2cm] at (2.7,-3.7) {\large $\RM{H}\ (\bar{\alpha}=4)$};
\node[text width=2cm] at (4.8,-3.7) {\large $\RM{H}\ (\bar{\alpha}=3)$};
\node[text width=2cm] at (0.58,-3.7) {\large ${\RM{H}}_{\rm T}$};
\node[text width=2cm] at (-1.35,-0.35) {\large $\alpha_0=0\Big{\{}$};
\node[text width=2cm] at (-1.35,-0.77) {\large $\alpha_1=2\Big{\{}$};
\node[text width=2cm] at (-1.35,-1.19) {\large $\alpha_2=1\Big{\{}$};
\node[text width=2cm] at (-1.35,-1.61) {\large $\alpha_3=0\Big{\{}$};
\node[text width=2cm] at (-1.35,-2.03) {\large $\alpha_4=1\Big{\{}$};
\node[text width=2cm] at (-1.35,-2.45) {\large $\alpha_5=0\Big{\{}$};
\node[text width=2cm] at (-1.35,-2.87) {\large $\alpha_6=2\Big{\{}$};

\node[text width=2cm] at (0.14,-0.04) {\large $----$};
\node[text width=2cm] at (0.28,0.125) {\large $L=4$};
\node[text width=2cm] at (0.35,-3.18) {\large $---$};
\node[text width=2cm] at (0.4,-3.35) {\large $L=3$};

\matrix [right of=mat1,node distance=4.5cm] (mat2) [table]
{ |[fill=4]| & |[fill=4]| & |[fill=4]|  \\
  |[fill=4]| & |[fill=4]| & |[fill=4]|  \\
 |[fill=4]| & |[fill=4]| & |[fill=4]| \\
 |[fill=4]| & |[fill=4]| & |[fill=4]| \\
 |[fill=4]| & |[fill=4]| & |[fill=4]| \\
 |[fill=4]| & |[fill=4]| & |[fill=4]| \\
 |[fill=4]| & |[fill=4]| & |[fill=4]| \\
 |[fill=4]| & |[fill=4]| & |[fill=4]| \\
 |[fill=4]| & |[fill=4]| & |[fill=4]| \\
 |[fill=4]| & |[fill=4]| & |[fill=4]| \\
 |[fill=4]| & |[fill=4]| & |[fill=4]| \\
 |[fill=4]| & |[fill=4]| & |[fill=4]| \\
 |[fill=4]| & |[fill=4]| & |[fill=4]| \\
 |[fill=4]| & |[fill=4]| & |[fill=4]| \\
};

\matrix [right of=mat1,node distance=8cm] (mat2) [table]
{ |[fill=4]| & |[fill=4]| & |[fill=4]| & |[fill=4]|  \\
  |[fill=4]|& |[fill=4]| & |[fill=4]| & |[fill=4]|  \\
  |[fill=4]| & |[fill=4]| & |[fill=4]| & |[fill=4]| \\
  |[fill=4]| & |[fill=4]| & |[fill=4]| & |[fill=4]| \\
  |[fill=4]| & |[fill=4]| & |[fill=4]| & |[fill=4]| \\
  |[fill=4]| & |[fill=4]| & |[fill=4]| & |[fill=4]| \\
  |[fill=4]| & |[fill=4]| & |[fill=4]| & |[fill=4]| \\
  |[fill=4]| & |[fill=4]| & |[fill=4]| & |[fill=4]| \\
  |[fill=4]| & |[fill=4]| & |[fill=4]| & |[fill=4]| \\
  |[fill=4]| & |[fill=4]| & |[fill=4]| & |[fill=4]| \\
  |[fill=4]| & |[fill=4]| & |[fill=4]| & |[fill=4]| \\
  |[fill=4]| & |[fill=4]| & |[fill=4]| & |[fill=4]| \\
  |[fill=4]| & |[fill=4]| & |[fill=4]| & |[fill=4]| \\
  |[fill=4]| & |[fill=4]| & |[fill=4]| & |[fill=4]| \\
};
(0,2) node[fill=red!20,draw,double,rounded corners] {3rd node};
\end{tikzpicture}
\vspace{-0.2em}
\caption{Different observation windows for \ac{iot} DI ($K_{\rm{u}}=7$, $N_{\rm{s}}=7$,  $\alpha_{\rm{max}}=2$, $\alpha_{\rm{min}}=0$, $1\leq L \leq 4$). The purple color is employed to show the packet of the \ac{iot}
devices, which is zero for inactive and non-zero for active \ac{iot} devices.}\label{fig:undervr}
\vspace{-1.5em}
\end{figure}

\subsubsection{Squared $\ell_2$-Norm \ac{ssr} \ac{iot} DI}
The squared $\ell_2$-norm convex relaxation form of \eqref{EEqz1} is given by
\vspace{-0.2em}
\begin{equation}\label{RD2}
 \hat{\mathbf{h}}_j=\text{arg}\underset{\hspace{0.4em}{\mathbf{h}}_j} {\ \text{min}}
\ \  \frac{1}{2}\big{\|}{\mathbf{r}}_j-\mathbf{X}{\mathbf{h}}_j\big{\|}_{\rm 2}^2+\lambda \big{\|}{\mathbf{h}}_j\big{\|}_2^2,
\vspace{-0.2em}
\end{equation}
where ${\mathbf{h}}_j$ is given in \eqref{eq16x} and \eqref{eq:opip}.
The squared $\ell_2$-norm \ac{ssr} algorithm formulates the \ac{iot} identification problem as a \ac{rd} estimation problem as in \eqref{RD2} followed by $K_{\rm{u}}$ parallel binary hypothesis testing problems. This is because the \ac{rd} does not
set the coefficients of $\hat{\mathbf{h}}_j$ to zero.
The optimal solution of \eqref{RD2} is obtained as \cite{hoerl1970ridge}
\vspace{-0.3em}
\begin{equation}\label{RD3}
\hat{\mathbf{h}}_j=\Big{(}\mathbf{X}^\dag\mathbf{X}+2\lambda \mathbf{I}\Big{)}^{-1}\mathbf{X}^\dag{\mathbf{r}}_j,
\vspace{-0.3em}
\end{equation}
which is a simple linear estimator of ${\mathbf{r}}_j$ that
shrinks ordinary \ac{ls} estimates towards zero.
The tuning parameter $\lambda$ for \ac{ssr} can be obtained through cross-validation and GCV \cite{golub1979generalized,ward2009compressed}.
The latter is a method of model selection that is widely employed; in this case,
$\lambda$ is obtained as follows \cite{golub1979generalized}
\vspace{-0.3em}
\begin{equation}\label{GCV}
\begin{aligned}
{\lambda_{\rm{cv}}}= \underset{\ \ \ \  \lambda}{\text{arg  min}}
\,\ \frac{\big{\|}({\mathbf{I}}-\mathbf{Q}){\mathbf{r}}_j\big{\|}_2^2}
{\big{[}\tr\big{(}\mathbf{I}-\mathbf{Q}\big{)}\big{]}^2},
\end{aligned}
\vspace{-0.3em}
\end{equation}
where $\mathbf{Q} \triangleq \mathbf{X}\big{(}\mathbf{X}^\dag\mathbf{X}+2\lambda \mathbf{I}\big{)}^{-1}\mathbf{X}^\dag$.
In \cite{boonstra2015small}, it has been shown that the optimal tuning parameter of the \ac{rd} estimator for ${\mathbf{r}}_j=\mathbf{X}{\mathbf{h}}_j+{\mathbf{w}}_j$ in terms of  minimum mean squared error can be approximated as follows
\vspace{-0.4em}
\begin{align}\label{AA1}
\lambda_{j}^{\rm{op}}\approx\frac{\sigma_{\rm{w}}^2 \tr[{\bar{\mathbf{\Sigma}}}_{\mathbf{X}}^{-1}]}
{{\mathbf{h}}_j^{\rm{H}}{\bar{\mathbf{\Sigma}}}_{\mathbf{X}}^{-1}
{\mathbf{h}}_j+3 \tr [{\bar{\mathbf{\Sigma}}}_{\mathbf{X}}^{-2}]},
\vspace{-0.4em}
\end{align}
where $\bar{\mathbf{\Sigma}}_{\mathbf{X}} \triangleq \mathbf{X}^\dag \mathbf{X}$.
As observed, $\lambda_{j}^{\rm{op}}$ depends on ${\mathbf{h}}_j$
which is unknown and needs to be estimated by
the \ac{rd} estimator. In this case, for moderate and high \ac{snr} range,
an approximation of \eqref{AA1} can be
obtained by replacing ${\mathbf{h}}_j^{\rm{H}}{\bar{\V{\Sigma}}}_{\mathbf{X}}^{-1}
{\mathbf{h}}_j$ with
its expected value \cite{bhattacharya2005digital}.
Since the elements of ${\mathbf{h}}_{j}$ are uncorrelated, by employing
$\mathbb{E}\big{\{}|{{h}}_{k,j,0}|^2\big{\}}=\mathbb{E}\big{\{}|{{h}}_{k,j,1}|^2\big{\}}=P_{\rm{a}}\eta_k p_k(\sigma_k^2+|\mu_k|^2)$, $k \in \mathcal{X}_{\rm{u}}$, we can show that
$\mathbb{E}\big{\{} {\mathbf{h}}_j^{\rm{H}}{\bar{\mathbf{\Sigma}}}_{\mathbf{X}}^{-1}
{\mathbf{h}}_j\big{\}}
=P_{\rm{a}}(\mathbf{\Gamma}^\dag \otimes \mathbf{1}_2^\dag) \bar{\mathbf{\Lambda}}_{\mathbf{X}}$,
where ${\mathbf{\Gamma}}\triangleq [ \gamma_0 \ \gamma_1  \  \hdots$  $\gamma_{K_{\rm{u}}-1}]^\dag$, $\gamma_k \triangleq   \eta_k p_k(\sigma_k^2+|\mu_k|^2)$, $\V{1}_2=[1 \ 1]^\dag$,
and $\bar{\mathbf{\Lambda}}_{\mathbf{X}}\triangleq \text{diag}(\bar{\V{\Sigma}}_{\mathbf{X}}^{-1})$.
Substituting $\mathbb{E}\big{\{} {\mathbf{h}}_j^{\rm{H}}{\bar{\V{\Sigma}}}_{\mathbf{X}}^{-1}
{\mathbf{h}}_j\big{\}}
=P_{\rm{a}}(\V{\Gamma}^\dag \otimes \V{1}^\dag) \bar{\V{\Lambda}}_{\mathbf{X}}$ into \eqref{AA1}, results in
\vspace{-0.2em}
\begin{equation}\label{RD4y}
\lambda^{\rm{opt}}\approx\frac{\sigma_{\rm{w}}^2\text{tr}\big{[}{\bar{\mathbf{\Sigma}}}_{\mathbf{X}}^{-1}\big{]}}
{P_{\rm{a}}(\V{\Gamma}^\dag \otimes \V{1}^\dag) \bar{\V{\Lambda}}_{\mathbf{X}}+3\text{tr}\big{[}{\bar{\mathbf{\Sigma}}}_{\mathbf{X}}^{-2}\big{]}}.
\vspace{-0.2em}
\end{equation}
  As seen in \eqref{RD4y},
$\lambda^{\rm{opt}}$ is inversely proportional to $P_{\rm{a}}$.

By substituting ${\mathbf{r}}_j=\mathbf{X}{\mathbf{h}}_j+{\mathbf{w}}_j$ in \eqref{E13} into \eqref{RD3}, $\hat{\mathbf{h}}_j$ can be written as a linear function of ${\mathbf{h}}_j$ as
\begin{equation}\label{RD4}
\hat{\mathbf{h}}_j=\mathbf{\Omega}{\mathbf{h}}_j+{\mathbf{w}}'_j,
\vspace{-0.3em}
\end{equation}
where
\begin{align}
\mathbf{\Omega} & \triangleq
\begin{bmatrix}
{\Omega}_{0,0} & {\Omega}_{0,1} & \dots & {\Omega}_{0,2K_{\rm{u}}-1} \\
{\Omega}_{1,0} &  {\Omega}_{1,1} & \dots & {\Omega}_{1,2K_{\rm{u}}-1} \\
\vdots & \vdots & \ddots & \vdots \\
{\Omega}_{2K_{\rm{u}}-1,0}  & {\Omega}_{2K_{\rm{u}}-1,1} & \dots  & {\Omega}_{2K_{\rm{u}}-1,2K_{\rm{u}}-1}
\end{bmatrix} \\  \nonumber
\vspace{1em}
&= \mathbf{I}-2\lambda^{\rm{opt}} \big{(}\bar{\mathbf{\Sigma}}_{\mathbf{X}}+2\lambda^{\rm{opt}}\mathbf{I}\big{)}^{-1},
\end{align}
and
\begin{align}\label{eq:OPI01}
\mathbf{w}'_j & \triangleq
\begin{bmatrix}
{{w}}'_{0,j,0}  \\
{{w}}'_{0,j,1} \\
\vdots  \\
{{w}}'_{K_{\rm{u}}-1,j,0} \\
{{w}}'_{K_{\rm{u}}-1,j,1} \\
\end{bmatrix}= \big{(}\bar{\mathbf{\Sigma}}_{\mathbf{X}} +2\lambda^{\rm{opt}} \mathbf{I}\big{)}^{-1}\mathbf{X}^\dag {\mathbf{w}_j}.
\end{align}
In \eqref{eq:OPI01},
$\mathbf{w}'_j$ is zero-mean complex Gaussian colored noise vector with covariance matrix given by
\begin{align} \nonumber
\mathbf{\Sigma}^{{{w'}}} & \triangleq
\begin{bmatrix}
{\Sigma}_{0,0}^{{{w}}'} & {\Sigma}_{0,1}^{{{w}}'} & \dots & {\Sigma}_{0,2K_{\rm{u}}-1}^{{{w}}'}\\
{\Sigma}_{1,0}^{{{w}}'} & {\Sigma}_{1,1}^{{{w}}'} & \dots & {\Sigma}_{1,2K_{\rm{u}}-1}^{{{w}}'}\\
\vdots & \vdots & \ddots & \vdots \\ \label{Eq40}
{\Sigma}_{2K_{\rm{u}}-1,0}^{{{w}}'} & {\Sigma}_{2K_{\rm{u}}-1,1}^{{{w}}'} & \dots & {\Sigma}_{2K_{\rm{u}}-1,2K_{\rm{u}}-1}^{{{w}}'}\\
\end{bmatrix} \\
&=\mathbb{E}
\big{\{}{\mathbf{w}}'_j\big{(}{{\mathbf{w}}'_{j}}\big{)}^{\rm{H}}\big{\}}=\sigma_{\rm{w}}^2\big{(}\bar{\mathbf{\Sigma}}_{\mathbf{X}} +2\lambda^{\rm{opt}}\mathbf{I}{)}^{-2}\bar{\mathbf{\Sigma}}_{\mathbf{X}},
\end{align}
where ${\Sigma}_{2k_1+f_1,2k_2+f_2}^{{{w}}'}=\mathbb{E}\big{\{}{w}'_{k_1,j,f_1}({w}'_{k_2,j,f_2})^*\big{\}}$.

\begin{figure}
\vspace{-2em}
  \centering
  \includegraphics[width=7cm]{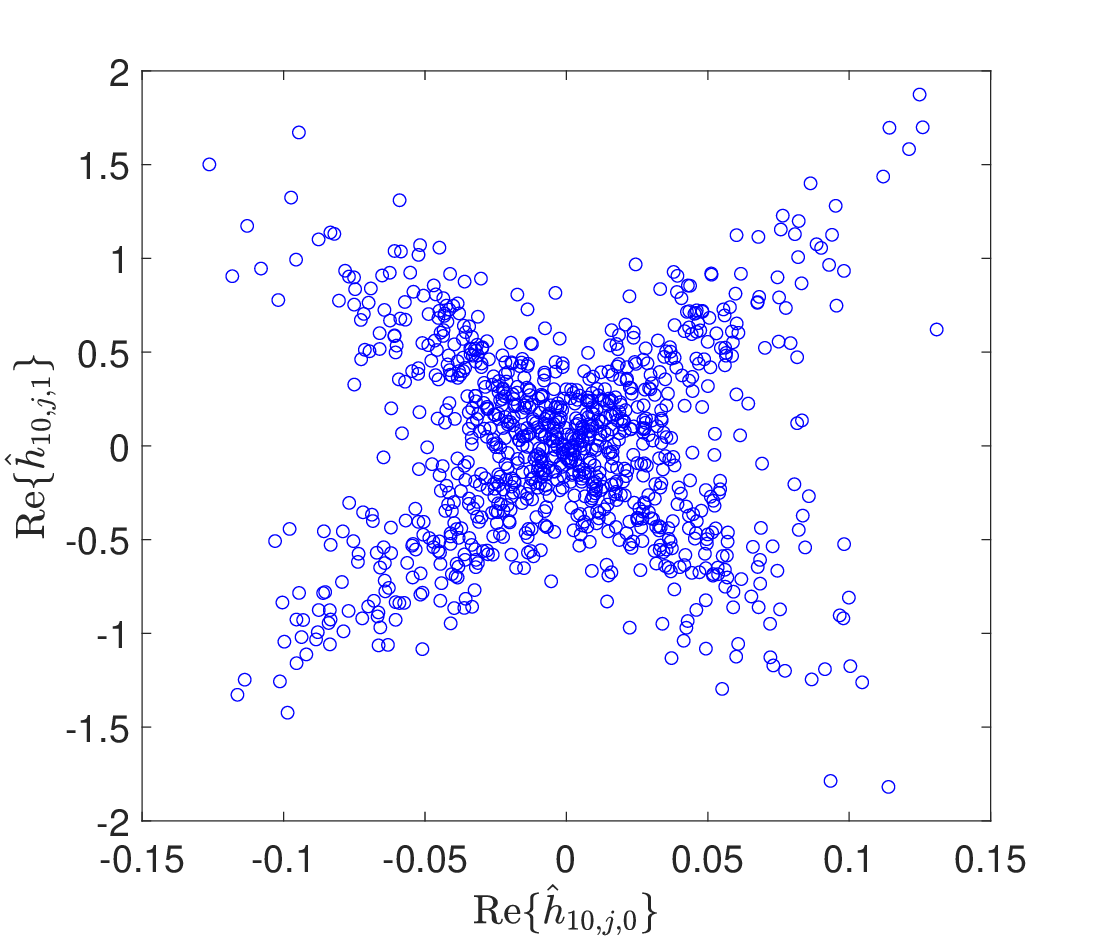}
  \caption{Scatter plot of $\Re\{\hat{{h}}_{10,j,1}\}$ and $\Re\{\hat{{h}}_{10,j,0}\}$ when device $k=$ $ 10$ is active, $K_{\rm{u}}=1536$, $N_{\rm{c}}=512$, $P_{\rm{a}}=0.01$, and SNR$=16$ dB.}\label{yuyuoere3}
\end{figure}

The elements of $\hat{\mathbf{h}}_j$ in \eqref{RD4} associated with the $k$th \ac{iot} device, i.e.,  $\hat{{h}}_{k,j,0}$ and $\hat{{h}}_{k,j,1}$ can be written as follows
\begin{align}\label{RD5}
\hat{{h}}_{k,j,f}
&={\Omega}_{2k+f,2k+f}{h}_{k,j,f}+{\Omega}_{2k+f,2k+\bar{f}}{h}_{k,j,\bar{f}} \\ \nonumber
&\hspace{-1em}+\sum_{n\neq k}\Big{\{}
{\Omega}_{2k+f,2n+f}{h}_{n,j,f}
 +{\Omega}_{2k+f,2n+\bar{f}}{h}_{n,j,\bar{f}}\Big{\}}+{w}'_{k,j,f},
\end{align}
where $f,\bar{f}  \in \{0,1\}$ and $\bar{f} \triangleq f+(-1)^f$. The second term on the right-hand side of \eqref{RD5} represents the effect of the multiuser interference caused by the active \ac{iot} devices in the network.
Due to the central limit theorem (CLT), $\hat{{h}}_{k,j,f}$, $f \in \{0,1\}$, in \eqref{RD5} given hypothesis $H_{0k}$ and
$H_{1k}$  can be accurately approximated by complex Gaussian random variables for sufficiently small values of $K$-factor ${\kappa_k}\triangleq |\mu_k|^2 / \sigma_k^2$ and large enough $P_{\rm{a}}K_{\rm{u}}$. Simulation results show that for ${\kappa_k}\triangleq |\mu_k|^2 / \sigma_k^2 <0.2$, Gaussian assumption is valid. In fact, the lower ${\kappa_k}$, the more reliable the Gaussian assumption is. It should be mentioned that
the random variables $\hat{{h}}_{k,j,0}$ and  $\hat{{h}}_{k,j,1}$ are not joint Gaussian random variables as shown in Fig. \ref{yuyuoere3}.
The mean, variance, and cross-correlation of $\hat{{h}}_{k,j,0}$ and  $\hat{{h}}_{k,j,1}$ are given in Lemma~\ref{lm_uu}.
\begin{customlemma}{1}\label{lm_uu}
First and second order statistics of the reconstructed signal for the $k$th \ac{iot} device in \eqref{RD5}, i.e.,
$\hat{{h}}_{k,j,0}$ and  $\hat{{h}}_{k,j,1}$, are given as follows
\vspace{-0.3em}
\begin{align}
\mathbb{E}\Big{\{}\hat{{h}}_{k,j,0}\big{|}H_{tk}\Big{\}} =\mathbb{E}\Big{\{}\hat{{h}}_{k,j,1}\big{|}H_{tk}\Big{\}}=0, \,\,\,\,\,\,\ t \in \{0,1\}
\end{align}
\begin{align}\label{RD7_1}
&{\Sigma}_{f,f}^{tk}  \triangleq \mathbb{V}{\rm{ar}} \Big{\{}\hat{{h}}_{k,j,f}\big{|} H_{tk} \Big{\}}=\mathbb{E}\Big{\{}\big{|}\hat{{h}}_{k,j,f}\big{|}^2\big{|}H_{tk}\Big{\}} \\  \nonumber
& =t \gamma_k\Big{(}{\Omega}_{2k+f,2k+f}^2
+{\Omega}_{2k+f,2k+\bar{f}}^2\Big{)}
\\ \nonumber
~~~&+P_{\rm{a}}\sum_{n\neq k}\gamma_n\Big{(}{\Omega}_{2k+f,2n+f}^2
+{\Omega}_{2k+f,2n+\bar{f}}^2\Big{)}
+{\Sigma}_{2k+f,2k+f}^{{{w}}'},
\end{align}
\vspace{-0.3em}
and
\begin{align}\nonumber
&{\Sigma}_{0,1}^{tk}=
\mathbb{C}{\rm{ov}}\Big{\{}\hat{{h}}_{k,j,0},\hat{{h}}_{k,j,1}\big{|} H_{tk} \Big{\}}=\mathbb{E}\Big{\{}\hat{{h}}_{k,j,0}\hat{{h}}_{k,j,1}^*  \big{|} H_{tk} \Big{\}}  \\ \nonumber
&\,\,\,\,\,\,\,\ =t\gamma_k\Big{(}{\Omega}_{2k,2k}{\Omega}_{2k+1,2k}+{\Omega}_{2k+1,2k+1}{\Omega}_{2k,2k+1}\Big{)} \\ \nonumber
&+P_{\rm{a}}\sum_{n\neq k} \gamma_n \Big{(} {\Omega}_{2k,2n}{\Omega}_{2k+1,2n}+{\Omega}_{2k+1,2n+1}{\Omega}_{2k,2n+1}\Big{)} \\ \label{eq:case1n}
& +\mathbf{\Sigma}_{2k,2k+1}^{\rm{w}'},
\end{align}
where $\gamma_k = \eta_k p_k(\sigma_k^2+|\mu_k|^2)$, ${\Sigma}_{1,0}^{tk}={\Sigma}_{0,1}^{tk}$, $t,f \in \{0,1\}$, $\bar{f} \triangleq f+(-1)^f$, and  ${\Sigma}_{2k+f,2k+f}^{{{w}}'}$
is given in \eqref{Eq40}  ({\it{Proof}} in Appendix~\ref{ditribution}) \qed
\vspace{-0.5em}
\end{customlemma}

Since the joint \ac{pdf} of $\hat{{h}}_{k,j,0}$ and  $\hat{{h}}_{k,j,1}$ given $H_{tk}$, i.e.,
$p\big{(}\hat{{h}}_{k,j,0}$, $\hat{{h}}_{k,j,1}|H_{tk}\big{)}$, $t\in \{0,1\}$,
cannot be expressed in a tractable mathematical form and since there is high correlation between
$\hat{{h}}_{k,j,0}$ and  $\hat{{h}}_{k,j,1}$, we can either use $\hat{{h}}_{k,j,0}$ or $\hat{{h}}_{k,j,1}$ to identify
the transmission state of the $k$th \ac{iot} device.
Moreover, the in-phase and quadrature components of $\hat{{h}}_{k,j,f}$ can be accurately approximated by correlated joint Gaussian random variables due to the CLT for sufficiently small values of $K$-factor ${\kappa_k}\triangleq |\mu_k|^2 / \sigma_k^2$ and large enough $P_{\rm{a}}K_{\rm{u}}$. To verify the credibility of Gaussian assumption, we evaluate the kurtosis and skewness for $\Re \{\hat{{h}}_{k,j,0}\}$ and $\Re \{\hat{{h}}_{k,j,1}\}$ in Table \ref{TFV}.

 Similar to the proof of Lemma 1, we can show that
the distribution of the reconstructed signal for the $k$th \ac{iot} device is given as follows
\vspace{-0.5em}
\begin{equation}\label{eq:functionbbn}
\begin{bmatrix}
\Re\{\hat{{h}}_{k,j,0}\} \\
\Im\{\hat{{h}}_{k,j,0}\}
\end{bmatrix} \sim
    \begin{cases}
      {\cal{N}}\big{(}\mathbf{0},{\mathbf{C}}_{0,0}^{0k} \big{)} , & \,\,\ H_{0k} \\
       {\cal{N}}\big{(}\mathbf{0},{\mathbf{C}}_{0,0}^{1k} \big{)} , & \,\,\ H_{1k}
    \end{cases},
  \end{equation}
and
\begin{equation}\label{eq:functionbbnnm}
\begin{bmatrix}
\Re\{\hat{{h}}_{k,j,1}\} \\
\Im\{\hat{{h}}_{k,j,1}\}
\end{bmatrix} \sim
    \begin{cases}
      {\cal{C}\cal{N}}\left(\mathbf{0},{\mathbf{C}}_{1,1}^{0k} \right) , & \,\,\ H_{0k} \\
       {\cal{C}\cal{N}}\left(\mathbf{0},{\mathbf{C}}_{1,1}^{1k} \right) , & \,\,\ H_{1k}
    \end{cases},
  \end{equation}
  where
\begin{align}\label{uuirnnmccc}
\mathbf{C}_{f,f}^{tk}=
\begin{bmatrix}
\bar{{\Sigma}}_{f,f}^{tk}  & \rho_{f,f}^{tk} \\
\rho_{f,f}^{tk} & \tilde{{\Sigma}}_{f,f}^{tk}
\end{bmatrix},
\end{align}

\begin{align}
\rho_{f,f}^{tk}&=\mathbb{E}\Big{\{}\Re\{\hat{{h}}_{k,j,f}\}\Im\{\hat{{h}}_{k,j,f}\}\big{|}H_{tk}\Big{\}} \\ \nonumber
&=t\bar{\mu}_k\tilde{\mu}_k\eta_k p_k\Big{(}{\Omega}_{2k+f,2k+f}^2
+{\Omega}_{2k+f,2k+\bar{f}}^2\Big{)}
\\ \nonumber
&~~~+P_{\rm{a}}\sum_{n\neq k}\bar{\mu}_n\tilde{\mu}_n \eta_n p_n\Big{(}{\Omega}_{2k+f,2n+f}^2
+{\Omega}_{2k+f,2n+\bar{f}}^2\Big{)},
\end{align}
\begin{align} \nonumber
&\bar{{\Sigma}}_{f,f}^{tk}  \triangleq \mathbb{V}\text{ar} \Big{\{}\Re\{\hat{{h}}_{k,j,f}\}\big{|} H_{tk} \Big{\}}=\mathbb{E}\Big{\{}\big{(}\Re\{\hat{{h}}_{k,j,f}\}\big{)}^2\big{|}H_{tk}\Big{\}} \\ \nonumber
& =t(\sigma_k^2/2+|\bar{\mu}_k|^2) \eta_k p_k\Big{(}{\Omega}_{2k+f,2k+f}^2
+{\Omega}_{2k+f,2k+\bar{f}}^2\Big{)}
\\ \nonumber
~~~&+P_{\rm{a}}\sum_{n\neq k}(\sigma_n^2/2+|\bar{\mu}_n|^2) \eta_n p_n\Big{(}{\Omega}_{2k+f,2n+f}^2
+{\Omega}_{2k+f,2n+\bar{f}}^2\Big{)}\\
~~~&+{\Sigma}_{2k+f,2k+f}^{{{w}}'}/2,
\end{align}
and
\begin{align} \nonumber
&\tilde{{\Sigma}}_{f,f}^{tk}  \triangleq \mathbb{V}\text{ar} \Big{\{}\Im\{\hat{{h}}_{k,j,f}\}\big{|} H_{tk} \Big{\}}=\mathbb{E}\Big{\{}\big{(}\Im\{\hat{{h}}_{k,j,f}\}\big{)}^2\big{|}H_{tk}\Big{\}} \\ \nonumber
& =t(\sigma_k^2/2+|\tilde{\mu}_k|^2) \eta_k p_k\Big{(}{\Omega}_{2k+f,2k+f}^2
+{\Omega}_{2k+f,2k+\bar{f}}^2\Big{)}
\\ \nonumber
~~~&+P_{\rm{a}}\sum_{n\neq k}(\sigma_n^2/2+|\tilde{\mu}_n|^2) \eta_n p_n\Big{(}{\Omega}_{2k+f,2n+f}^2
+{\Omega}_{2k+f,2n+\bar{f}}^2\Big{)}\\
~~~&+{\Sigma}_{2k+f,2k+f}^{{{w}}'}/2,
\end{align}
with $\bar{\mu}_k \triangleq \Re\{\mu_k\}$ and $\tilde{\mu}_k \triangleq \Im\{\mu_k\}$.

\begin{table}\label{yuybnm}
\centering
\caption{Credibility of Gaussian assumption for $\hat{{h}}_{k,j,0}$ and $\hat{{h}}_{k,j,1}$.} \label{TFV}
\begin{tabular}{ |p{2.1cm}||p{1cm}|p{1cm}|p{1.8cm}| }
 \hline
 ~~~~~ Variable & Kurtosis    & Skewness & ~~~ Variance     \\
 \hline
Gaussian (theory) & ~~~ $3$    & ~~~ 0& 0.002 \{0.2657\} \\
$~~~~ \Re\{\hat{{h}}_{k,j,0}\}$   & ~  $3.105$  & \hspace{0.031em} $0.0224$ & ~~~~ 0.001\\
$~~~~\Re\{\hat{{h}}_{k,j,1}\}$  &  ~ $3.014$  &  $-0.0233$ & ~~ \{0.2643\} \\
 \hline
\end{tabular}
\vspace{-1em}
\end{table}

The larger the ratio of the variances in \eqref{eq:functionbbn} and \eqref{eq:functionbbnnm}, i.e.,  ${\Sigma}_{0,0}^{1k}/{\Sigma}_{0,0}^{0k}$, and  ${\Sigma}_{1,1}^{1k}/{\Sigma}_{1,1}^{0k}$, the better identification performance. Accordingly, we use the reconstructed signal in \eqref{RD5} for the identification of the $k$th \ac{iot} device as follows
\begin{align}\label{dd44444444}
\breve{\bf{h}}_{k,j}\triangleq \begin{bmatrix}
\Re\{\hat{{h}}_{k,j,1}\} \\
\Im\{\hat{{h}}_{k,j,1}\}
\end{bmatrix} \mathbb{I} \big{\{}\varrho_k < 0\big{\}}+
\begin{bmatrix}
\Re\{\hat{{h}}_{k,j,0}\} \\
\Im\{\hat{{h}}_{k,j,0}\}
\end{bmatrix} \mathbb{I} \big{\{}\varrho_k \geq 0\big{\}},
\end{align}
where $\breve{\mathbf{h}}_{k,j}=[\breve{{h}}_{k,j,0},\breve{{h}}_{k,j,1}]^\dag$, $k \in \mathcal{X}_{\rm{u}}$, and
\begin{align}
\varrho_k=
\Bigg{(}\frac{{\Sigma}_{0,0}^{1k}}{{\Sigma}_{0,0}^{0k}}-\frac{{\Sigma}_{1,1}^{1k}}{{\Sigma}_{1,1}^{0k}}\Bigg{)}.
\end{align}
Using the reconstructed signal in the form of \eqref{dd44444444} enables us to derive closed-form expressions for the correct identification and false alarm rates.
In order to identify the transmission state of the $k$th \ac{iot} device, $k \in \mathcal{X}_{\rm{u}}$, based on $\breve{\mathbf{h}}_{k,j}$, the
\ac{mlr} test can be used \cite{kay1998fundamentals}.
\begin{customlemma}{2}\label{one}
The optimal \ac{mlr} decision rule for \ac{iot} DI
based on the reconstructed signal $\breve{\mathbf{h}}_{k,j}$, $k \in \mathcal{X}_{\rm{u}}$, in \eqref{dd44444444} is given by
\begin{equation}\label{eq:function}
   {d}_k =
    \begin{cases}
      H_{1k}, & \,\,\ \phi\big{(}\breve{\mathbf{h}}_{k,j} \big{)} \geq \theta_k \\
      H_{0k}, &  \,\,\  \phi\big{(}\breve{\mathbf{h}}_{k,j} \big{)} < \theta_k
    \end{cases},
  \end{equation}
where
\vspace{-0.8em}
\begin{align}\label{eq:io9xzb7120m}
\phi\big{(}\breve{\mathbf{h}}_{k,j} \big{)}= \sum_{n=0}^{1}\chi_{f,f}[n]z_{k,j}^2[n],
\end{align}
\begin{align}\label{u899kooD}
\chi_{f,f}[n] \triangleq \frac{\lambda_{f,f}[n]}{\lambda_{f,f}[n]+1}.
\end{align}
In \eqref{u899kooD}, $\lambda_{f,f}[0]$ and $\lambda_{f,f}[1]$ are the eigenvalues of the symmetric matrix $\mathbf{B}_{f,f}^{1k} \triangleq (\mathbf{A}_{f,f}^{0k})^\dag{\mathbf{C}}_{f,f}^{1k}\mathbf{A}_{f,f}^{0k}$ (${\mathbf{C}}_{f,f}^{tk}$ is defined in \eqref{uuirnnmccc}), and
\begin{equation}
[z_{k,j}[0] , z_{k,j}[1]]^\dag \triangleq (\mathbf{V}_{f,f}^{1k})^{\dag}(\mathbf{A}_{f,f}^{0k})^\dag \breve{\mathbf{h}}_{k,j},
\vspace{-0.2em}
\end{equation}
where ${\bf{V}}_{f,f}^{1k}$ is the modal matrix  of ${\bf{B}}_{f,f}^{1k}$, i.e., $\big{(}{\bf{V}}_{f,f}^{1k}\big{)}^\dag{{\bf{B}}}_{f,f}^{1k}{\bf{V}}_{f,f}^{1k}$ $={\bf{\Lambda}}_{f,f}^{1k}$ (matrix of eigenvalues for ${\bf{B}}_{f,f}^{1k}$),
${\bf{A}}_{f,f}^{0k} \triangleq {\bf{V}}_{f,f}^{0k}\big{(}{\bf{\Lambda}}_{f,f}^{0k}\big{)}^{\frac{-1}{2}}$, $\big{(}\mathbf{V}_{f,f}^{0k}\big{)}^\dag{\mathbf{C}}_{f,f}^{0k}\mathbf{V}_{f,f}^{0k}=\mathbf{\Lambda}_{f,f}^{0k}$,
and $f=1$ for $\varrho_k < 0$, and $f=0$ for $\varrho_k \geq 0$.
The threshold value for the $k$th  \ac{iot} device, i.e.,  $\theta_k$,  is set
for a desirable false alarm rate $P_k^{{\rm{(f)}}} \triangleq \mathbb{P}\big{\{}{d}_k= {H}_{1k}|{H}_{0k}\big{\}}$  as follows
\begin{align}\label{flase}
P_k^{(\rm{f})}=\frac{1}{2\pi}\int_{\theta_k}^{+\infty} \int_{-\infty}^{+\infty}
\prod_{n=0}^{1}\frac{\exp(-j\omega x)}{\sqrt{1-2j\chi_{f,f}[n] \omega}} {\rm d} \omega {\rm d}x.
\end{align}
By using \eqref{eq:io9xzb7120m}, we can obtain the correct identification rate for the $k$th \ac{iot} device as follows
\begin{align}\label{correct}
P_k^{(\rm{c})}=\frac{1}{2\pi}\int_{\theta_k}^{+\infty} \int_{-\infty}^{+\infty} \prod_{n=0}^{1}\frac{\exp(-j\omega x)}{\sqrt{1-2j\lambda_{f,f}[n] \omega}}{\rm d} \omega {\rm d}x.
\end{align}
({\it{Proof}} in Appendix~\ref{ap:beyesian ruledf}) \qed
\end{customlemma}

\begin{algorithm}[t]
    \caption{Squared $\ell_2$-norm \ac{ssr} \ac{iot} DI}\label{Table1xrrrr}
    \begin{algorithmic}[1]
      \Statex \textbf{Input:} $\mathbf{X}$, $\mathbf{R}$, $P_k^{(\rm{f})}$, $n_k$, $ k \in \mathcal{X}_{\rm{u}}$
      \Statex \textbf{Output:} Active \ac{iot} set $\hat{\Set{X}}_{\rm{a}}$
      \Statex \textbf{Initialization:} $\hat{\Set{X}}_{\rm{a}}=\emptyset$
      \For {$k=0,1,\dots,K_{\rm{u}}-1$}
      \State Obtain $\theta_k$ by using \eqref{flase}
      \State Obtain $\breve{\mathbf{h}}_{k,j}$ by employing \eqref{RD3} and \eqref{dd44444444}
      \State Compute $\phi\big{(}\breve{\mathbf{h}}_{k,j} \big{)}$ for $j=0,1,\dots,L-1$ using \eqref{eq:io9xzb7120m}
      \State Identify the transmission state of the $k$th \ac{iot} device
       \Statex \ \ \ \ \ by employing \eqref{eq:function} and then \eqref{eq:61c}
      \If {${D}_k=H_{1k}$}
      \State  $\hat{\Set{X}}_{\rm{a}}\leftarrow\{\hat{\Set{X}}_{\rm{a}},k\}$
      \EndIf
      \EndFor
    \end{algorithmic}
  \end{algorithm}

The decisions corresponding to the $L$ measurements for the $k$th \ac{iot} device, i.e., $d_{k,\ell}$, $\ell=1,2,\dots,L$,  can be fused together as
\begin{equation}\label{eq:61c}
   D_k =
    \begin{cases}
      H_{1k}, & \,\,\ \sum_{\ell=1}^{L}d_{k,\ell} \geq n_k \\
      H_{0k}, &  \,\,\  \sum_{\ell=1}^{L}d_{k,\ell} < n_k,
    \end{cases},
  \end{equation}
  where $n_k$ is an integer value \cite{al2019decision}.
A formal description of the proposed squared $\ell_2$-norm \ac{ssr} \ac{iot} DI algorithm is given in Algorithm
\ref{Table1xrrrr}. 

\subsection{\ac{iot} DI for Unknown Activity Rate}
Here, we develop an \ac{iot} DI algorithm for the case of unknown activity rate $P_{\rm{a}}$  at the \ac{bs}.
The convex relaxation through $\ell_1-\ell_2$ mixed-norm is employed for signal reconstruction, which can directly identify the active \ac{iot} devices through the non-zero elements of the reconstructed signal.
This is attributed to the $\ell_1-\ell_2$ mixed-norm ability to
provide sparse estimates.
Note that the proposed $\ell_1-\ell_2$  mixed-norm \ac{sssr} \ac{iot} DI algorithm does not require knowledge of transmit power by \ac{iot} devices.
\subsubsection{{\ac{bic} $\ell_1-\ell_2$  Mixed-Norm \ac{sssr} \ac{iot} DI Algorithm}}
Let us consider $P_{\rm{a}} \in [0 , \ P_{\rm{max}}]$, where
$P_{\rm{a}}$ and the maximum activity rate $P_{\rm{max}}$ are unknown at the \ac{bs}.
Since in-phase and  quadrature  components  tend  to
be  either  zero  or  non-zero  simultaneously, this provides additional grouping in \ac{sssr}.
By stacking the in-phase and quadrature components ($\mathbf{X}$ is a real-valued matrix), we can write \eqref{u8i02xngo} as
\begin{align}\label{wqacvbiolp}
{\bf{Y}}={\bf{X}}{\bf{U}}+{\bf{V}},
\end{align}
where
\begin{subequations}
\begin{align}
{\bf{Y}}& \triangleq \big{[}\Re{{\bf{R}}} \ \Im{{\bf{R}}}\big{]},
\\
{\bf{U}} &\triangleq \big{[}\Re{{\bf{H}}} \ \Im{{\bf{H}}}\big{]},
\\
{\bf{V}} &   \triangleq \big{[}\Re{{\bf{W}}} \  \Im{{\bf{W}}}\big{]}.
\end{align}
\end{subequations}

$\Re\{rt\}$

For block-sparse matrix $\bf{U}$ in \eqref{wqacvbiolp}, the $\ell_1-\ell_2$  mixed-norm \ac{sssr} is given as follows
\begin{equation}\label{elastic}
\begin{aligned}
{\hat{ {\bf{U}}}}= \underset{\hspace{2em}{{\bf{U}}}} {\text{arg  min}}
 \,\ \frac{1}{2}\big{\|}{\bf{Y}}-{\bf{X}}{\bf{U}}\big{\|}_{\rm{F}}^2
+N_{\rm{d}}\lambda_{\rm{g}} \sum_{k=0}^{K_{\rm{u}}-1} {\big{\|}\mathbf{u}_{\rm{G}_{k}}\big{\|}_2},
\end{aligned}
\end{equation}
where $N_{\rm{d}} \triangleq 2LN_{\rm{c}}$
\begin{align}\label{990i}
{\bf{u}}_{{\rm{G}}_{k}} \triangleq \big{[}{\bf{U}}_{2k,\cdot} \ {\bf{U}}_{2k+1,\cdot} \big{]},
\end{align}
and ${\bf{U}}_{k,\cdot} $ is the $k$th row of $\bf{U}$. In \eqref{elastic}, $\lambda_{\rm{g}}$ represents the tuning parameter which is unknown and time-varying. The degrees of sparsity depends on $\lambda_{\rm{g}}$; the larger
$\lambda_{\rm{g}}$ is, the sparser the estimate is.
For unknown $\lambda_{\rm{g}}$,
model order selection methods can be employed to identify active \ac{iot} devices.
By extending the \ac{bic} model order selection method in \cite{yuan2006model} to multiple measurement vectors, the reconstructed matrix ${\hat{ {\mathbf{U}}}}$ is given by
\begin{align}\label{yyiyy44y77712az}
{\hat{ {\bf{U}}}}={\hat{ {\bf{U}}}}^{({\hat{\lambda}})},
\end{align}
 where
\begin{align}\label{E1823288io}
\hat{\lambda} = \underset{\lambda \in [\lambda_{\rm L},\lambda_{\rm U}]}{\text{arg  min}} C_{\rm{BIC}}(\lambda),
\end{align}
\begin{align}\label{777690897450025}
C_{\rm{BIC}}(\lambda) \triangleq {\rm{log}}\Big{(}\frac{1}{N_{\rm{d}}}{\Big{\|}{\mathbf{Y}}-\mathbf{X}\hat{\mathbf{U}}^{(\lambda)} \big{\|}_{\rm{F}}^2}\Big{)}+{\rm{log}}(N_{\rm{d}})\frac{df}{N_{\rm{d}}},
\end{align}
\begin{align}\label{uerrrbvqw905}
\hspace{-1em}{\hat{ {\mathbf{U}}}}^{({\lambda})}&= \underset{\hspace{1.6em}{{\mathbf{U}}}} {\text{arg  min}}
 \,\ \frac{1}{2}\big{\|}{\mathbf{Y}}-\mathbf{X}{\mathbf{U}}\big{\|}_{\rm{F}}^2
+N_{\rm{d}}\lambda \sum_{k=0}^{K_{\rm{u}}-1} {\big{\|}\mathbf{u}_{{\rm{G}}_k}\big{\|}_2},
\end{align}
and ${d}{f}$ is the degree of freedom which is given as follows
\begin{align}\label{qqqqqqqqqxcvc}
{d}{f}=\sum_{k=0}^{K_{\rm{u}}-1}\mathbb{I}\big{\{} {\big{\|}\hat{\mathbf{u}}_{{\rm{G}}_k}\big{\|}_2}>0\big{\}}+(2L-1)\sum_{k=0}^{K_{\rm{u}}-1}\frac{{\big{\|}\hat{\mathbf{u}}_{{\rm{G}}_k}\big{\|}_2}}
{{\big{\|}\hat{\mathbf{u}}_{{\rm{G}}_k}^{\rm{LS}}\big{\|}_2}},
\end{align}
where $\hat{\mathbf{u}}_{{\rm{G}}_k}^{\rm{LS}}$ is the \ac{ls} estimate for the $k$th \ac{iot} device signal. 

The Karush–Kuhn–Tucker (KKT) optimality conditions of the optimization problem in \eqref{uerrrbvqw905} is given as
\begin{subequations}\label{kktuu0obvcty0009}
\begin{align} \label{hin1x90olpvvvv}
-{\mathbf{\Psi}_{k}}+ N_{\rm{d}} \lambda \frac{{\mathbf{u}}_{{\rm{G}}_k}}{\big{\|}{\mathbf{u}}_{{\rm{G}}_k}\big{\|}_2}&=\mathbf{0} \,\,\,\,\,\,\,\,\,\,\,\,\,\,\ \text{if}\ {\mathbf{u}}_{{\rm{G}}_k} \neq \mathbf{0}^\dag
\\  \label{hin1x90olpvvvvb}
\,\,\,\,\,\,\,\,\,\,\,\,\,\,\,\,\,\,\,\,\,\,\,\,\,\,\,\,\,\,\,\,\,\,\,\,\,\,\,\,\,\,\,\
\big{\|}{\mathbf{\Psi}_{k}}\big{\|}_2& \leq N_{\rm{d}} {\lambda} \,\,\,\,\,\ \text{if}\ {\mathbf{u}}_{{\rm{G}}_k} = \mathbf{0}^\dag,
\end{align}
\end{subequations}
where
\begin{align} \label{hin3nnnz}
\mathbf{\Psi}_{k}& \triangleq \nabla_{{\mathbf{u}}_{{\rm{G}}_k}}\frac{1}{2}\big{\|}{\mathbf{Y}}-\mathbf{X}{\mathbf{U}}\big{\|}_{\rm{F}}^2 \\ \nonumber
&=\big{[}{\mathbf{X}}_{\cdot,2k}^\dag({\mathbf{Y}}-\mathbf{X}{\mathbf{U}})\ \ {\mathbf{X}}_{\cdot,2k+1}^\dag({\mathbf{Y}}-\mathbf{X}{\mathbf{U}})\big{]},
\end{align}
and $\mathbf{X}_{\cdot,k} $ is the $k$th column of $\mathbf{X}$.
Let us write $\mathbf{\Psi}_{k}$ as
\begin{align} \label{hin3nnnz}
\mathbf{\Psi}_{k}
=\mathbf{\varphi}_{k}-{\mathbf{u}}_{{\rm{G}}_k}\mathbf{\Lambda}_k
\end{align}
where
\begin{align}\label{eq:iiiiossssssssssssssxxx0}
\mathbf{\varphi}_{k}=\big{[}{\mathbf{X}}_{\cdot,2k}^\dag({\mathbf{Y}}-\mathbf{X}{\mathbf{U}_{-\{2k,\cdot\}}})\ \ {\mathbf{X}}_{\cdot,2k+1}^\dag({\mathbf{Y}}-\mathbf{X}{\mathbf{U}_{-\{2k+1,\cdot\}}})\big{]},
\end{align}
and
\begin{align}
\mathbf{\Lambda}_k \triangleq
{\rm{diag}}& \big{\{}{\mathbf{X}}_{\cdot,2k}^\dag {\mathbf{X}}_{\cdot,2k}, \dots, {\mathbf{X}}_{\cdot,2k}^\dag {\mathbf{X}}_{\cdot,2k}, \\ \nonumber
& {\mathbf{X}}_{\cdot,2k+1}^\dag {\mathbf{X}}_{\cdot,2k+1}, \dots, {\mathbf{X}}_{\cdot,2k+1}^\dag {\mathbf{X}}_{\cdot,2k+1}\big{\}}
\end{align}
with ${\mathbf{U}}_{-\{i,\cdot\}}$ as the matrix ${\mathbf{U}}$ with the $i$th row
being set to $\mathbf{0}^\dag$. The dimension of the diagonal matrix $\mathbf{\Lambda}_k$ is $4L \times 4L$.
\begin{algorithm}[t!]\label{mpl1l2}
    \caption{\ac{bic} $\ell_1-\ell_2$  mixed-norm \ac{sssr} \ac{iot} DI algorithm}
    \begin{algorithmic}[1]
    \Statex \textbf{Input:} $\mathbf{Y}$, $\mathbf{X}$, $\mathbf{\Lambda}_k$, $k=0,1,\dots,K_{\rm{u}}-1$, $\lambda \in [\lambda_{\rm{L}}, \lambda_{\rm{U}}]$, $M_{\rm{C}}$, and $M_{\rm{G}}$
      \Statex \textbf{Output:} Active \ac{iot} set $\hat{\mathcal{X}}_{\rm{a}}$
      \Statex \textbf{Initialization:} $\hat{\mathcal{X}}_{\rm{a}}=\emptyset$, $i=1$, $Golden=1$, $\lambda_1=\lambda_{\rm L}$
      \While{$Golden=1$}
      \Statex $\hat{\mathbf{U}}^{[0]}=\bf{0}$,
       $t=1$, $SSSR=1$
        \While{$SSSR=1$}
        \For{$k=0,1,\dots,K_{\rm{u}}-1$}
        \Statex  \hspace{4em} Obtain $\bm{\varphi}_{k}^{[t]}$ by employing \eqref{eq:iiiiossssssssssssssxxx077800nnbm}
        \State \textbf{if} ${\|\bm{\varphi}_{k}^{[t]}\|_2} \leq N_{\rm{c}} \lambda_i$ \textbf{then}
        \Statex \ \ \ \ \ \ \ \ \ \ \ \ \ \ \ \ \  $\hat{\mathbf{u}}_{{\rm{G}}_k}^{[t]}=\V{0}$
        \Statex \ \ \ \ \ \ \ \ \ \ \ \ \   \textbf{else}
        \State \ \ \ \ \ Update ${\mathbf{u}}_{{\rm{G}}_k}^{[t]}$ as in \eqref{eq:tttttttttttttcccccyuyu}
       \Statex \ \ \ \ \ \ \ \ \ \ \ \ \textbf{end if}
        \EndFor\label{euclidendwhilev}
        \State \ \ \ \ \textbf{if} $\big{(}\|\hat{\mathbf{U}}^{[t]}-\hat{\mathbf{U}}^{[t-1]}\| \geq \epsilon_{\rm c} \big{)}\cap (t < M_{\rm C})$ \textbf{then}
        \State   \ \ \ \ \ \ \ \ $t\leftarrow t+1$
         \State \ \ \  \textbf{else}
        \Statex \hspace{5em} $\hat{\RM{U}}^{(\lambda_i)}=\hat{\mathbf{U}}^{[t]}$,  $SSSR=0$
        \Statex \ \ \ \ \ \ \ \ \ \ \ \ \textbf{end if}
         \EndWhile\label{euclidendwhil}
         \Statex \ \ \ \ \ \ \ \ \ \ \ \ $i \leftarrow i+1$
          \State \textbf{if} $i=2$ \textbf{then}
        \Statex \ \ \ \ \ \ \ \ \ \ \ \ \   $\lambda_i=\lambda_{\rm U}$, $Golden=1$
       \State \textbf{else if} $\big{(}|\lambda_{i-1}-\lambda_{i-2}| \geq \epsilon_{\rm g}) \cap (i< M_{\rm{G}}+1)$  \textbf{then}
       \Statex \ \ \ \ \ \ \ \ \ \ \ \ \ \  Find $\lambda_{i}$ using the Golden selection search
       \Statex \ \ \ \ \ \ \ \ \ \ \ \ \ \  $Golden=1$
       \State \textbf{else}
       \Statex \ \ \ \ \ \ \ \ \ $\hat{\lambda} = \underset{\lambda \in \{\lambda_{i-1},\lambda_{i-2}\}}{\text{arg  min}} C_{\rm{BIC}}(\lambda)$,
       \Statex \ \ \ \ \ \ \ \ \  $\hat{\mathbf{U}}=\hat{\mathbf{U}}^{(\hat{\lambda})}$, $Golden=0$,
        \State \textbf{end if}
        \EndWhile
        \State $\hat{\Set{X}}_{\rm{a}}=\big{\{}\forall k\in \{0, \dots K_{\rm{u}}-1\}\big{|} \|\hat{\mathbf{u}}_{{\rm{G}}_k}\|_2\neq0 \big{\}}$.
\end{algorithmic} \label{euclidendwhilemmmmmeqqqq}
  \end{algorithm}

From \eqref{hin1x90olpvvvv} and \eqref{hin3nnnz}, we have
${\mathbf{u}}_{{\rm{G}}_k} \big{(}{N_{\rm{d}}{\lambda}\mathbf{I}}/{{\big{\|}{{\mathbf{u}}_{{\rm{G}}_k}}\big{\|}_2}}+\mathbf{\Lambda}_k\big{)}$ $= \bm{\varphi}_{k}$
when the $k$th \ac{iot} device is active. In contrast, when the $k$th \ac{iot} device is inactive, $\mathbf{\Psi}_{k}=\bm{\varphi}_{k}$.
Hence, we can write
\begin{align}\label{eq:tttttttttttttccccc}
{\mathbf{u}}_{{\rm{G}}_k}={\mathbb{I}\Big{\{}{\|\mathbf{\varphi}_{k}\|_2}> N_{\rm{d}} \lambda\Big{\}}\bm{\varphi}_{k}}{\Bigg{(}\frac{N_{\rm{d}} \lambda}{\big{\|}{{\mathbf{u}}_{{\rm{G}}_k}}\big{\|}_2}\mathbf{I}+\mathbf{\Lambda}_k\Bigg{)}^{-1}}.
\end{align}

To solve the optimization \eqref{uerrrbvqw905}, we can use block-coordinate descent algorithm, where
consists of solving each $\mathbf{u}_{{\rm{G}}_{k}}$ in \eqref{990i}
at a time. By starting from a sparse solution like, ${\hat{ {\mathbf{U}}}}=\mathbf{0}$, at each iteration, we check for a given $k$ whether $\mathbf{u}_{{\rm{G}}_{k}}$ is optimal or not based on the conditions in \eqref{kktuu0obvcty0009}.
If ${\|\bm{\varphi}_{k}\|_2} \leq N_{\rm{d}} \lambda$, $\hat{\mathbf{u}}_{{\rm{G}}_k}=\V{0}$; otherwise, ${\mathbf{u}}_{{\rm{G}}_k}$ at the $t$th iteration is iteratively updated as
\begin{align}\label{eq:tttttttttttttcccccyuyu}
\hspace{-0.5em}\hat{\mathbf{u}}_{{\rm{G}}_k}^{[t]}={\mathbb{I}\Big{\{}{\|\bm{\varphi}_{k}^{[t]}\|_2}> N_{\rm{d}}\lambda\Big{\}}\bm{\varphi}_{k}^{[t]}}{\Bigg{(}\frac{N_{\rm{d}}\lambda}{\big{\|}{\hat{\mathbf{u}}_{{\rm{G}}_k}^{[t-1]}}\big{\|}_2}\mathbf{I}+\mathbf{\Lambda}_k\Bigg{)}^{-1}},
\end{align}
where
\begin{align}\label{eq:iiiiossssssssssssssxxx077800nnbm}
\bm{\varphi}_{k}^{[t]}=\big{[}{\mathbf{X}}_{\cdot,2k}^\dag({\mathbf{Y}}-\mathbf{X}{\mathbf{U}_{-\{2k,\cdot\}}^{[t-1]}})\ \ {\mathbf{X}}_{\cdot,2k+1}^\dag({\mathbf{Y}}-\mathbf{X}{\mathbf{U}_{-\{2k+1,\cdot\}}^{[t-1]}})\big{]}.
\end{align}
This procedure continues until the absolute difference of successive iterations becomes smaller than the tolerance value $\epsilon_{\rm c}$.

\subsubsection{Efficient One-dimensional Search}
Efficient one dimensional
iterative search algorithms can be used to solve the optimization problem in \eqref{E1823288io}.
In an iterative search method, the interval $[\lambda_{\rm{L}}, \lambda_{\rm{U}}]$ is repeatedly reduced on the basis of function evaluations
until a reduced bracket $[\lambda_{\rm{L}}, \lambda_{\rm{U}}]$ is achieved which is sufficiently small. These
methods can be applied to any function and differentiability of the function is not
essential.
An iterative search method in which iterations can be performed
until the desired accuracy in either the minimizer or the minimum value of the
objective function is achieved is the golden-section search method \cite{antoniou2007practical}.

{\it Convergence of the Optimization Problem in \eqref{uerrrbvqw905}}: It has been shown  that  for an optimization  problem whose objective  function is  the  sum  of a smooth and convex function and a non-smooth but block-separable convex function, block-coordinate descent optimization converges towards the global minimum of the problem \cite{tseng2001convergence}. In \eqref{uerrrbvqw905}, ${\|}{\mathbf{Y}}-\mathbf{X}{\mathbf{U}}{\|}_{\rm{F}}^2$ is a  smooth  and  differentiable convex function and $\sum_{k=0}^{K_{\rm{u}}-1} {{\|}\mathbf{u}_{{\rm{G}}_k}{\|}_2}$ is a separable penalty function, where ${{\|}\mathbf{u}_{{\rm{G}}_k}{\|}_2}$
is a continuous and convex function with respect to $\mathbf{u}_{{\rm{G}}_k}$. Thus, block-coordinate descent converges to the global minimum.

A formal description of the $\ell_1-\ell_2$  mixed-norm \ac{sssr} \ac{iot} identification algorithm is summarized in
Algorithm~\ref{euclidendwhilemmmmmeqqqq}. In Algorithm~\ref{euclidendwhilemmmmmeqqqq}, $M_{\rm{G}}$ and $M_{\rm{C}}$, denote the maximum number of iterations for the Golden selection search and the block-coordinate descent optimization, respectively.

\vspace{-0.9em}
\section{Data Detection}\label{sec:sm00000m}
The next step after \ac{iot} DI is to detect the data of devices identified as active.
Since CSI is unknown, the existing \ac{mud} algorithms, such as SIC cannot be employed.
In this section, we propose a new nonlinear \ac{mud}
algorithm which does not require CSI for data detection.

\vspace{-1em}
\subsection{{\ac{2mc}-\ac{mud} Algorithm}}
The output of the \ac{iot} DI algorithm is a set of \ac{iot} devices $\hat{\Set{X}}_{\rm{a}}$. Since
the delay of the \ac{iot} devices are known, we can apply sequence matched filtering to the small set of active \ac{iot} devices.
Without loss of generality, we assume that $\hat{\Set{X}}_{\rm{a}} \triangleq \{k_0,k_1,\hdots, k_{\hat{K}_{\rm{a}}-1}\}$ and $\tau_{k_0}\leq \tau_{k_1} \leq \dots \leq \tau_{k_{\hat{K}_{\rm{a}}-1}}$, where $\hat{K}_{\rm{a}} \triangleq \mathbf{card}(\hat{{\Set{X}}_{\rm{a}}})$. 

We consider a bank of $\hat{K}_{\rm{a}}$ single-user
MFs for the identified active \ac{iot} devices in $\hat{\mathcal{X}}_{\rm{a}}$.
The output of the MF after synchronized sampling and normalization by $N_{\rm{c}}$ for the $k_n$th \ac{iot} device is expressed as \cite{verdu1998multiuser}
\begin{align}\label{matfilt}
\hspace{-0.1em}
{y}_{{k_n},i}&\triangleq \frac{1}{N_{\rm{c}}} \int_{{{\tau}}_{k_n}+iT_{\rm{s}}}^{{{\tau}}_{k_n}+(i+1)T_{\rm{s}}} {r}(t)s_{k_n}(t-iT_{\rm{s}}-{{\tau}}_{k_n})dt \\  \nonumber
&={g}_{k_n}{b}_{{k_n},i}
\hspace{-0.1em}+\hspace{-0.5em}\sum_{k_j<k_n}{g}_{k_j}{b}_{{k_j},i+1}\rho_{k_nk_j}
\hspace{-0.1em}+\hspace{-0.5em}\sum_{k_j<k_n}{g}_{k_j}{b}_{{k_j},i}\rho_{k_jk_n}\\ \nonumber
&+\sum_{k_j>k_n}{g}_{k_j}{b}_{{k_j},i}\rho_{k_nk_j}
 +\sum_{k_j>k_n}{g}_{k_j}{b}_{{k_j},i-1}\rho_{k_jk_n}+{w}_{{k_n},i},
\end{align}
where
$
{w}_{{k_n},i}\triangleq {\sigma_{\rm{w}}}\int_{{{\tau}}_{k_n}+iT_{\rm{s}}}^{{{\tau}}_{k_n}+(i+1)T_{\rm{s}}} {w}(t)s_{k_n}(t-iT_{\rm{s}}-{{\tau}}_{k_n}){\rm d}t
$,
$\rho_{k_nk_j} \hspace{-0.2em}\triangleq \hspace{-0.2em}
\frac{1}{N_{\rm{c}}}  \int_{\tau_{k_j}}^{T_{\rm{s}}} s_{k_n}(t)s_{k_j}(t-\tau_{k_j}){\rm d}t$,
and
$
\rho_{k_jk_n} \hspace{-0.2em} \triangleq \hspace{-0.2em}
\frac{1}{N_{\rm{c}}}  \int_{0}^{\tau_{k_j}}\hspace{-0.3em} s_{k_n}(t)$ $s_{k_j}(t+T_{\rm{s}}-\tau_{k_j}){\rm d}t$.

The output of the single-user MF in \eqref{matfilt} for the $k_n$th \ac{iot} device can be written as
\begin{align}\label{ioyui}
{y}_{{k_n},i}={g}_{k_n}{b}_{{k_n},i}+{v}_{{k_n},i},\,\,\,\,\,\,\,\,\ i=0,1,\dots,N_{\rm{s}}-1,
\end{align}
where ${v}_{{k_n},i}$ represents the effect of noise and multiuser interference on the $k_n$th \ac{iot} device, and
${b}_{{k_n},i} \in \{-1,+1\}$.

For data detection without any sign ambiguity, the phase of ${g}_{k_n}$, $k_n \in \hat{\mathcal{X}}_{\rm{a}}$, is leastwise required to be known at the \ac{bs}. However, by employing differential coding at \ac{iot} devices, a \ac{mud} algorithm can be developed which removes the need for such {\it a priori} knowledge.
Differential coding is a coding technique used for non-coherent data detection.
Instead of encoding a bit sequence directly, it encodes the
difference between the bit sequence as  \cite{bhattacharya2005digital}
\begin{align}
{b}_{{k_n},i}^{\rm c}={b}_{{k_n},i-1}^{\rm c}\oplus {b}_{k_n,i}^{{\rm{d}}}\,\,\,\,,\,\,\,\,\,\ k_n \in \Set{X}_{\rm{a}},
\end{align}
where $\oplus$ is the modulo-2 addition and ${b}_{k_n,i}^{\rm{c}}\in \{0,1\}$ is the $i$th bit at the output of the channel encoder of the $k_n$th \ac{iot} device as shown  in Fig.~\ref{fig:iot_tx}. The \ac{bpsk} modulated data for the $k_n$th \ac{iot} device in \eqref{ioyui}, i.e.,
${b}_{{k_n},i}$ can be mathematically expressed based on the differentially coded bit ${b}_{k_n,i}^{\rm{d}}\in \{0,1\}$ as ${b}_{{k_n},i}=(-1)^{{b}_{k_n,i}^{{\rm{d}}}}$.

Since ${g}_{k_n}$, $k_n \in \mathcal{X}_{\rm{a}}$, remains unchanged
during the short packet, the received symbols of
the active \ac{iot} device $k_n$ in \eqref{ioyui} form two clusters corresponding
 to the transmitted bits $1$ and $0$.
The main idea behind the proposed \ac{mud} algorithm is to extract these two clusters
regardless of which cluster is labeled $1$ or $0$. By extracting the two clusters, the data stream of the active \ac{iot} device $k_n$ can be detected without any prior knowledge about the CSI and CP due to
differential coding.

To extract these two clusters for each active \ac{iot} device, the \ac{2mc} algorithm can be employed.
By applying the \ac{2mc} algorithm to ${y}_{{k_n},i}$, $i=0,1,\dots,N_{\rm{s}}-1$, in \eqref{ioyui}, the two clusters are separated based on the nearest mean criterion disregard to the label.
The \ac{2mc} minimizes the \ac{wcss}, i.e, the sum of the squared Euclidean distance
\cite{teknomo2006k}.
\begin{figure}[t!]
\vspace{-1.4em}
\centering
  \includegraphics[width=6.5cm]{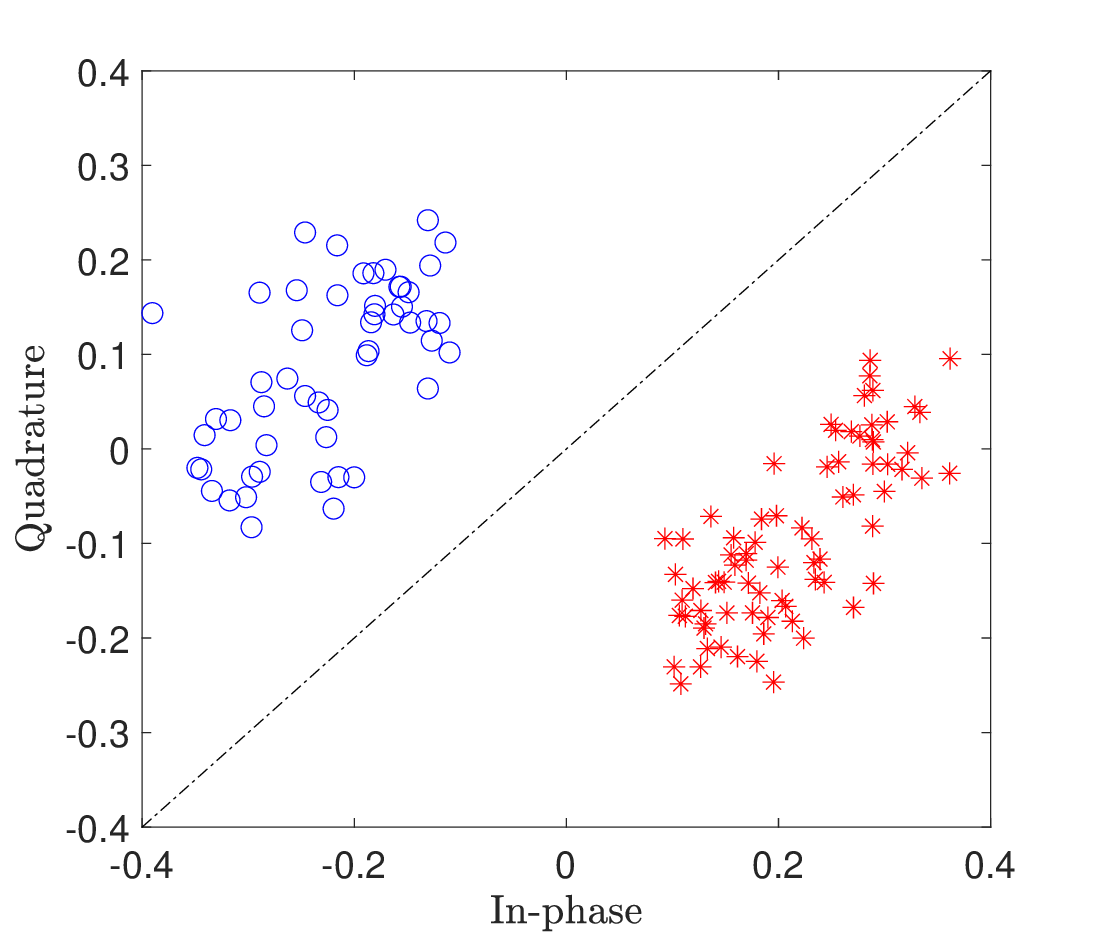}%
  \label{fig:evaluation:revenue}
\caption{Separated symbols by the proposed 2-MC-MUD algorithm for an active \ac{iot} device at 10 dB \ac{snr}, $K_{\rm{u}}=768$, and $N_{\rm{c}}=512$.}\label{fig:mul776501}
\vspace{-2em}
\end{figure}
Let us define $\Set{U} \triangleq \{0,1,\dots,N_{\rm{s}}-1\}$. The \ac{2mc} algorithm partitions $\Set{U}$ into two sets $\Set{U}_{k_n,0}$ and $\Set{U}_{k_n,1}$ by minimizing the \ac{wcss} as follows
\begin{align} \nonumber
& \underset{\hspace{2em}\Set{U}}{\text{arg min}} \
 \sum_{i \in \Set{U}_{k_n,0} }\big{|}{y}_{{k_n},i}-{\mu}_{k_n,0}\big{|}^2
\vspace{-0.5em}+\sum_{i \in \Set{U}_{k_n,1} }\big{|}{y}_{{k_n},i}-{\mu}_{k_n,1}\big{|}^2, \\ \nonumber
& \text{subject to}
\,\,\,\,\,\  {\mu}_{k_n,0}=\frac{1}{\mathbf{card}(\Set{U}_{k_n,0})}\sum_{i \in \Set{U}_{k_n,0} }{y}_{{k_n},i}, \\ \label{N0u}
&\,\,\,\,\,\,\,\,\,\,\,\,\,\,\,\,\,\,\,\,\,\,\,\,\,\,\,\,\,\ {\mu}_{k_n,1}=\frac{1}{\mathbf{card}(\Set{U}_{k_n,1})}\sum_{i \in \Set{U}_{k_n,1} }{y}_{{k_n},i}.
\end{align}

The minimization problem in \eqref{N0u} can be solved by different methods. One of the most common algorithm is the
Lloyd's algorithm which uses an iterative refinement technique \cite{tang2016lloyd}.
Given initial mean values\footnote{The Forgy method is used for intialization, where two observations from the dataset are used  as the initial means \cite{{teknomo2006k}}.}
${\mu}_{k_n,0}^{[0]}$ and ${\mu}_{k_n,1}^{[1]}$ for ${\mu}_{k_n,0}$ and ${\mu}_{k_n,1}$ in \eqref{N0u}, the Lloyd's algorithm
proceeds by alternating between the assignment and updating steps as follows:

{\it{Assignment Step:}}
The element of $\Set{U}$ at iteration $t$, i.e., $\Set{U}^{[t]}$ is assigned to $\Set{U}_{k_n,0}^{[t]}$ when
\begin{align}\label{cluster}
\Set{U}_{k_n,0}^{[t]}=\Big{\{}i:  \big{|}{y}_{{k_n},i}-{\mu}_{k_n,0}^{[t]}\big{|}^2
 \leq \big{|}{y}_{{k_n},i}-{\mu}_{k_n,1}^{[t]}\big{|}^2 \Big{\}}.
\end{align}
Otherwise, it is assigned to $\Set{U}_{k_n,1}^{[t]}$.

{\it{Updating Step:}}
The mean of the the clusters $\Set{U}_{k_n,0}^{[t]}$ and $\Set{U}_{k_n,1}^{[t]}$ are updated as
\begin{subequations}
\begin{align} \label{opiy2}
{\mu}_{k_n,1}^{[t+1]}=\frac{1}{\mathbf{card}\big{(}\Set{U}_{k_n,1}^{[t]}\big{)}}\sum_{i \in \Set{U}_{k_n,1}^{[t]} }{y}_{{k_n},i},
\\ \label{opiy3}
{\mu}_{k_n,0}^{[t+1]}=\frac{1}{\mathbf{card}\big{(}\Set{U}_{k_n,0}^{[t]}\big{)}}\sum_{i \in \Set{U}_{k_n,0}^{[t]}  }{y}_{{k_n},i}.
\end{align}
\end{subequations}
The \ac{2mc} algorithm converges when the assignment step does not change. Fig. \ref{fig:mul776501} shows the output of  the \ac{2mc} algorithm for an active \ac{iot} device. As seen, the sequence at the output of the MF is portioned into two clusters regardless of the label.

  \begin{algorithm}[!t]
    \caption{: \ac{2mc}-\ac{mud} algorithm}\label{euclid}
    \begin{algorithmic}[1]
    \Statex \textbf{Input:} ${r}(t)$, $\hat{\Set{X}}_{\rm{a}}$, $\hat{K}_{\rm{a}} = \mathbf{card}(\hat{{\Set{X}}_{\rm{a}}})$
      \Statex \textbf{Output:} $\hat{\mathbf{b}}_{k_n}$, $k_n \in \hat{\Set{X}}_{\rm{a}}$
      \For {$n=0,1,\dots,\hat{K}_{\rm{a}}-1$}
\State  \text{Set initial value for $\Set{U}_{k_n,1}^{[0]}$ and $\Set{U}_{k_n,0}^{[0]}$}
\State  \text{Obtain ${y}_{k_n,i}$, $i=0,1,\dots,N_{\rm{s}}-1$, by employing \eqref{matfilt}}
       \While{$\Set{U}_{k_n,1}^{[t+1]} \neq \Set{U}_{k_n,1}^{[t]}$}
       \State \text{obtain $\Set{U}_{k_n,1}^{[t]}$ and $\Set{U}_{k_n,0}^{[t]}$ by employing \eqref{cluster}}
       \State \text{${\mu}_{k_n,1}^{[t+1]}\leftarrow \Set{U}_{k_n,1}^{[t]}$  by employing
       \eqref{opiy2}}
       \State \text{${\mu}_{k_n,0}^{[t+1]}\leftarrow \Set{U}_{k_n,0}^{[t]}$  by employing
       \eqref{opiy3}}
       \EndWhile
\State \text{Obtain the binary mapped sequence ${\mathbf{b}}_{k_n}^{\rm{m}}$ through \eqref{mapseq}}
\State \text{Apply differential decoding  in
\eqref{cdecod} to ${\mathbf{b}}_{k_n}^{\rm{m}}$ to obtain $\hat{\mathbf{b}}_{k_n}^{\rm{c}}$}
\State \text{Apply channel decoding to $\hat{\mathbf{b}}_{k_n}^{\rm{c}}$ to obtain $\hat{\mathbf{d}}_{k_n}$}
\EndFor
    \end{algorithmic}\label{2mcmult}
  \end{algorithm}

After partitioning $\Set{U}$ into two clusters $\Set{U}_{k_n,0}$ and $\Set{U}_{k_n,1}$, ${y}_{{k_n},i}$, $i=0,1,\dots,N_{\rm{s}}-1$, is mapped into a binary sequence ${\mathbf{b}}_{k_n}^{\rm{m}} \triangleq {[}{{b}}_{k_n,0}^{\rm{m}} \ {{b}}_{k_n,1}^{\rm{m}} \ \dots \ {{b}}_{k_n,N_{\rm{s}}-1}^{\rm{m}}{]}^\dag$ with elements as
\begin{align}\label{mapseq}
{{b}}_{k_n,i}^{\rm{m}}=\mathbb{I}\big{\{}i \in \Set{U}_{k_n,1}\big{\}}.
\end{align}
Then, by applying differential decoding to the mapped binary sequence ${\mathbf{b}}_{k_n}^{\rm{m}}$, the channel coded data stream for the active \ac{iot} device $k_n$ is obtained as follows
\vspace{-0.3em}
\begin{equation}\label{cdecod}
\hat{{b}}_{k_n,i}^{\rm{c}}={b}_{k_n,i}^{\rm{m}}\oplus {b}_{k_n,i-1}^{\rm{m}}.
\vspace{-0.3em}
\end{equation}
Finally, $\hat{\mathbf{b}}_{k_n}^{\rm{c}} \triangleq \big{[}\hat{{b}}_{k_n,0}^{\rm{c}} \ \hat{{b}}_{k_n,1}^{\rm{c}} \ \dots \ \hat{{b}}_{k_n,N_{\rm{s}}-2}^{\rm{c}}\big{]}^\dag $ is decoded by the channel decoder, and the data stream of the active \ac{iot} device $k_n$ is obtained. The proposed \ac{2mc}-\ac{mud} algorithm is summarized in Algorithm~\ref{2mcmult}.

\subsection{Complexity Analysis}
The complexity of the proposed squared $\ell_2$-norm \ac{ssr} \ac{iot} DI algorithm is
$\mathcal{O}(K_{\rm u} L N_{\rm{c}}^2 +L N_{\rm{c}}^3)$.
The complexity of the proposed \ac{bic} $\ell_1-\ell_2$  mixed-norm \ac{sssr} \ac{iot} DI algorithm per each iteration is $\mathcal{O}(N_{\rm{c}} L  K_{\rm u}^2)$, where the maximum number of iterations is  $M_{\rm{G}}M_{\rm{C}}$ ($M_{\rm{G}}$ and $M_{\rm{C}}$ are the maximum number of iterations for the Golden selection search and the block-coordinate descent optimization, respectively.
The complexity of the proposed $2$-MC-MUD for single user matched filtering and clustering is $\mathcal{O}(N_{\rm c}^2 N_{\rm{s}}{k}_{\rm{a}})$ and $\mathcal{O}(K_{\rm u} N_{\rm{s}}{k}_{\rm{a}})$, respectively, where
${k}_{\rm{a}}=\mathbf{card}(\RS{X}_{\rm{a}})$.

\begin{figure*}[h!]
\vspace{-3.3em}
\centering
\subfloat[$P_{\rm{C}}$ versus \ac{snr}]{%
  \includegraphics[width=7cm]{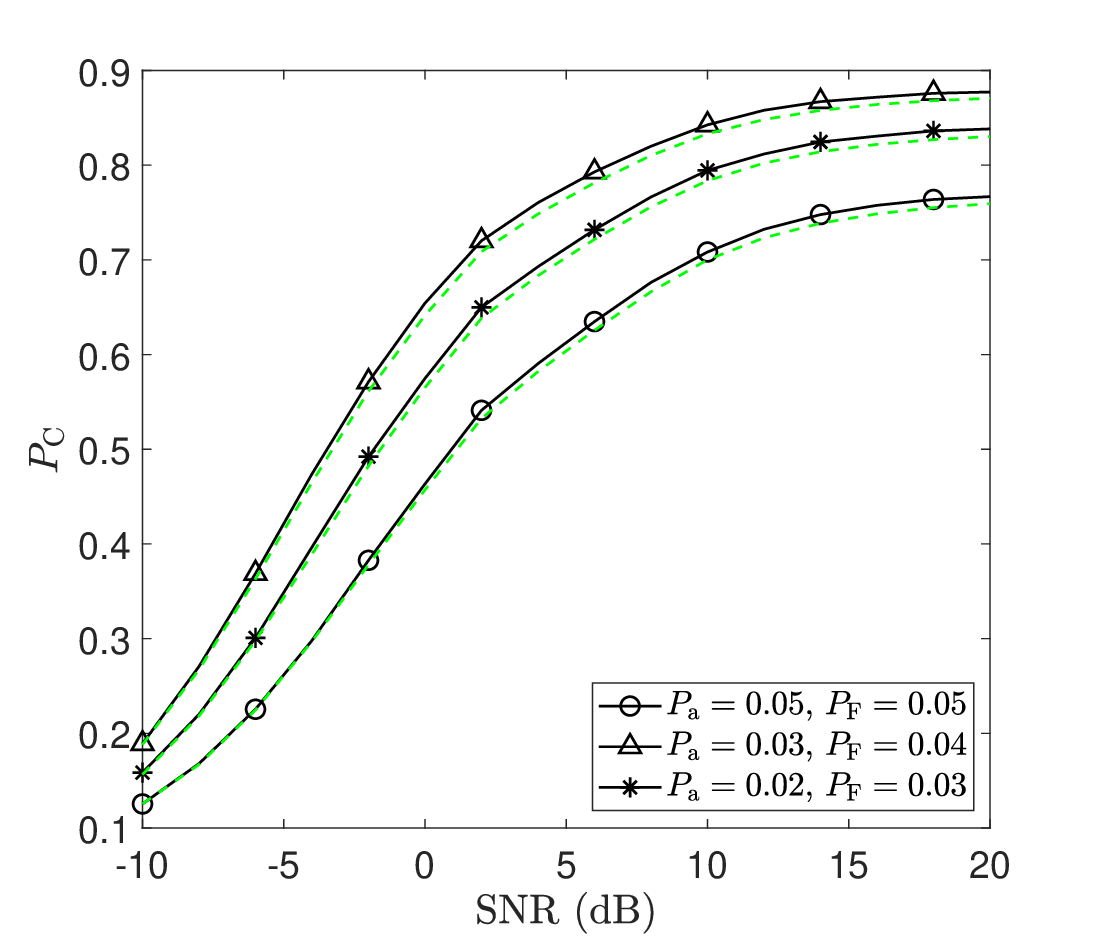}%
  \label{fig:evaluation:revenue}%
}\qquad
\subfloat[$P_{\rm{F}}$ versus \ac{snr}]{%
  \includegraphics[width=7cm]{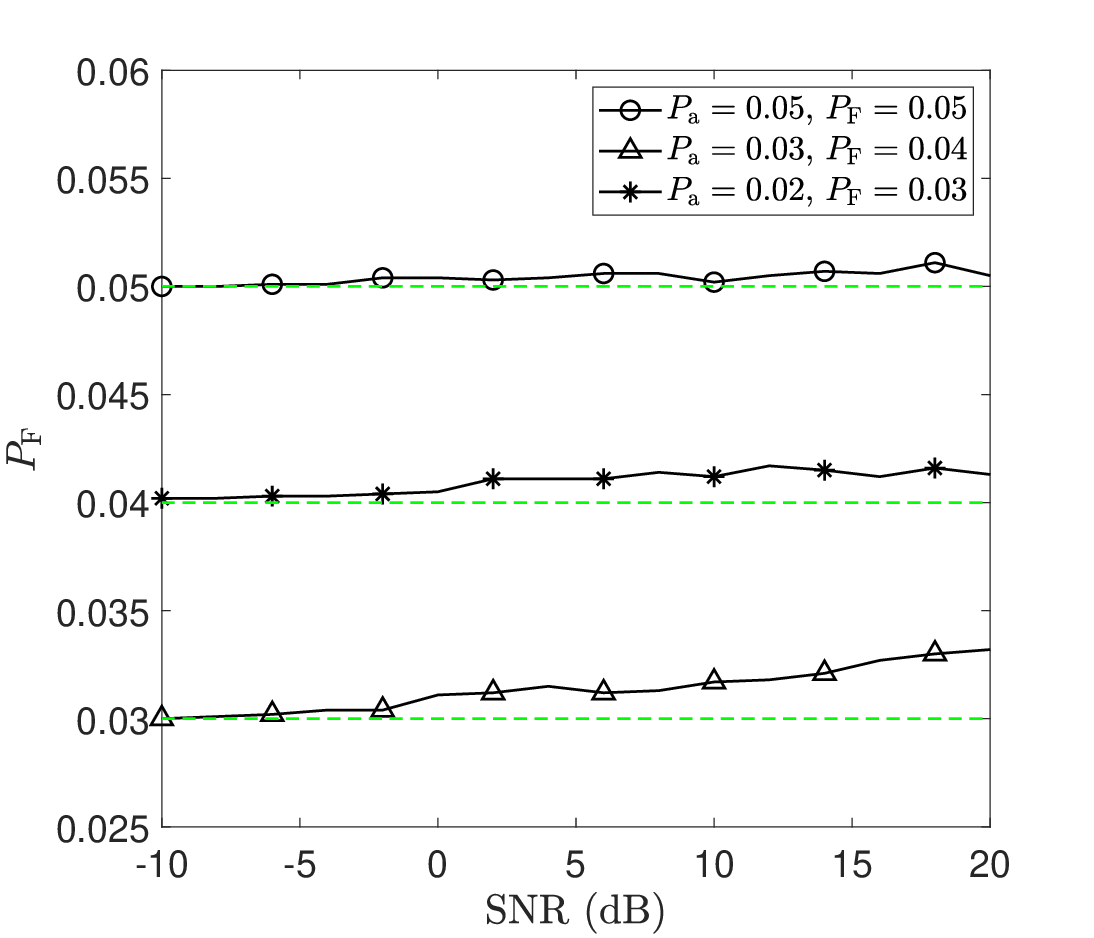}%
  \label{fig:evaluation:avgPrice}%
}
\caption{The system correct identification rate, $P_{\rm{C}}$, and false alarm rate, $P_{\rm{F}}$, of the proposed
squared $\ell_2$-norm \ac{ssr} \ac{iot} DI algorithm (Algorithm \ref{Table1xrrrr})
versus \ac{snr} for different values of $P_{\rm{a}}$, $P_{\rm{f}}^{(k)}\in \{0.03,0.04,0.05\}$, $K_{\rm{u}}=1024$, and $L=1$.}\label{fig:nioio}
\end{figure*}
\begin{figure*}[t!]
\vspace{-1.8em}
\centering
\subfloat[$P_{\rm{M}}$ versus \ac{snr}.]{%
  \includegraphics[width=7cm]{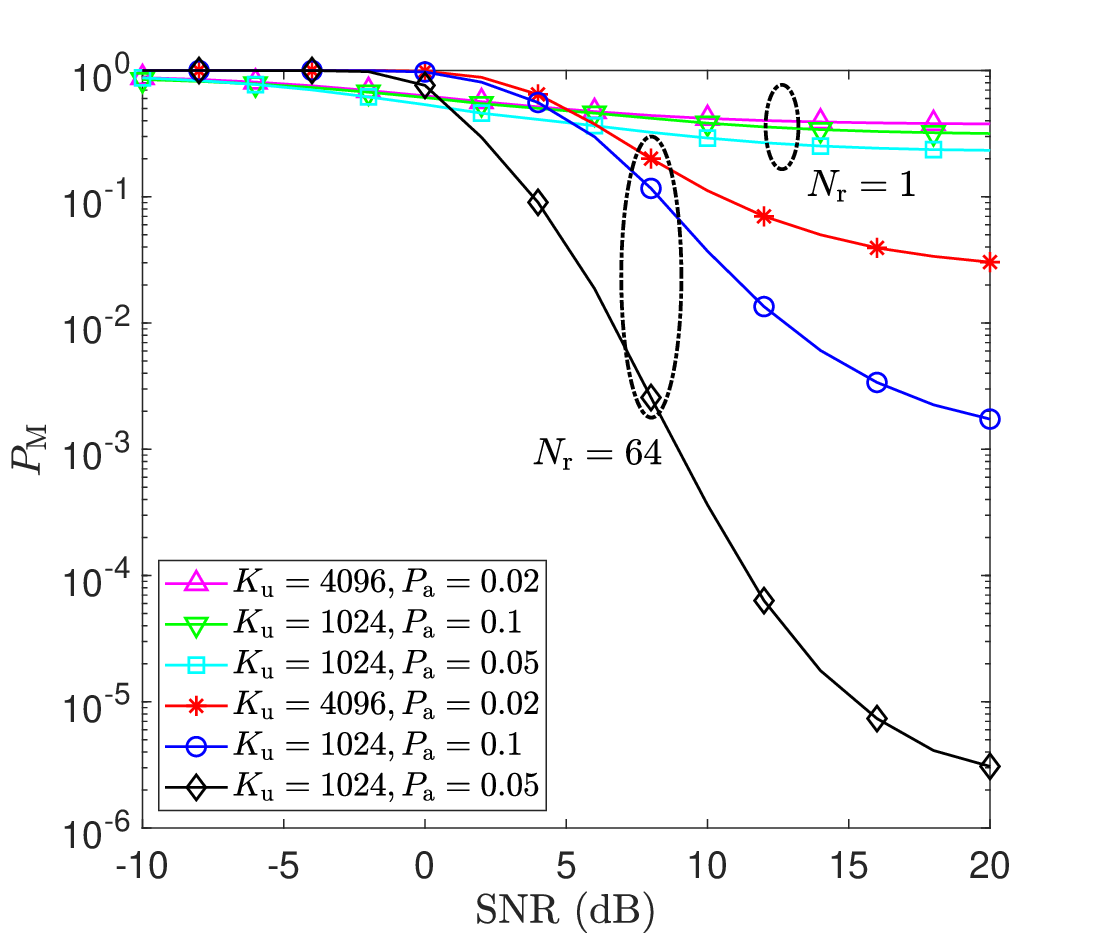}%
  \label{fig:1a}%
}\qquad
\subfloat[$P_{\rm{F}}$ versus \ac{snr}.]{%
  \includegraphics[width=7cm]{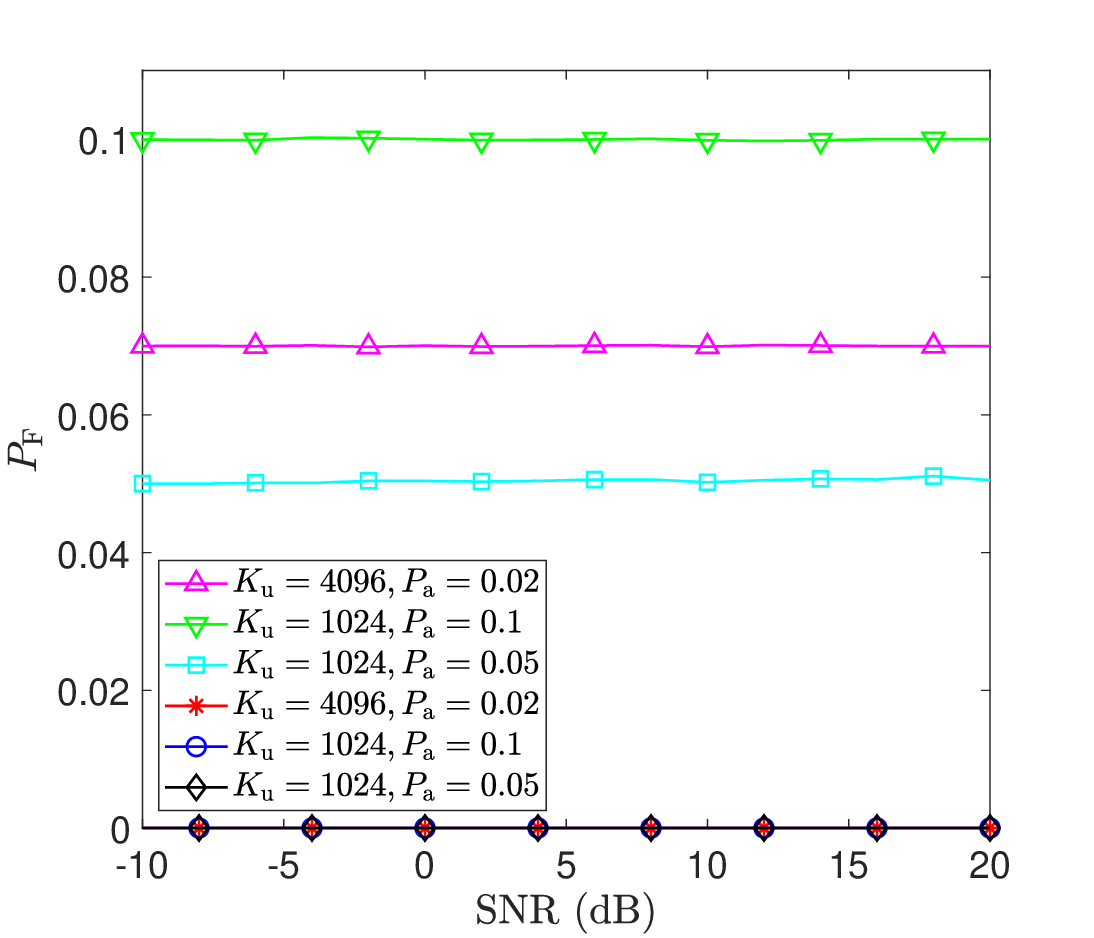}%
  \label{fig:1b}%
}
\caption{Performance of the proposed $\ell_2$-norm \ac{ssr} \ac{iot} DI algorithm for $P_{\rm{f}}^{(k)}\in \{0.05,0.07,0.1\}$, $K_{\rm{u}}\in \{1024,4096\}$, $L=1$, and $N_{\rm{r}}=\{1,64\}$ with the majority hard decision combining.}\label{Fig_mimo}
\end{figure*}

\begin{figure*}[h!]
\vspace{-2.1em}
\centering
\subfloat[$P_{\rm{C}}$ versus \ac{snr}]{%
  \includegraphics[width=7cm]{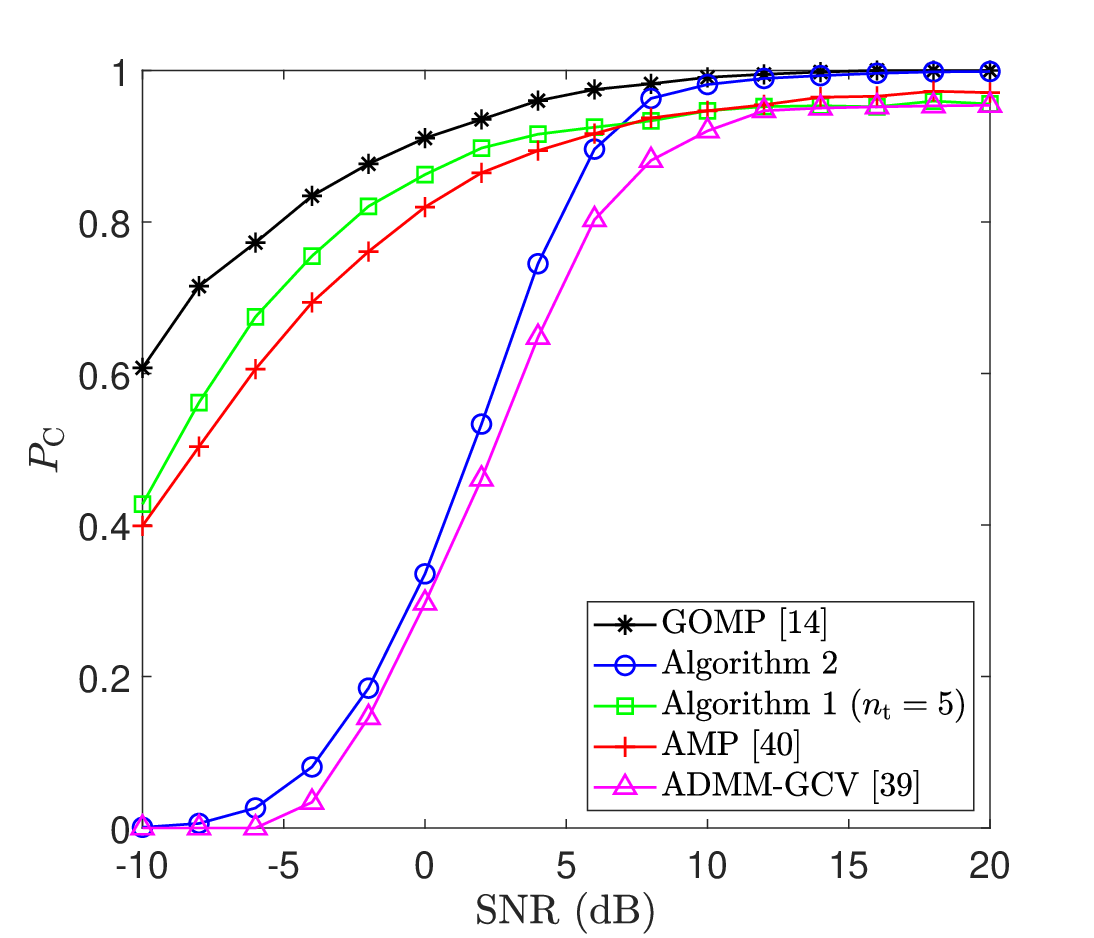}%
  \label{fig:evaluation:revenue}%
}\qquad
\subfloat[$P_{\rm{F}}$ versus \ac{snr}]{%
  \includegraphics[width=7cm]{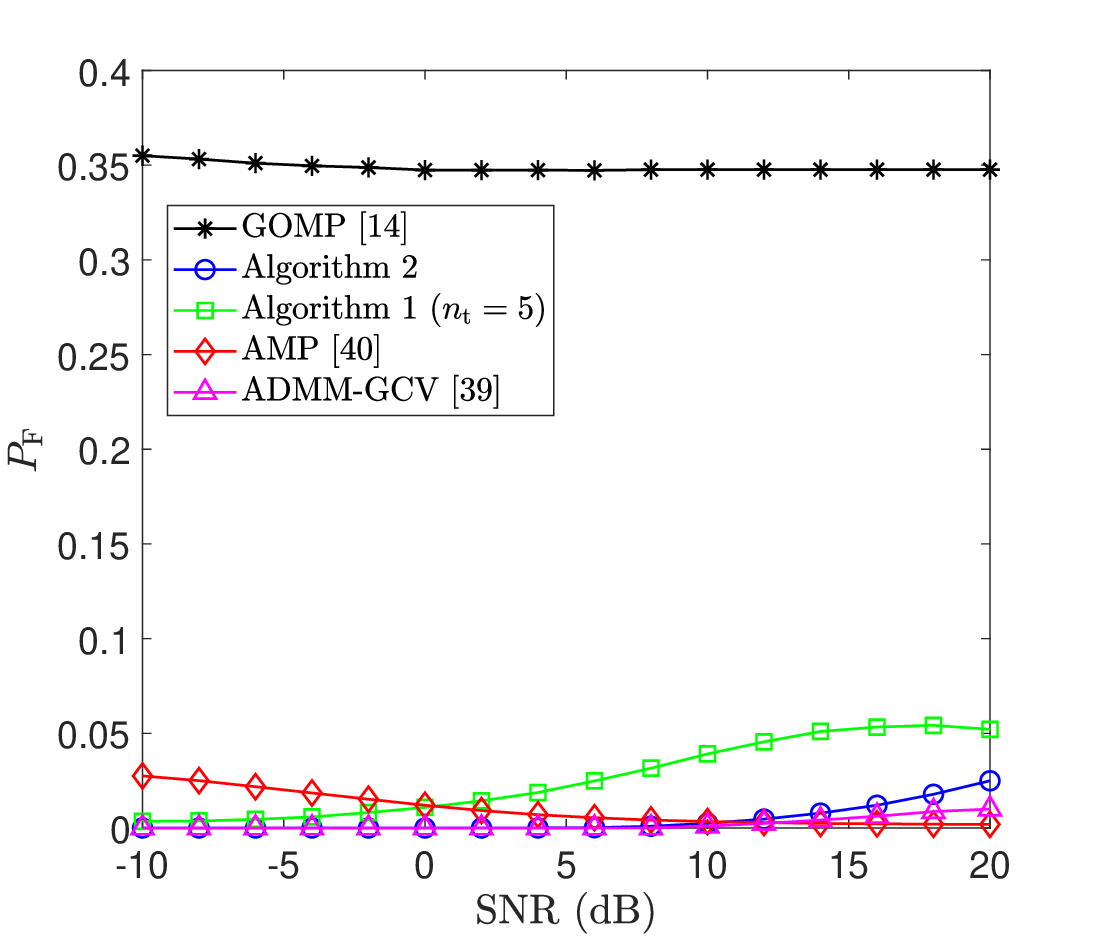}%
  \label{fig:evaluation:avgPrice}%
}
\caption{Performance comparison of the proposed \ac{iot} DI algorithms with the ADMM-GCV in \cite{boyd2011distributed},
the AMP-based non-coherent activity detection algorithm in \cite{senel2018grant} and GOMP method in \cite{schepker2012compressive}
for $P_{\rm{a}}=0.02$,  $K_{\rm{u}}=1024$, $N_{\rm{r}}=1$, $L=21$, and $P_k^{(\rm{f})}=0.05$ (Algorithm 1).}\label{fig:12}
\vspace{-1.2em}
\end{figure*}

\begin{figure*}[h!]
\vspace{-2.75em}
\centering
\subfloat[$P_{\rm{M}}$ versus \ac{snr}]{%
  \includegraphics[width=7cm]{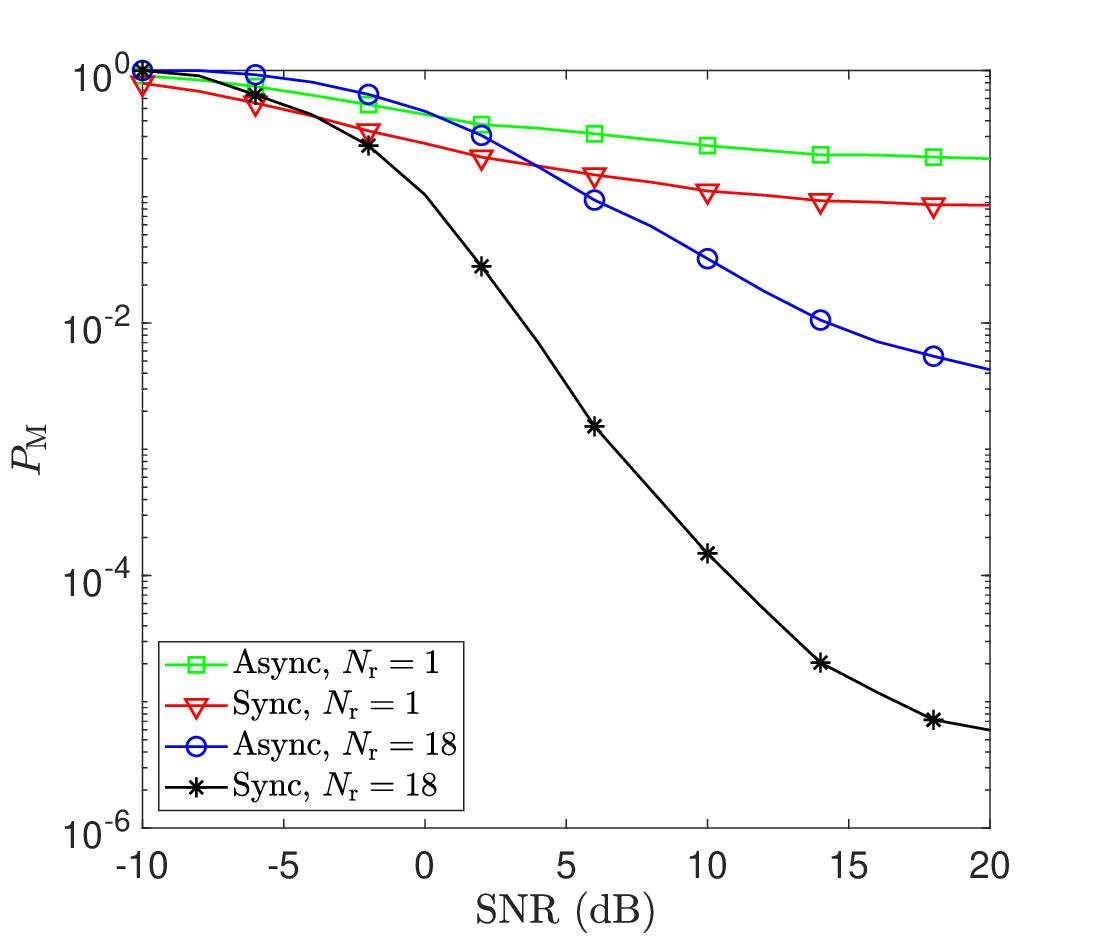}%
  \label{fig:evaluation:revenue}%
}\qquad
\subfloat[$P_{\rm{F}}$ versus \ac{snr}]{%
  \includegraphics[width=7cm]{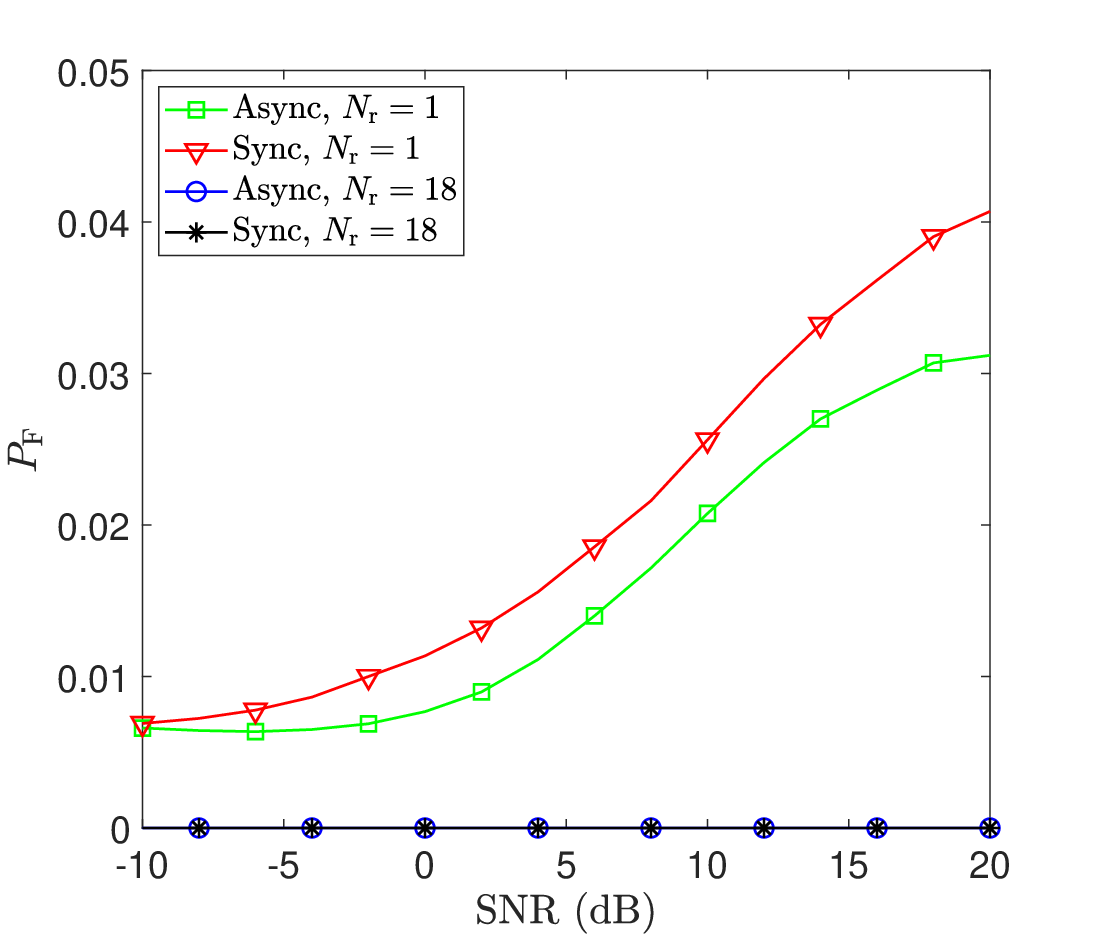}%
  \label{fig:evaluation:avgPrice}%
}
\caption{Effect of uncertainty in delay $\tau_k$, $k \in \Set{X}_{\rm{u}}$, on the performance of the proposed $\ell_2$-norm \ac{ssr} \ac{iot} DI algorithm for $P_{\rm{a}}=0.05$, $K_{\rm{u}}=1024$, $N_{\rm{r}}=\{1,18\}$, $L=21$, and $P_{\rm{f}}^{(k)}=0.06$.}\label{fig:13}
\end{figure*}

\section{Multiple Receive Antennas}\label{mimo_dsi}
The performance of the proposed \ac{iot} DI algorithms drastically improve by employing multiple receive antennas at the \ac{bs} due to spatial diversity.
Let us consider that the \ac{bs} is equipped with $N_{\rm{r}}$ receive antennas and
fuses all 1-bit activity decisions made by each receive antennas according to following logic rule\
\begin{equation}\label{eq:comb}
Z_k=
    \begin{cases}
      H_{1k}, & \,\,\ \sum_{i=1}^{N_{\rm{r}}}D_{k,i} \geq m_k \\
      H_{0k}, &  \,\,\  \sum_{i=1}^{N_{\rm{r}}}D_{k,i} < m_k,
    \end{cases},
  \end{equation}
where $D_{k,i}\in \{0,1\}$ is the decision made by the $i$th receive antenna on the transmission state of the $k$th \ac{iot} device, and $m_k$ is an arbitrary integer for the $k$th \ac{iot} device.
For the suboptimal detector in \eqref{eq:comb} and $m_k=m$, $k=0,1,\dots,K_{\rm{u}}-1$, the correct identification and false alarm rates are given by
\begin{equation}
{Q_{\rm{C}}} = \sum\limits_{l = m}^{{N_{\rm{r}}}} \binom{N_{\rm{r}}}{l}
 P_{\rm{C}}^l{\left( {1 - P_{\rm{C}}^l} \right)^{{N_{\rm{r}}} - l}},
\end{equation}
and
\begin{equation}\label{87uiiio}
{Q_{\rm{F}}} = \sum\limits_{l = m}^{{N_{\rm{r}}}} \binom{N_{\rm{r}}}{l} P_{\rm{F}}^l{\left( {1 - P_{\rm{F}}^l} \right)^{{N_{\rm{r}}} - l}},
\end{equation}
where $P_{\rm{C}}$ and $P_{\rm{F}}$ are the correct identification and false alarm rates for a single receive antennas.
Similarly, 1-bit decision fusion can be used for the 2-MC-MUD algorithm to improve data detection performance.

\vspace{-0.7em}
\section{Simulation Results}\label{nn:iop}
In this section, we examine the performance of the proposed
\ac{iot} DI algorithms and the \ac{2mc}-\ac{mud} algorithm through several simulation experiments.
\vspace{-0.95em}
\subsection{Simulation Setup}
Unless otherwise mentioned,
we considered an \ac{iot} network with $K_{\rm{u}}=1024$ \ac{iot} devices.
It is assumed that the spreading sequences of the \ac{iot} devices are random binary codes with spreading factor $N_{\rm{c}}=512$.
Each \ac{iot} packet is $128$ bits with payload length of 40 bits.
The delay of the \ac{iot} devices was generated as uniform distributions $\alpha_k \sim {\cal{U}}_{\rm{d}}\big{[}0,5\big{]}$, $\beta_k \sim  {\cal{U}}_{\rm{d}}\big{[}0,511\big{]}$, and $\xi_k  \sim {\cal{U}}_{\rm{c}}\big{[}0,1\big{)}$.
The effect of the unknown CSI and CP for each \ac{iot} device
was modeled as independent complex Gaussian
random variables with mean $\mu_k=\sqrt{0.1}+j\sqrt{0.1}$ and variance $\sigma_k^2=1$, $k \in \Set{X}_{\rm{u}}$, i.e., Rician fading with $K$-factor $0.2$ was considered.
 The average
system \ac{snr} was defined as
$
\vartheta \triangleq   {\bar{P}_{\rm{a}}\sum\limits(|\mu_k|^2+\sigma_{{\rm}_k}^2)p_k \eta_k}/{\sigma_{\rm{w}}^2},
$
where $p_k=\varsigma /\eta_k$ ($\varsigma$ changes according to $\vartheta$), $\bar{P}_{\rm{a}}={P}_{\rm{a}}$ for Algorithm \ref{Table1xrrrr}, $\bar{P}_{\rm{a}}={P}_{\rm{max}}/2$ for Algorithm \ref{euclidendwhilemmmmmeqqqq} (for the case of time-varying $P_{\rm{a}}$), and $\sigma_{{\rm{w}}}^2=1$ is the variance of the additive noise.
The range of tuning parameter for the \ac{bic} minimization in \eqref{E1823288io} was set as $\lambda=[0 \ 500]$, $\epsilon_{\rm{g}}=2$, and $M_{\rm{G}}=50$.
The performance of the proposed \ac{iot} DI algorithms were evaluated in terms of
system correct identification $P_{\rm{C}}$ and system false alarm $P_{\rm{F}}$ rates for $10^6$ Monte Carlo trials.
Also,
the performance of the proposed \ac{2mc}-\ac{mud} algorithm was evaluated in terms of average \ac{per} in the presence \ac{iot} DI error.

\vspace{-0.8em}
\subsection{Simulation Results}
\vspace{-0.2em}
Fig.~\ref{fig:nioio} depicts $P_{\rm{C}}$ and $P_{\rm{F}}$  of the proposed
squared $\ell_2$-norm \ac{ssr} \ac{iot} DI algorithm (Algorithm \ref{Table1xrrrr})
versus \ac{snr} for different values of $P_{\rm{a}}$ and $P_{\rm{f}}^{(k)}$, $K_{\rm{u}}=1024$, and $L=1$. The threshold values $\theta_k$, $k \in \Set{X}_{\rm{u}}$, are set by using \eqref{flase}.
As seen, Algorithm \ref{Table1xrrrr} can offer high correct identification error rate even for a single observation vector.
Also, there is an insignificant gap between $P_{\rm{C}}$ obtained in the simulation experiment and the theoretical result in \eqref{correct}. Similarly, $P_{\rm{F}}$ matches the preset false alarm rate, i.e., $P_{\rm{f}}^{(k)}\in \{0.03,0.04,0.05\}$, $k \in \Set{X}_{\rm{u}}$.
We notice from Fig.~\ref{fig:nioio} that the theoretical results  more accurately match the simulation results at higher $P_{\rm{a}}$ and at lower \ac{snr}s since the CLT is more reliable.

In Fig. \ref{Fig_mimo}, we illustrate the performance of Algorithm \ref{Table1xrrrr} for  $K_{\rm{u}} \in \{1024,4096\}$, $P_{\rm{a}} \in \{0.02, 0.05, 0.1\} $, and $P_{\rm{f}}^{(k)} \in \{0.05, 0.07, 0.1\}$
when multiple receive antennas are employed at the \ac{bs}.
Here, we consider the majority rule hard decision combining as a suboptimal detector for $N_{\rm{r}}=64$ receive antennas. As observed,
miss identification ($P_{\rm{M}}=1-P_{\rm{C}}$) and false alarm rates substantially decrease by employing multiple receive antennas at the \ac{bs}. Here, for $10^{5}$ Monte Carlo trials, we obtain $P_{\rm{F}}=0$; this value is $4.43 \times 10^{-26}$ for the theoretical result given in \eqref{87uiiio}.

\begin{figure*}[t!]
\vspace{-2em}
\centering
\subfloat[$P_{\rm{M}}$ versus \ac{snr}]{%
  \includegraphics[width=7cm]{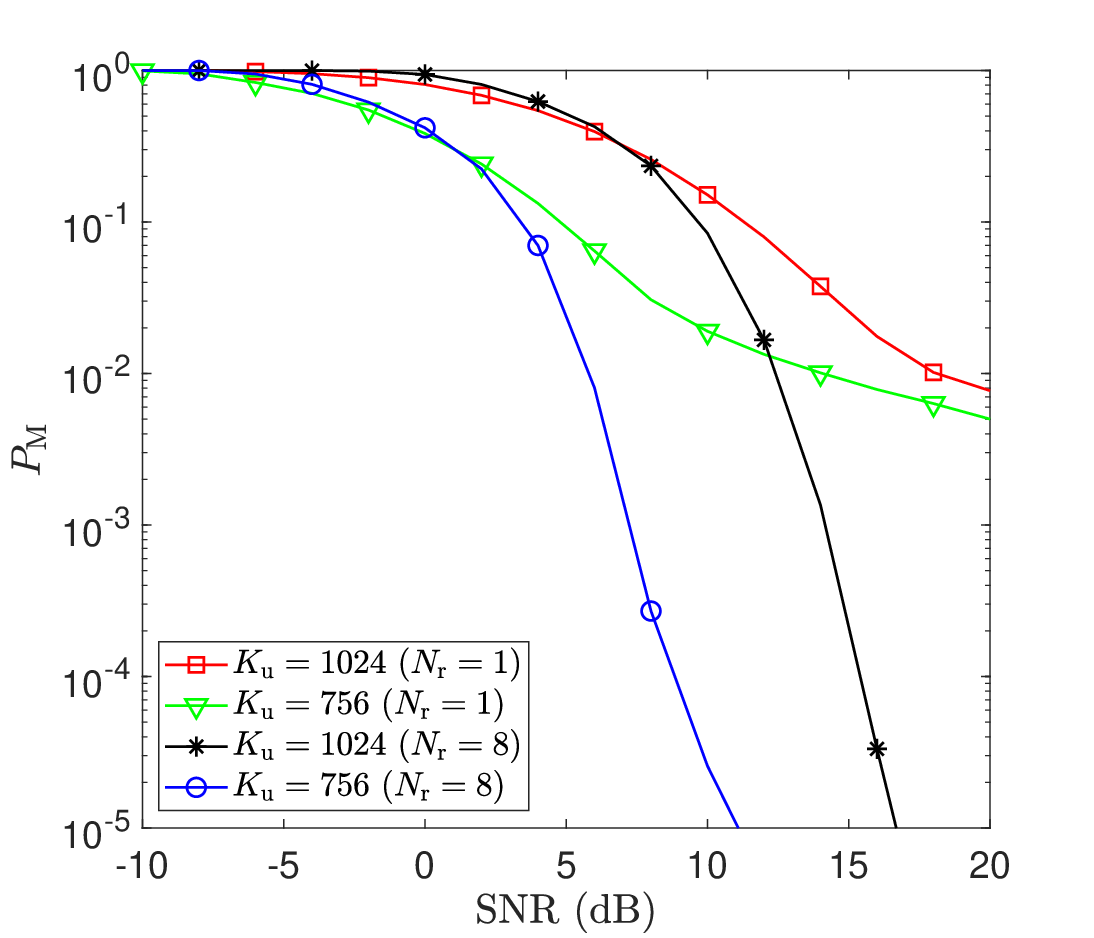}%
  \label{fig:evaluation:revenue}%
}\qquad
\subfloat[$P_{\rm{F}}$ versus \ac{snr}]{%
  \includegraphics[width=7cm]{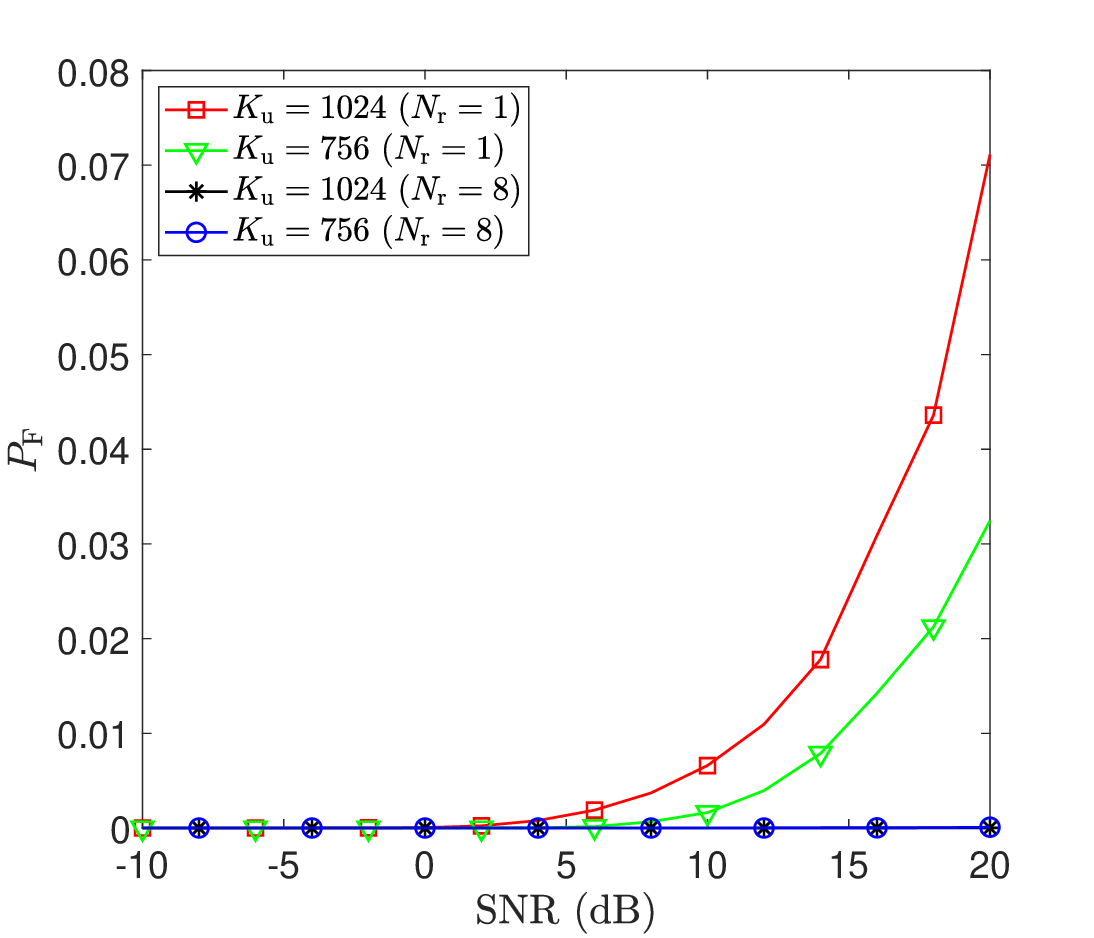}%
  \label{fig:evaluation:avgPrice}%
}
\caption{The system miss identification rate, $P_{\rm{M}}$, and false alarm rate, $P_{\rm{F}}$, of the proposed
\ac{bic} $\ell_1-\ell_2$  mixed-norm \ac{sssr} \ac{iot} DI algorithm (Algorithm \ref{euclidendwhilemmmmmeqqqq})
versus \ac{snr} for random activity rate $P_{\rm{a}} \in [0 \ 0.06]$ and $L=21$ with the majority rule
hard decision combining.}\label{fig:14}
\vspace{-1.5em}
\end{figure*}

In Fig.~\ref{fig:12}, we compare the performance of Algorithm \ref{Table1xrrrr} and Algorithm \ref{euclidendwhilemmmmmeqqqq} with the ADMM algorithm using GCV (ADMM-GCV) in \cite{boyd2011distributed},
the AMP-based non-coherent activity detection algorithm in \cite{senel2018grant} and the GOMP method in \cite{schepker2012compressive}
 for  $P_{\rm{a}}=0.02$,  $K_{\rm{u}}=1024$, $L=21$, and $N_{\rm{r}}=1$.
 For Algorithm \ref{Table1xrrrr}, the threshold values $\theta_k$, $k \in \Set{X}_{\rm{u}}$, are set for $P^{(\rm{f})}_k=0.05$ in \eqref{flase}, and \textcolor{blue}{$n_k=5$} are used for hard decision combining in \eqref{eq:61c}.
As seen, our proposed algorithms offer acceptable correct identification and false alarm rates for a single receive antenna at the \ac{bs}. Moreover, we observe that
GOMP exhibits a larger correct identification rate at the expense of significantly higher false alarm rate. This is different for the algorithm in \cite{senel2018grant}, which offers a lower false alarm rate compared with Algorithm 2 in the range of $[10 , 20]$ dB \ac{snr}
 at the cost of reduced correct identification rate.
  We also notice that the performance improvement of Algorithm \ref{Table1xrrrr} for $L=21$ is not very high compared to $L=1$ in Fig.~\ref{fig:nioio} since $p\big{(}\hat{{h}}_{k,j_1,m}$, $\hat{{h}}_{k,j_2,m}$ $|H_{tk}\big{)}\neq p\big{(}\hat{{h}}_{k,j_1,m}|H_{tk}\big{)} p\big{(}\hat{{h}}_{k,j_2,m}|H_{tk}\big{)}$, $j_1\neq j_2$, $t\in \{0,1\}$.
 Moreover, as seen, the false alarm rate of Algorithm 1 increases slowly as the \ac{snr} increases.
 There are two reasons for this behaviour: 1) the validity of joint Gaussian \ac{pdf} assumption decreases, and 2)
 the effectiveness of the hard decision
combining decreases due to the high correlation among the reconstructed vectors.


 In Fig.~\ref{fig:13}, we show the effect of uncertainty in delay ${\tau}_k \triangleq {\alpha}_k{T_{\rm{s}}}+{\beta}_k{T_{\rm{c}}} + {\xi}_k$, $k \in \Set{X}_{\rm{u}}$, on the performance of
Algorithm \ref{Table1xrrrr}
for  $P_{\rm{a}}=0.05$,  $K_{\rm{u}}=1024$, $L=21$, $n_{\rm{t}}=5$, and $P_{\rm{f}}^{(k)}=0.06$.
We consider that packets arrive at the \ac{bs} with chip delay uncertainty ${\xi}_k/T_{\rm{c}} \in {\cal{U}}_{\rm{c}}{[}0,1)$ while the identification and detection are performed by employing the estimated delay based on distance, which is a priori known and fixed at the \ac{bs}. For multiple receive antennas, we consider the majority rule hard decision combining as a suboptimal detector for $N_{\rm{r}}=18$ receive antennas.
  As observed,  the performance of Algorithm \ref{Table1xrrrr} degrades in the presence of the round-trip delay estimation error. However,
by employing multiple receive antennas at the \ac{bs}, this performance degradation can be reduced.

Fig.~\ref{fig:14} illustrates  $P_{\rm{C}}$  and $P_{\rm{F}}$ of the proposed \ac{bic}
 $\ell_1-\ell_2$ mixed-norm \ac{sssr} \ac{iot} DI algorithm  versus \ac{snr} when the activity rate varies uniformly in the range $P_{\rm{a}} \in [0,  \ 0.06]$.
As seen, the proposed algorithm exhibits high correct identification rate for high overloading factors, such as $OF=2$ (1024 devices), when $P_{\rm{a}}$ is unknown and time-varying.
Also, the false alarm rate of the proposed algorithm is significantly low for the \ac{snr} values lower than $15$ dB.

\begin{figure}[t!]
  \centering
\includegraphics[width=7cm]{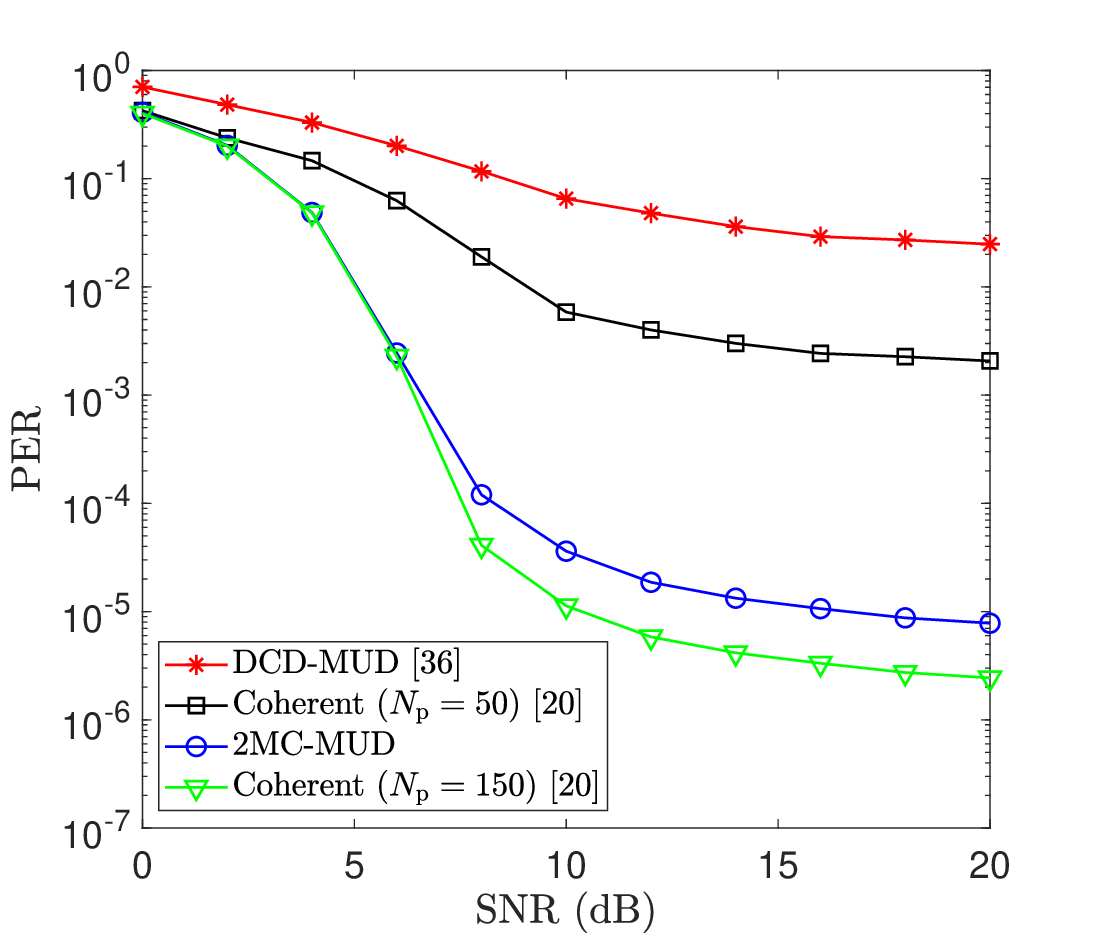}
\vspace{-0.2em}
  \caption{PER of the proposed \ac{ma} scheme when \ac{2mc}-\ac{mud} and \ac{dcd}-\ac{mud} \cite{verdu1998multiuser} algorithms are employed for non-coherent data detection at the \ac{bs} ($K_{\rm{u}}=768$, $OF=1.5$, and $P_{\rm{a}} \in [0, \ 0.06]$). }\label{fig:5633vvvxcxxvxvvvcdsda3333332}
\vspace{-1.4em}
\end{figure}

Fig. \ref{fig:5633vvvxcxxvxvvvcdsda3333332} compares the performance of the developed \ac{ma} scheme when
the proposed \ac{2mc}-\ac{mud} and \ac{dcd}-\ac{mud} \cite{verdu1998multiuser} algorithms are employed for $N_{\rm{r}}=12$.
We also show the performance with estimated CSI using the joint activity and channel estimation method in \cite{liu2018massive}.
 As seen, the proposed \ac{mud} algorithm outperforms the \ac{dcd}-\ac{mud} \cite{verdu1998multiuser}.
This superiority in performance is related to the capability of the \ac{2mc} algorithm to accurately separate the two clusters of data.
We also observe that data detection with estimated CSI can offer lower \ac{per} compared with our proposed non-coherent detection
when a sufficient number of pilots $N_{\rm{p}}$ is used for joint activity detection and channel estimation.
This is achieved at the expense of lower spectral efficiency and higher latency.

\section{Conclusion}\label{sec:conclude}
A new uncoordinated uplink \ac{ma} for \ac{mmtc} with short-packet and sporadic traffic was proposed in this paper.
The proposed \ac{ma} scheme reduces the control signaling associated with the \ac{mac} and \ac{phy} layers.
Instead of transmitting the device identifier using a portion of bits in a packet,
the squared $\ell_2$-norm \ac{ssr} and \ac{bic} $\ell_1-\ell_2$  mixed-norm \ac{sssr} \ac{iot} DI algorithms were proposed to identify active \ac{iot} devices through the assigned unique non-orthogonal spreading code to each \ac{iot} device.
To further reduce the overhead, we removed the preambles and pilots used for channel estimation by developing the non-coherent \ac{2mc}-\ac{mud} algorithm based on unsupervised machine learning.
\appendices
\section{}\label{ditribution}
By applying the statistical expectation to \eqref{RD5} and employing $\mathbb{E}\big{\{}{{h}}_{k,j,1}|H_{tk}\big{\}}=0$ and $\mathbb{E}\big{\{}{{h}}_{k,j,0}|H_{tk}\big{\}}=0$, we can write $\mathbb{E}\big{\{}\hat{{h}}_{k,j,f}|H_{tk}\big{\}}={0}$, $t,f \in\{{0,1}\}$.
To obtain the variance of $\hat{{h}}_{k,j,f}$ given $H_{tk}$, i.e., ${\Sigma}_{f,f}^{tk}$,
we use the variance sum law as
\begin{equation}\label{BB2}
\vspace{-0.3em}
\hspace{-0.9em}
\mathbb{V}{\text{ar}}\Big{\{} \hspace{-0.25em}\sum_{i} \hspace{-0.2em} a_i{z}_i\Big{\}}\hspace{-0.3em}=\hspace{-0.3em}
\sum_{i}\hspace{-0.2em}\Big{(}{|}a_i{|}^2\mathbb{V}{\text{ar}}\big{\{}\hspace{-0.1em} {z}_i \hspace{-0.2em} \big{\}}
\hspace{-0.25em}+\hspace{-0.3em}\sum_{j\neq i}
\hspace{-0.1em}a_ia_j^*\mathbb{C}{\text{ov}}\big{\{}\hspace{-0.1em}{z}_i,{z}_j \hspace{-0.1em}\big{\}}\hspace{-0.1em}\Big{)}.
\vspace{-0.3em}
\end{equation}
Since ${{h}}_{k_1,j,f}$, ${{h}}_{k_2,j,\bar{f}}$, and ${w}'_{k,j,f}$,  $k,k_1,k_2 \in \Set{X}_{\rm{u}}$,  in \eqref{RD5}, are zero-mean and
uncorrelated random variables, by applying \eqref{BB2} to \eqref{RD5}, we obtain \eqref{1qax78nv4} at the top of next page.
\begin{figure*}
\vspace{-1.6em}
\begin{align}\label{1qax78nv4}
&{\Sigma}_{f,f}^{tk}=\mathbb{V}\text{ar} \big{\{}\hat{{h}}_{k,j,f}  | H_{tk} \big{\}} =\mathbb{E}\big{\{}|\hat{{h}}_{k,j,f}{|}^2|H_{tk}\big{\}}
=t{\Omega}_{2k+f,2k+f}^2\mathbb{V}\text{ar} \big{\{}{h}_{k,j,f}|H_{tk}\big{\}}
+t{\Omega}_{2k+f,2k+\bar{f}}^2\mathbb{V}\text{ar} \big{\{}{h}_{k,j,\bar{f}}|H_{tk}\big{\}}\\ \nonumber
&\quad \quad +\sum\nolimits_{n\neq k}{\Omega}_{2k+f,2n+f}^2\mathbb{V}\text{ar} \big{\{}{h}_{n,j,f}|H_{tk}\big{\}}
+\sum\nolimits_{n\neq n}{\Omega}_{2k+f,2n+\bar{f}}^2\mathbb{V}\text{ar} \big{\{}{h}_{n,j,\bar{f}}|H_{tk}\big{\}}+\mathbb{V}\text{ar} \big{\{}{w}'_{k,j,f}\big{\}}.
\end{align}
\setcounter{equation}{89}
\vspace{-1.8em}
\\
\begin{align}\label{eq:case12}
 &{\Sigma}_{0,1}^{1k}=
\mathbb{C}\text{ov}\big{\{}\hat{{h}}_{k,j,0},\hat{{h}}_{k,j,1} {|} H_{tk} \big{\}}=\mathbb{E}\big{\{}\hat{{h}}_{k,j,0}\hat{{h}}_{k,j,1}^* {|} H_{tk} \big{\}}
=t\big{(}{\Omega}_{2k,2k}{\Omega}_{2k+1,2k}\big{)}\mathbb{E}\big{\{}{|}{h}_{k,j,0}{|}^2{|}H_{tk}\big{\}} \\ \nonumber
&\quad \quad +t\big{(}{\Omega}_{2k+1,2k+1}{\Omega}_{2k,2k+1}\big{)}\mathbb{E}\big{\{}{|}{h}_{k,j,1}{|}^2{|}H_{tk}\big{\}}
 +\sum\nolimits_{n\neq k}{\Omega}_{2k,2n}{\Omega}_{2k+1,2n}\mathbb{E}\big{\{}{|}{h}_{n,j,0}{|}^2{|}H_{tk}{\}} \\ \nonumber
&\quad \quad +\sum\nolimits_{n\neq k}{\Omega}_{2k+1,2n+1}{\Omega}_{2k,2n+1}\mathbb{E}\big{\{}{|}{h}_{n,j,1}{|}^2{|}H_{tk}\big{\}} +\mathbb{E}\big{\{}{w}'_{k,j,0}{(}{w}'_{k,j,1}{)}^*{|}H_{tk}\big{\}}.
\end{align}
\hrule
\hspace{-1em}
\end{figure*}

\setcounter{equation}{83}
By using \eqref{eq:opip},
$\mathbb{V}\text{ar} {\{}{h}_{k,j,f}{|}H_{1k}{\}}$, $f \in \{0,1\}$,  in \eqref{1qax78nv4} can be written as
\begin{align} \label{BB4}
\mathbb{V}\text{ar} &\big{\{}{h}_{k,j,f}{|}H_{tk}\big{\}}=
\mathbb{V}{\text{ar}}\big{\{}{g}_k{b}_{k,j-{\alpha}_k-1+f}{|}H_{tk}\big{\}}\\ \nonumber
&=
\mathbb{E}\Big{\{}\mathbb{V}{\text{ar}}\big{\{}{g}_k{b}_{k,j-{\alpha}_k-1+f}
{|}{g}_k,H_{tk}\big{\}}\Big{\}} \\ \nonumber
&{\quad}+\mathbb{V}{\text{ar}}\Big{\{}\mathbb{E}
\big{\{}{g}_k{b}_{k,j-{\alpha}_k-1+f}{|}{g}_k,H_{tk}\big{\}}\Big{\}}.
\end{align}
Since $\mathbb{V}{\text{ar}}\big{\{}{b}_{k,j-{\alpha}_k-1+f}{|}H_{tk}\big{\}}
\vspace{-0.2em}
=t$, $t,f \in \{0,1\}$, we can write
\begin{align}\label{iob5zq1}
\mathbb{E}\Big{\{}&\mathbb{V}{\text{ar}}\big{\{}{g}_k{b}_{k,j-{\alpha}_k-1+f}
{|}{g}_k,H_{tk}\big{\}}\Big{\}} \\ \nonumber
&=\mathbb{E}\big{\{}{|}{g}_k{|}^2\big{\}}\mathbb{V}{\text{ar}}\big{\{}{b}_{k,j-{\alpha}_k-1+{f}}
{|}H_{tk}\big{\}}=t(\sigma_k^2+|\mu_k|^2) \eta_k p_k.
\end{align}
By substituting \eqref{iob5zq1} and $\mathbb{E}
\big{\{}{g}_k{b}_{k,j-{\alpha}_k-1+f}{|}{g}_k,H_{tk}\big{\}}=0$, $f \in \{0,1\}$, into \eqref{BB4}, we obtain
\begin{align}\label{0m1sx7ub4602xo}
\mathbb{V}\text{ar} \big{\{}{h}_{k,j,f}{|}H_{tk}\big{\}}
=t(\sigma_k^2+|\mu_k|^2) \eta_k p_k.
\end{align}
Similar to \eqref{BB4}, for $n \neq k$ and $f \in \{0,1\}$, we can write
\begin{align} \label{BBxgt}
\mathbb{V}\text{ar} &\big{\{}{h}_{n,j,f}{|}H_{tk}\big{\}}=
\mathbb{V}{\text{ar}}\big{\{}{g}_n{b}_{n,j-{\alpha}_n-1+f}{|}H_{tk}\big{\}}\\ \nonumber
&=
\mathbb{E}\Big{\{}\mathbb{V}{\text{ar}}\big{\{}{g}_n{b}_{n,j-{\alpha}_n-1+f}
{|}{g}_n,H_{tk}\big{\}}\Big{\}} \\ \nonumber
&~~+\mathbb{V}{\text{ar}}\Big{\{}\mathbb{E}
\big{\{}{g}_n {b}_{n,j-{\alpha}_n-1+f}{|}{g}_n,H_{tk}\big{\}}\Big{\}}.
\end{align}
Because $\mathbb{V}{\text{ar}}\big{\{}{b}_{n,j-{\alpha}_n-1+f}{|}H_{tk}\big{\}}=\mathbb{V}{\text{ar}}\big{\{}{b}_{n,j-{\alpha}_n-1+f}\big{\}}
=P_{\rm{a}}$, $n \neq k$, $f \in \{0,1\}$, we can write
\vspace{-0.5em}
\begin{align}\label{iob5u403bx}
&\mathbb{E}\Big{\{}\mathbb{V}{\text{ar}}\big{\{}{g}_n{b}_{k,j-{\alpha}_n-1+f}
{|}{g}_n,H_{tk}\big{\}}\Big{\}} \\ \nonumber
&=\mathbb{E}\big{\{}{|}{g}_n{|}^2\big{\}}\mathbb{V}{\text{ar}}\big{\{}{b}_{n,j-{\alpha}_n-1+{f}}
{|}H_{tk}\big{\}}=P_{\rm{a}}(\sigma_n^2+|\mu_n|^2) \eta_n p_n.
\end{align}
By substituting \eqref{iob5u403bx} and $\mathbb{E}
\big{\{}{g}_n{b}_{k,j-{\alpha}_n-1+f}{|}{g}_n,H_{tk}\big{\}}=0$, $f \in \{0,1\}$, into \eqref{BBxgt}, we obtain
\begin{align}\label{0m1sx7ub4602xwwwo}
\mathbb{V}\text{ar}\big{\{}{h}_{n,j,f}{|}H_{tk}\big{\}}
=P_{\rm{a}}(\sigma_n^2+|\mu_n|^2) \eta_n p_n.
\end{align}
Finally, by substituting \eqref{0m1sx7ub4602xo}, \eqref{0m1sx7ub4602xwwwo},  $\mathbb{V}\text{ar} {\{}{w}'_{k,j,f}{\}}={\Sigma}_{2k+f,2k+f}^{{{w}}'}$
into \eqref{1qax78nv4}, \eqref{RD7_1}
is derived.
For the cross-correlation of $\hat{{h}}_{k,j,0}$ and $\hat{{h}}_{k,j,1}$, we obtain \eqref{eq:case12}  at the top of next page, where
by substituting \eqref{0m1sx7ub4602xo} and \eqref{0m1sx7ub4602xwwwo} into \eqref{eq:case12}, and then by using
$\mathbb{E}{\{}{w}'_{k,j,0}({{{w}}'_{k,j,1}})^* $ $ {|}H_{tk}{\}} = \Sigma_{2k,2k+1}^{{{w}}'}$, results in
\eqref{eq:case1n}.

\setcounter{equation}{90}
\section{}\label{ap:beyesian ruledf}
By employing the \ac{mlr} test, the transmission state of the $k$th \ac{iot} device
is identified as active, i.e., ${d}_k=H_{1k}$, if
\begin{align}\nonumber
\frac{p\big{(}\breve{\bf{h}}_{k,j}|H_{1k}
\big{)}}{p\big{(}\breve{\bf{h}}_{k,j}|H_{0k}\big{)}}	
&=\frac{{2\pi|{\bf{C}}_{f,f}^{0k}|^{\frac{1}{2}}}
\text{exp}\bigg{(}-\frac{1}{2}{\breve{\bf{h}}_{k,j}^\dag}({{\bf{C}}_{f,f}^{1k}})^{-1}{\breve{\bf{h}}_{k,j}}
\bigg{)}}{{2\pi|{\bf{C}}_{f,f}^{1k}|^{\frac{1}{2}}}
\text{exp}\bigg{(}-\frac{1}{2}{\breve{\bf{h}}_{k,j}^\dag}({{\bf{C}}_{f,f}^{0k}})^{-1}{\breve{\bf{h}}_{k,j}}
\bigg{)}}> \lambda,
\end{align}
where $\lambda=(1-P_{\rm{a}})/P_{\rm{a}}$.
A canonical form of the above detector is given by \cite{kay1998fundamentals}
\begin{align}\label{90iojk}
\breve{\bf{h}}_{k,j}^\dag ({\bf{C}}_{f,f}^{0k})^{-1} ({\bf{C}}_{f,f}^{1k}) ({{\bf{C}}_{f,f}^{1k}}+{{\bf{C}}_{f,f}^{0k}})^{-1}\breve{\bf{h}}_{k,j}> \theta_k,
\end{align}
where $\theta_k$ is determined based on desirable false alarm rate for the $k$th \ac{iot} device.
Let us write
$
{{\bf{C}}_{f,f}^{0k}}= {\bf{V}}_{f,f}^{0k} {\bf{\Lambda}}_{f,f}^{0k} ({\bf{V}}_{f,f}^{0k})^{-1},
$
where ${\bf{V}}_{f,f}^{0k}$ is an square matrix whose columns are eigenvectors of ${{\bf{C}}_{f,f}^{0k}}$, and
${\bf{\Lambda}}_{f,f}^{0k}$ is a diagonal matrix where its $i$th diagonal element is the eigenvalue associated with the $i$th column of ${{\bf{C}}_{f,f}^{0k}}$.
We define ${\bf{A}}_{f,f}^{0k}={\bf{V}}_{f,f}^{0k} ({{\bf{\Lambda}}_{f,f}^{0k}})^{-\frac{1}{2}}$ and
${\bf{B}}_{f,f}^{1k} \triangleq
({\bf{A}}_{f,f}^{0k})^\dag {\bf{C}}_{f,f}^{1k} {\bf{A}}_{f,f}^{0k}$. Taking into account $({\bf{V}}_{f,f}^{0k})^\dag{\bf{V}}_{f,f}^{0k}={\bf{I}}$, we can show that
$({\bf{A}}_{f,f}^{0k})^\dag {{\bf{C}}_{f,f}^{0k}} {\bf{A}}_{f,f}^{0k}={\bf{I}}$.
Then, using this result, the canonical detector in
 \eqref{90iojk} can be written as follows
\begin{align}\label{cononic}
\breve{\bf{h}}_{k,j}^\dag {\bf{A}}_{f,f}^{0k}{\bf{B}}_{f,f}^{1k}({\bf{B}}_{f,f}^{1k}+{\bf{I}})^{-1}({\bf{A}}_{f,f}^{0k})^\dag\breve{\bf{h}}_{k,j}> \theta_k.
\end{align}
Based on eigenvalue decomposition of ${\mathbf{B}_{f,f}^{0k}}$, we have
\begin{align}
{\bf{B}}_{f,f}^{1k} \triangleq
({\bf{A}}_{f,f}^{0k})^\dag {\bf{C}}_{f,f}^{1k} {\bf{A}}_{f,f}^{0k} = {\bf{V}}_{f,f}^{1k} {\bf{\Lambda}}_{f,f}^{1k} ({\bf{V}}_{f,f}^{1k})^{-1},
\end{align}
where ${\bf{V}}_{f,f}^{1k}$ and ${\bf{\Lambda}}_{f,f}^{1k}$ are the eigenvector and eigenvalue matrices of ${\bf{B}}_{f,f}^{1k}$, respectively.
Since ${\bf{B}}_{f,f}^{1k}$ is a symmetric matrix, we have $({\bf{V}}_{f,f}^{1k})^\dag {\bf{V}}_{f,f}^{1k} ={\bf{I}}$.
By letting ${\bf{z}}_{k,j} \triangleq  [{\bf{z}}_{k,j}[0] , {\bf{z}}_{k,j}[1]]^\dag \triangleq ({\bf{V}}_{f,f}^{1k})^{\dag}({\bf{A}}_{f,f}^{0k})^\dag \breve{\bf{h}}_{k,j}$ in \eqref{cononic},
we obtain
\begin{align}
{\bf{z}}_{k,j}^\dag {{\bf{\Lambda}}_{f,f}^{1k}}({{\bf{\Lambda}}_{f,f}^{1k}}+{\bf{I}})^{-1}{\bf{z}}_{k,j}> \theta_k,
\end{align}
which is equivalent to the test statistics in
\eqref{eq:io9xzb7120m} and \eqref{u899kooD}.
Note that the matrix ${\bf{A}}_{f,f}^{0k} {\bf{V}}_{f,f}^{1k}$ diagonalizes both ${{\bf{C}}_{f,f}^{0k}}$ and ${{\bf{C}}_{f,f}^{1k}}$. 

By employing \eqref{eq:io9xzb7120m} and \eqref{u899kooD}, the false alarm rate for
the $kt$h \ac{iot} device is derived as follows
\begin{align}\label{uux56}
{P}_{k}^{(\rm{f})}&= \mathbb{P}\big{\{}{d}_k=H_{1k}\big{|}H_{0k}\big{\}} \\ \nonumber
&=\mathbb{P}\bigg{\{}
 \sum_{n=0}^{1}\chi_{f,f}[n]{z}_{k,j}^2[n] \geq \theta_{k}\big{|}H_{0k}\bigg{\}}.
\end{align}
To obtain the \ac{pdf} of
${U} \triangleq \sum_{n=0}^{1}\chi_{f,f}[n]{z}_{k,j}^2[n]$ in \eqref{uux56},
we need to derive its characteristic function (CF) and then express the \ac{pdf} as the inverse Fourier transform.
Since ${z}_{k,j}^2[n]$ in \eqref{eq:io9xzb7120m} under hypothesis $H_{0k}$
follows the central Chi-squared ($\chi^2$) distribution with $1$ degrees of freedom and the fact that ${z}_{k,j}^2[0]$ and
${z}_{k,j}^2[1]$ are independent random variables, we obtain the CF of ${U}$ as follows
\begin{align}
\phi_{U}(\omega)\triangleq \mathbb{E}\{\exp(j\omega {U})\}=\prod_{n=0}^{1}\frac{1}{\sqrt{1-2j\chi_{f,f}[n] \omega}},
\end{align}
where $\chi_{f,f}[n]$ is given \eqref{u899kooD}.
Taking the inverse Fourier transform of $\phi_{U}(\omega)$, we have
\begin{align}\label{7674494}
p_{{U}}(u|H_{0k})=\frac{1}{2 \pi} \int_{-\infty}^{+\infty}  \prod_{n=0}^{1}\frac{\exp(-j\omega u)}{\sqrt{1-2j\chi_{f,f}[n] \omega}}  {\rm{d}}\omega.
\end{align}
Using \eqref{7674494}, the false alarm rate for the $k$th \ac{iot} device is obtained as in \eqref{flase}.
Following the same procedure, the correct identification rate for the $k$th \ac{iot} device in \eqref{correct} is derived.

\end{document}